\documentclass[12pt]{article}
\usepackage[utf8]{inputenc}
\usepackage{amsmath}
\usepackage{amssymb}
\usepackage{enumitem}
\usepackage{amsfonts}
\usepackage{eurosym}
\usepackage{makeidx}
\usepackage{times}
\usepackage{anysize}
\usepackage{hyperref}
\usepackage{tikz}
\usepackage{pgfplots}

\makeatletter
\marginsize{2cm}{2cm}{1cm}{1cm}
\newtheorem{theorem}{Theorem}[section]

\newtheorem{corollary}[theorem]{Corollary}

\newtheorem{lemma}[theorem]{Lemma}

\newtheorem{proposition}[theorem]{Proposition}
\newtheorem{remark}[theorem]{Remark}

\newenvironment{proof}[1][Proof]{\noindent\textbf{#1.} }{\ \rule{0.5em}{0.5em}}
\makeatother

\begin{document}

\title{Non-Linear Operator-valued Elliptic Flows with Application to Quantum
Field Theory}
\author{J.-B. Bru \and N. M\'{e}traud}
\date{\today }
\maketitle

\begin{abstract}
Differential equations on spaces of operators are very little developed in
Mathematics, being in general very challenging. Here, we study a novel
system of such (non-linear) differential equations. We show it has a unique
solution for all times, for instance in the  Schatten norm
topology . This system presents remarkable ellipticity properties that
turn out to be crucial for the study of the infinite-time limit of its
solution, which is proven under relatively weak, albeit probably not
necessary, hypotheses on the initial data. This system of differential
equations is the elliptic counterpart of an hyperbolic flow applied to
quantum field theory to diagonalize Hamiltonians that are quadratic in the
bosonic field. In a similar way, this elliptic flow, in particular its
asymptotics, has application in quantum field theory: it can be used to
diagonalize Hamiltonians that are quadratic in the fermionic field while
giving new explicit expressions and properties of these pivotal Hamiltonians
of quantum field theory and quantum statistical mechanics.\bigskip

\bigskip{}

\end{abstract}

\tableofcontents%

\section{Introduction}

Ordinary differential equations (ODEs), and more generally, partial
differential equations (PDEs) have been studied for decades and form a
well-established, albeit still very active, research field of Mathematics.
For certain types of evolution equations involving $L^{p}$-valued functions
many good results are available.

Differential equations on non-commutative spaces are much less known. The
same situation appears in probability theory. An example is given by the
celebrated Polchinski equation on Grassmann algebras, which can meanwhile be
related to a fresh development of Grassmannian stochastic analysis \cite%
{Grassmannianstochasticanalysis} (cf. algebraic probability /
non-commutative or quantum probability). Probability theory on operator
spaces usually refers to the celebrated random Schr\"{o}dinger operators 
\cite{RandomShrodingeroperators}, but differential equations on
non-commutative algebras of operators (like the Brockett-Wegner flow \cite%
{Brockett1,Wegner1}) are very little developed in Mathematics, apart
from non-autonomous evolution equations on Banach spaces (see, e.g., \cite%
{Katobis,Caps,Schnaubelt1,Pazy,Neidhardt-zagrebnov}). In fact, differential
equations on operators involve a priori a high degree of mathematical
complexity, in particular for unbounded ones\footnote{%
For instance, for bounded generators, the existence, uniqueness and even an
explicit form of the solution to non-autonomous evolution equations are
given via the Dyson series, as is well-known. It is much more delicate for
unbounded generators. Yet, after decades of studies, no unified theory of
such linear evolution equations that gives a complete characterization
analogously to the Hille-Yosida generation theorems is known.}. Hence, any
mathematical result on relevant differential equations for operators can be useful for developing this field.

In this paper, we study a novel (non-autonomous, non-linear) system of ODEs
that involves operator-valued functions $\left( \Upsilon _{t}\right) _{t\geq
0}$ and $\left( D_{t}\right) _{t\geq 0}$, defined by%
\begin{equation}
\left\{ 
\begin{array}{llllll}
\partial _{t}\Upsilon _{t}=16D_{t}D_{t}^{\ast } & , & \quad \Upsilon
_{t=0}\doteq \Upsilon _{0} & , & \quad t\in \mathbb{R}_{0}^{+}\doteq \lbrack
0,\infty ) & , \\ 
\partial _{t}D_{t}=-2\left( \Upsilon _{t}D_{t}+D_{t}\Upsilon _{t}^{\top
}\right) & , & \quad D_{t=0}\doteq D_{0} & , & \quad t\in \mathbb{R}%
^{+}\doteq (0,\infty ) & ,%
\end{array}%
\right.  \label{flow equation-quadratic delta}
\end{equation}%
in the strong (operator) topology (on the dense domain $\mathcal{D}(\Upsilon
_{0}^{\top })$ for the second evolution equation), where $\Upsilon _{0}\geq
-\mu \mathbf{1}$, with $\mu \in \mathbb{R}$, and $D_{0}=\pm D_{0}^{\top }$
are two operators acting on a separable (complex) Hilbert space $\mathfrak{h}
$, $D_{0}$ being bounded (i.e., $D_{0}\in \mathcal{B}\left( \mathfrak{h}%
\right) $). Here, $A^{\top }$ is the transpose of an operator $A$. Note in
passing that this flow is intrinsically related to non-autonomous evolution
equations (see, e.g., Sections \ref{Diagonalization of Quadratic} and \ref%
{Section tech1}).

The \emph{positive} constants $16$ and $2$ of the non-autonomous, non-linear
system (\ref{flow equation-quadratic delta}) of ODEs are unimportant for the
study of solutions to (\ref{flow equation-quadratic delta}), thanks to a
simple rescalling argument\footnote{%
For instance, if $(\Upsilon _{t})_{t\geq 0}$ and $(D_{t})_{t\geq 0}$ solve (%
\ref{flow equation-quadratic delta}), then, for any $\alpha ,\beta \in 
\mathbb{R}^{+}$, the operators $\tilde{\Upsilon}_{t}\doteq 2\beta
^{-1}\Upsilon _{t}$ and $\tilde{D}_{t}\doteq \sqrt{32\beta ^{-1}\alpha ^{-1}}%
D_{t}$ for $t\geq 0$ satisfy $\partial _{t}\tilde{\Upsilon}_{t}=\alpha 
\tilde{D}_{t}\tilde{D}_{t}^{\ast }$ and $\partial _{t}\tilde{D}_{t}=-\beta (%
\tilde{\Upsilon}_{t}\tilde{D}_{t}+\tilde{D}_{t}\tilde{\Upsilon}_{t}^{\top })$%
.}. They are only taken for convenience, in view of our applications to
quantum field theory. Indeed, besides its elegant properties that are
mathematically interesting in their own right, the flow (\ref{flow
equation-quadratic delta}) have salient applications to the diagonalization
of (generally unbounded) many-body Hamiltonians which are quadratic in the
quantum fields, by using the formal Brockett-Wegner flow \cite%
{Brockett1,Wegner1,bach-bru} as a guideline. For instance, we show in \cite%
{QuadraticFermionic} that the well-posedness and asymptotics of the flow (%
\ref{flow equation-quadratic delta}) for Hilbert-Schmidt initial data $D_{0}$
can lead to the $\mathrm{N}$--diagonalization of fermionic quadratic
Hamiltonians under much more general conditions than before (cf. Theorem \ref%
{thm3}). In Section \ref{Diagonalization of Quadratic}\ we shortly explain
this application, starting, for the reader's convenience, with elementary
considerations on quantum field theory and quantum statistical mechanics.

Our analysis is reminiscent of the monograph \cite{bach-bru-memo}, which
studies a very similar system of ODEs on two families $(\Omega _{t})_{t\geq
0}$ and $(B_{t})_{t\geq 0}$ of operators acting on a separable Hilbert space 
$\mathfrak{h}$.\ The aim in \cite{bach-bru-memo} is the $\mathrm{N}$%
--diagonalization of bosonic quadratic Hamiltonians, in contrast with the
fermionic case \cite{QuadraticFermionic}. In this situation, it is defined
by 
\begin{equation}
\left\{ 
\begin{array}{llllll}
\partial _{t}\Omega _{t}=-16B_{t}B_{t}^{\ast } & , & \quad \Omega
_{t=0}\doteq \Omega _{0} & , & \quad t\in \mathbb{R}_{0}^{+} & , \\ 
\partial _{t}B_{t}=-2\left( \Omega _{t}B_{t}+B_{t}\Omega _{t}^{\top }\right)
& , & \quad B_{t=0}\doteq B_{0} & , & \quad t\in \mathbb{R}^{+} & ,%
\end{array}%
\right.  \label{flow equation-quadratic}
\end{equation}%
in the strong topology (on the dense domain $\mathcal{D}(\Omega _{0}^{\top
}) $ for the second evolution equation), where $\Omega _{0}=\Omega
_{0}^{\ast }\geq 0$ and $B_{0}=B_{0}^{\top }$ are two operators acting on $%
\mathfrak{h}$. Note that $\Omega _{0}$ is a possibly unbounded self-adjoint
operator, while $B_{0}$ is a Hilbert-Schmidt operator in \cite{bach-bru-memo}%
.

The global existence of a solution to (\ref{flow equation-quadratic})
requires a positive operator $\Omega _{0}$, by \cite[Theorem 11]%
{bach-bru-memo}. This is not anymore necessary for the global existence of a
solution to (\ref{flow equation-quadratic delta}) and the initial
self-adjoint operator $\Upsilon _{0}$ only needs to be lower semibounded,
i.e., $\Upsilon _{0}\geq -\mu \mathbf{1}$ with $\mu \in \mathbb{R}$. This is
proven in Theorem \ref{lemma existence 2 copy(1)} for semibounded,
self-adjoint, operators $\Upsilon _{0}\ $and (non-zero) bounded operator $%
D_{0}$, both acting on the separable complex Hilbert space $\mathfrak{h}$.
See also Equation (\ref{sdssds}). In the case the initial data $D_{0}$ has
finite Schatten $2p$--norm for some $p\in \lbrack 1,\infty ]$, the
operator-valued flow (\ref{flow equation-quadratic delta}) can also be seen
as a well-posed system of ODEs on the (noncommutative) $L^{p}$--spaces
constructed from Schatten norms, with in this case $(D_{t})_{t\in \mathbb{R}%
_{0}^{+}}\in \mathcal{L}^{2p}(\mathfrak{h})$ and $(\Delta _{t}\doteq
\Upsilon _{t}-\Upsilon _{0})_{t\in \mathbb{R}_{0}^{+}}\in \mathcal{L}^{p}(%
\mathfrak{h})$, thanks to Theorems \ref{lemma existence 2 copy(4)} and \ref%
{Corollary existence copy(1)}.

The systems (\ref{flow equation-quadratic delta}) and (\ref{flow
equation-quadratic}) only differ by a sign in the first evolution equations
on $\Upsilon _{t}$ and $\Omega _{t}$, respectively. This difference is
pivotal because it determines the elliptic or hyperbolic natures of the
flow: By \cite[Theorem 12]{bach-bru-memo}, the flow (\ref{flow
equation-quadratic}) is an \emph{hyperbolic} system of evolution equations
in the sense that if the initial data satisfy\footnote{%
In this condition, the operator $\Omega _{0}^{\top }$ is assumed to be
invertible on $\mathrm{Ran}B_{0}^{\ast }$.} 
\begin{equation}
\Omega _{0}\geq (4+\mathrm{r})B_{0}(\Omega _{0}^{\top })^{-1}B_{0}^{\ast
}+\mu \mathbf{1}\ ,  \label{assumption bis}
\end{equation}%
for some $\mu ,\mathrm{r}\geq 0$, then the relation is conserved for all
times by the flow: 
\begin{equation}
\Omega _{t}\geq (4+\mathrm{r})B_{t}(\Omega _{t}^{\top })^{-1}B_{t}^{\ast
}+\mu \mathbf{1}\ ,\qquad t\in \mathbb{R}_{0}^{+}\ .
\label{condition hyperbol}
\end{equation}%
Note that Assumption (\ref{assumption bis}) for $\mu =\mathrm{r}=0$ is
crucial for the global existence of a solution to the system (\ref{flow
equation-quadratic}), see \cite[Theorem 11 and Condition A3]{bach-bru-memo}.
This condition (formally) defines a region of the space of operators $%
(\Omega ,B)$ that is enclosed by an \emph{hyperbola}. For instance, in the
one-dimensional case, $\Omega _{t}=x\in \mathbb{R}^{+}$, $B_{0}B_{0}^{\ast
}=y^{2}\in \mathbb{R}^{+}$ ($y\neq 0$ being the non-trivial case for the
flow) and (\ref{condition hyperbol}) reads as 
\begin{equation*}
\left( 2x-\mu \right) ^{2}-4(4+\mathrm{r})y^{2}\geq \mu ^{2}
\end{equation*}%
for some fixed positive constants $\mu ,\mathrm{r}\geq 0$.

The operator flow (\ref{flow equation-quadratic delta}) is the \emph{elliptic%
} counterpart of the system (\ref{flow equation-quadratic}) of evolution
equations. See Equations (\ref{constant of motion eq1})--(\ref{constant of
motion eq2}), Theorems \ref{lemma constante of motion copy(6)}, \ref{theorem
sympa0}, \ref{lemma constante of motion copy(3)} and \ref{Theorem sympa} as
well as Corollary \ref{lemma constante of motion copy(4)}. For instance, we
show in Theorem \ref{lemma constante of motion copy(6)} that the condition 
\begin{equation}
\Upsilon _{0}-\mu \mathbf{1}+4D_{0}\left( \Upsilon _{0}^{\top }+\mu \mathbf{1%
}\right) ^{-1}D_{0}^{\ast }\leq \mathrm{C}\mathbf{1}  \label{ssssss}
\end{equation}%
for some fixed constant $\mathrm{C}\geq 0$ and appropriate $\mu \in \mathbb{R%
}$, is conserved by the flow (\ref{flow equation-quadratic delta}):\ 
\begin{equation}
\Upsilon _{t}-\mu \mathbf{1}+4D_{t}\left( \Upsilon _{t}^{\top }+\mu \mathbf{1%
}\right) ^{-1}D_{t}^{\ast }\leq \mathrm{C}\mathbf{1}\ ,\qquad t\in \mathbb{R}%
_{0}^{+}\ .  \label{dddsdssdsdbis}
\end{equation}%
This is the equivalent of the interior of an \emph{ellipse} for operators $%
\Upsilon _{0}$ and $D_{0}$. For example, in the (non-trivial)
one-dimensional case, $\Upsilon _{0}=x\in \mathbb{R}$, $D_{0}D_{0}^{\ast
}=y^{2}\in \mathbb{R}^{+}$ and (\ref{dddsdssdsdbis}) for $\mu >-x$ reads as 
\begin{equation}
\left( 2x-\mathrm{C}\right) ^{2}+16y^{2}\leq \left( 2\mu +\mathrm{C}\right)
^{2}\ .  \label{eliptic condition}
\end{equation}%
Note that (\ref{ssssss}) implies the boundedness of the initial operator $%
\Upsilon _{0}$, but more general results can be obtained for lower
semibounded $\Upsilon _{0}$, see Theorem \ref{lemma constante of motion
copy(6)} (ii) (or Theorem \ref{theorem sympa0}) as well as Theorems \ref%
{lemma constante of motion copy(3)} (ii) and \ref{Theorem sympa}. 

An important issue in such flows concerns their asymptotics for infinite
times. This analysis can be highly non-trivial for the hyperbolic case \cite%
{bach-bru-memo} and we perform an analogous study\ for the elliptic
situation analyzed here. In this case, Theorem \ref{lemma asymptotics1
copy(4)} demonstrates that, if $D_{0}^{\top }=\pm D_{0}$, the operator
families $\left( D_{t}\right) _{t\geq 0}$ and $\left( \Upsilon _{t}\right)
_{t\geq 0}$ can converge in various topologies\footnote{%
I.e., the weak operator topology, the norm topology, the Hilbert-Schmidt
topology and the trace norm topology.} to respectively $D_{\infty }\doteq 0$
and 
\begin{equation*}
\Upsilon _{\infty }\doteq \Upsilon _{0}+16\int_{0}^{\infty }D_{\tau }D_{\tau
}^{\ast }\mathrm{d}\tau \ .
\end{equation*}%
The most important situation in view of applications to quantum field theory
concerns the case of Hilbert-Schmidt initial data $D_{0}$. In this case, by
Theorem \ref{lemma asymptotics1 copy(4)} (iii), we show that $\left(
D_{t}\right) _{t\geq 0}$ and $\left( \Upsilon _{t}\right) _{t\geq 0}$ can
converge exponentially fast in the Hilbert-Schmidt topology to $D_{\infty
}\doteq 0$ and $\Upsilon _{\infty }$, as $t\rightarrow \infty $. In this
case, $(\Upsilon _{\infty }-\Upsilon _{0})$ is a Hilbert-Schmidt operator.

These last asymptotic results with Hilbert-Schmidt initial data $D_{0}$ are
based on the additional assumption 
\begin{equation}
\Upsilon _{0}-\mu \mathbf{1}+4D_{0}\left( \Upsilon _{0}^{\top }+\mu \mathbf{1%
}\right) ^{-1}D_{0}^{\ast }\geq 0\ .  \label{kkkkkk}
\end{equation}%
This is the equivalent of the outside of an ellipse for operators $\Upsilon
_{0}$ and $D_{0}$ (up to the singular case $\mu =0$). See Theorem \ref{lemma
constante of motion copy(3)} (ii). Therefore, elliptic properties of the
flow are pivotal to get its asymptotics.

Last but not least, if we also have $\Upsilon _{0}D_{0}=D_{0}\Upsilon
_{0}^{\top }$ together with either a Hilbert-Schmidt initial data $D_{0}$
satisfying (\ref{kkkkkk}) or $\Upsilon _{0}\geq \alpha \mathbf{1}$ for some
strictly positive parameter $\alpha >0$, then $\Upsilon _{\infty }$ can be 
\emph{fully} characterized since in this case%
\begin{equation*}
\Upsilon _{\infty }=\sqrt{\Upsilon _{0}^{2}+4D_{0}D_{0}^{\ast }}
\end{equation*}%
on the domain $\mathcal{D}(\Upsilon _{0})$. See below Equation (\ref%
{constant of motion eq2bis}).

To conclude, our main results are Theorems \ref{lemma existence 2 copy(1)}, %
\ref{lemma existence 2 copy(4)} on the well-posedness of the elliptic flow
as well as Theorem \ref{lemma asymptotics1 copy(4)} on its asymptotics.\
 Theorem \ref{Corollary existence copy(1)} on the well-posedness of
the flow within $L^{p}$--spaces of operators endowed with Schatten norms is
also interesting, but it uses a bounded initial operator $\Upsilon _{0}$,
which is a restrictive assumption.  Ellipticity properties of the flow
are demonstrated in various places, as in Theorem \ref{lemma constante of
motion copy(6)} and \ref{Theorem sympa}, and constant of motions can be
obtained, see, e.g., Equations (\ref{constant of motion eq1})--(\ref%
{constant of motion eq2}).

The paper is organized as follows: Section \ref{Eliptic flow} analyzes the
elliptic operator-valued flow, followed by a concise explanation of its
application to many-fermion theory. Most of technical results on the
elliptic operator-valued flow explained in Section \ref{Eliptic flow} are
gathered in Section \ref{Technical Results}.

\section{The Elliptic Operator-Valued Flow\label{Eliptic flow}}

\subsection{Notation and Preliminary Definitions}

 We denote the set of non-negative real numbers by $\mathbb{R}%
_{0}^{+}\doteq \lbrack 0,\infty )$, while $\mathbb{R}^{+}\doteq (0,\infty )$%
.  Let $\mathfrak{h}$ be any separable complex Hilbert space $\mathfrak{h}
$ endowed with a complex conjugation $\mathcal{C}$, which is an antiunitary
involution on $\mathfrak{h}$, i.e., an antilinear mapping $\mathfrak{h}%
\rightarrow \mathfrak{h}$ such that $\mathcal{C}^{2}=\mathbf{1}$ and 
\begin{equation*}
\left\langle \mathcal{C}f,\mathcal{C}g\right\rangle _{\mathfrak{h}%
}=\left\langle g,f\right\rangle _{\mathfrak{h}}\ ,\qquad f,g\in \mathfrak{h}%
\ .
\end{equation*}%
See, e.g., \cite[Chapter 7]{Bru-Pedra-livre}. Then, for any operator $X$ on $%
\mathfrak{h}$, we define its transpose $X^{\top }$ and its complex conjugate 
$\bar{X}$ by 
\begin{equation*}
X^{\top }=\mathcal{C}X^{\ast }\mathcal{C}\qquad \text{and}\qquad \overline{X}%
=\mathcal{C}X\mathcal{C}\ .
\end{equation*}%
Note that $X\geq 0$ iff\footnote{%
For instance, $X\geq 0$ for some operator $X$ with domain $\mathcal{D}\left(
X\right) \subseteq \mathfrak{h}$ implies that $X^{\top }=(X^{\top })^{\ast }$
and for any $f\in \mathcal{D}(X^{\top })$,%
\begin{equation*}
\left\langle f|X^{\top }f\right\rangle _{\mathfrak{h}}=\left\langle f|%
\mathcal{C}X^{\ast }\mathcal{C}f\right\rangle _{\mathfrak{h}}=\left\langle
X^{\ast }\mathcal{C}f|\mathcal{C}f\right\rangle _{\mathfrak{h}}=\left\langle 
\mathcal{C}f|X\mathcal{C}f\right\rangle _{\mathfrak{h}}\geq 0\mathbf{\ .}
\end{equation*}%
In other words, $X\geq 0$ iff $X^{\top }\geq 0$.} $X^{\top }\geq 0$ and the
adjoint of an operator $X$ equals 
\begin{equation*}
X^{\ast }=\overline{X^{\top }}=\overline{X}^{\top },
\end{equation*}%
where it exists.

For instance, $\mathfrak{h}$ can be realized as a space $L^{2}(\mathcal{M})$
of square-integrable, complex-valued functions on a measure space $(\mathcal{%
M},\mathfrak{a})$ and $\mathcal{C}$ can be the antiunitary involution
defined for each square-integrable function $f\in L^{2}(\mathcal{M})$ by its
complex conjugate $\mathcal{C}f=\bar{f}$, where $\bar{f}\left( x\right)
\doteq \overline{f\left( x\right) }$ for all $x\in \mathcal{M}$. In this
case, for any bounded operator $X$ on $L^{2}(\mathcal{M})$, its transpose $%
X^{\top }$ and its complex conjugate $\bar{X}$ can be defined by the
conditions 
\begin{equation*}
\langle f,X^{\top }g\rangle _{\mathfrak{h}}=\langle \bar{g},X\bar{f}\rangle
_{\mathfrak{h}}\qquad \text{and}\qquad \langle f,\bar{X}g\rangle _{\mathfrak{%
h}}=\overline{\langle \bar{f},X\bar{g}\rangle _{\mathfrak{h}}}
\end{equation*}%
for square-integrable functions $f,g\in L^{2}(\mathcal{M})$.

The Banach space of bounded operators acting on $\mathfrak{h}$ is denoted by 
$\mathcal{B}(\mathfrak{h})\equiv \mathcal{L}^{\infty }(\mathfrak{h})$,
while $\mathcal{L}^{p}(\mathfrak{h})$, $p\in \lbrack 1,\infty )$, are the
(noncommutative) $L^{p}$--spaces constructed from Schatten $p$--norms. In
particular, $\mathcal{L}^{1}(\mathfrak{h})$\ and $\mathcal{L}^{2}(\mathfrak{h%
})$ are the spaces of trace-class and Hilbert-Schmidt operators,
respectively. Norms in $\mathcal{B}(\mathfrak{h})$ and $\mathcal{L}^{p}(%
\mathfrak{h})$ for $p\in \lbrack 1,\infty )$ are denoted by 
\begin{equation*}
\left\Vert X\right\Vert _{\mathrm{op}}\equiv \left\Vert X\right\Vert
_{\infty }\doteq \sup_{f\in \mathfrak{h}:\left\Vert f\right\Vert _{\mathfrak{%
h}}=1}\left\Vert Xf\right\Vert _{\mathfrak{h}}\mathrm{\quad }\text{and}%
\mathrm{\quad }\left\Vert X\right\Vert _{p}\doteq \left( \mathrm{tr}\left(
\left\vert X\right\vert ^{p}\right) \right) ^{1/p},
\end{equation*}%
for $X$ in $\mathcal{B}(\mathfrak{h})$ and $\mathcal{L}^{p}(\mathfrak{h})$,
respectively, where $|X|\doteq \sqrt{X^{\ast }X}$ and $\mathrm{tr}(\cdot )$
denotes the usual trace for operators.   The set of positive operators is
denoted by 
\begin{equation*}
\mathcal{B}\left( \mathfrak{h}\right) ^{+}\doteq \left\{ X\in \mathcal{B}%
\left( \mathfrak{h}\right) :X\geq 0\right\} =\left\{ YY^{\ast }:Y\in 
\mathcal{B}\left( \mathfrak{h}\right) \right\} \ .
\end{equation*}%
The notation $\mathbf{1}$ is always the identity operator and $C\left( I;%
\mathcal{Y}\right) $ denotes the set of continuous functions from a closed
set $I\subseteq \mathbb{R}$ to a closed subset $\mathcal{Y\subseteq X}$ of a
Banach space $\left( \mathcal{X},\left\Vert \cdot \right\Vert _{\mathcal{X}%
}\right) $, along with the norm 
\begin{equation*}
\left\Vert X\right\Vert _{\infty }\doteq \sup_{t\in I}\left\Vert
X_{t}\right\Vert _{\mathcal{X}}\ ,\qquad X\in C\left( I;\mathcal{X}\right)
\supseteq C\left( I;\mathcal{Y}\right) \ .
\end{equation*}%
As is well-known, $C\left( I;\mathcal{X}\right) $ is in this case a Banach
space. Finally, $\rho (X)\subseteq \mathbb{C}$ and $\sigma (X)=\mathbb{C}%
\backslash \rho (X)$ denote respectively the resolvent set and the spectrum
of an operator $X$.

\begin{remark}
\mbox{}\newline
We use below a complex Hilbert space $\mathfrak{h}$ that is separable for
convenience. In fact, a large part (but not all) of results on the flow
obtained here can be proven without the separability of $\mathfrak{h}$. By
contrast, in the application to quantum field theory (see Section \ref%
{Diagonalization of Quadratic}) we generally need this property via the use
of the trace and Hilbert-Schmidt norms.
\end{remark}

\subsection{Well-Posedness of the Operator-Valued Flow\label{Well--Posedness
of the Flow}}

We study the system (\ref{flow equation-quadratic delta}) of differential
equations for semibounded, self-adjoint, operators $\Upsilon _{0}\geq -\mu 
\mathbf{1}$, with $\mu \in \mathbb{R}$, and (non-zero) bounded operator $%
D_{0}\in \mathcal{B}(\mathfrak{h})$, both acting on the Hilbert space $%
\mathfrak{h}$. In this case, keeping in mind that the family $(\Upsilon
_{t})_{t\geq 0}$ of possibly unbounded operators in (\ref{flow
equation-quadratic delta}) can be written as 
\begin{equation}
\Upsilon _{t}=\Upsilon _{0}+\Delta _{t}\ ,\qquad t\in \mathbb{R}_{0}^{+}\ ,
\label{sdssds}
\end{equation}%
instead of (\ref{flow equation-quadratic delta}), it is more convenient to
consider the following system of differential equations: 
\begin{equation}
\left\{ 
\begin{array}{llll}
\partial _{t}\Delta _{t}\varphi =16D_{t}D_{t}^{\ast }\varphi \ , & \Delta
_{t=0}\doteq 0\ , & t\in \mathbb{R}_{0}^{+}\ , & \varphi \in \mathfrak{h}\ .
\\ 
\partial _{t}D_{t}\varphi =-2\left( \left( \Delta _{t}+\Upsilon _{0}\right)
D_{t}+D_{t}\left( \Delta _{t}+\Upsilon _{0}\right) ^{\top }\right) \varphi \
, & D_{t=0}\doteq D_{0}\ , & t\in \mathbb{R}^{+}\ , & \varphi \in \mathcal{D}%
(\Upsilon _{0}^{\top })\ .%
\end{array}%
\right.  \label{flow equation-quadratic deltabis}
\end{equation}%
Here, $\mathcal{D}(\Upsilon _{0}^{\top })\subseteq \mathfrak{h}$ is the
(dense) domain of the self-adjoint operator $\Upsilon _{0}^{\top }$. Note
that the use of (\ref{sdssds}) is advantageous since $(\Delta _{t})_{t\geq
0}\subseteq \mathcal{B}(\mathfrak{h})$ for $D_{t}\in \mathcal{B}(\mathfrak{h}%
)$, in contrast with $(\Upsilon _{t})_{t\geq 0}$ which is only bounded from
below, in general.

The local existence of a solution to the operator-valued flow (\ref{flow
equation-quadratic deltabis}) is proven like in the hyperbolic case studied
in \cite[Lemma 34 and Corollary 37]{bach-bru-memo}. However, its global
existence requires different arguments: In the hyperbolic flow (\ref{flow
equation-quadratic}), one uses the conservation of Assumption (\ref%
{assumption bis}) at $\mu =\mathrm{r}=0$ for all times in order to ensure
the positivity of the operators $\Omega _{t}\geq 0$ and the decrease of the
mapping $t\mapsto \left\Vert B_{t}\right\Vert _{\mathrm{op}}$ on positive
times. This last property is important for the global existence since the
contraction mapping principle works a priori only for small times. These
studies were complicated in the hyperbolic case and they have to be adapted
to the elliptic case.

Here, we trivially have $\Upsilon _{t}\geq \Upsilon _{0}$ for all times $%
t\in \mathbb{R}_{0}^{+}$, by positivity of the operator $\Delta _{t}$, and
thus, $\Upsilon _{0}\geq 0$ directly yields $\Upsilon _{t}\geq 0$. However,
our assumptions are now more general in the sense that the initial operator,
here $\Upsilon _{0}$,\ is \emph{not} necessarily positive like in the
hyperbolic case (where $\Omega _{0}\geq 0$), but only lower semibounded,
i.e., $\Upsilon _{0}\geq -\mu \mathbf{1}$ with $\mu \in \mathbb{R}$. The
mapping $t\mapsto \left\Vert D_{t}\right\Vert _{\mathrm{op}}$ on positive
times is therefore not always decreasing, as one can explicitly see already
in dimension 1. See Section \ref{Flows on Scalar Fields} and Figure \ref{fig}%
. However, its (possibly exponential) increase can still be controlled by
using Gr\"{o}nwall's lemma. The well-posedness of the operator-valued flow
then follows for all times:

\begin{theorem}[Well-posedness of the flow -- Strong topology]
\label{lemma existence 2 copy(1)}\mbox{}\newline
Assume $\Upsilon _{0}=\Upsilon _{0}^{\ast }\geq -\mu \mathbf{1}$ with $\mu
\in \mathbb{R}$ and $D_{0}\in \mathcal{B}\left( \mathfrak{h}\right) $ ($%
D_{0}\neq 0$). Then, there exists a unique solution $\Delta \equiv (\Delta
_{t})_{t\geq 0},D\equiv (D_{t})_{t\geq 0}$ of strongly continuous mappings
on $\mathcal{B}\left( \mathfrak{h}\right) $ to the initial value problem (%
\ref{flow equation-quadratic deltabis}). These operator-valued functions
additionally satisfy the following properties:

\begin{enumerate}
\item[\emph{(i)}] $\Delta \in C\left( \mathbb{R}_{0}^{+};\mathcal{B}(%
\mathfrak{h})^{+}\right) $ is locally Lipschitz continuous.

\item[\emph{(ii)}] $\left( D_{t}\right) _{t>0}\in C\left( \mathbb{R}^{+};%
\mathcal{B}(\mathfrak{h})\right) $ is locally Lipschitz continuous with $%
D_{t>0}\mathfrak{h}\subseteq \mathcal{D}(\Upsilon _{0})$. If $D_{0}^{\top
}=\pm D_{0}$, then $D_{t}^{\top }=\pm D_{t}$ for any $t\in \mathbb{R}%
_{0}^{+} $.
\end{enumerate}
\end{theorem}

\begin{proof}
This assertion refers to Theorem \ref{section extension gap}, the complete
proof of which is performed in Sections \ref{Section tech1}--\ref{Section
tech2}. In particular, the local existence of a solution to (\ref{flow
equation-quadratic delta}) is proven in Lemmata \ref{lemma existence 2
copy(2)} and \ref{lemma existence 2}, while the global existence uses the
blow-up alternative expressed in the proof of Theorem \ref{section extension
gap} together with Gr\"{o}nwall's lemma. Note that the uniqueness is given
 by Corollary \ref{coro uniqueness}. 
\end{proof}

In other words, Theorem \ref{lemma existence 2 copy(1)} gives a unique
solution to the operator-valued flow (\ref{flow equation-quadratic deltabis}%
) or (\ref{flow equation-quadratic delta}), keeping in mind (\ref{sdssds})
for any lower semibounded operator $\Upsilon _{0}$ acting on the Hilbert
space $\mathfrak{h}$ and each bounded operator $D_{0}$. In the case the
initial data $D_{0}$ is a Hilbert-Schmidt operator we can strongly
strengthen the well-posedness of the operator-valued flow by seeing it as a
system of differential equations on the Hilbert space $\mathcal{L}^{2}(%
\mathfrak{h})$ of Hilbert-Schmidt operators acting on $\mathfrak{h}$: \ 

\begin{theorem}[Well-posedness of the flow -- Hilbert-Schmidt topology]
\label{lemma existence 2 copy(4)}\mbox{}\newline
Assume $\Upsilon _{0}=\Upsilon _{0}^{\ast }\geq -\mu \mathbf{1}$ with $\mu
\in \mathbb{R}$ and $D_{0}\in \mathcal{L}^{2}(\mathfrak{h})$ ($D_{0}\neq 0$%
). Then, there exists a unique solution  $(\Delta ,D)\in C(\mathbb{R}%
_{0}^{+};\mathcal{L}^{1}(\mathfrak{h})\times \mathcal{L}^{2}(\mathfrak{h}))$
to the system of differential equations 
\begin{equation*}
\left\{ 
\begin{array}{llllll}
\partial _{t}\Delta _{t}=16D_{t}D_{t}^{\ast } & , & \quad \Delta
_{t=0}\doteq 0 & , & \quad t\in \mathbb{R}_{0}^{+} & , \\ 
\partial _{t}D_{t}=-2\left( \left( \Delta _{t}+\Upsilon _{0}\right)
D_{t}+D_{t}\left( \Delta _{t}+\Upsilon _{0}\right) ^{\top }\right) & , & 
\quad D_{t=0}\doteq D_{0} & , & \quad t\in \mathbb{R}^{+} & ,%
\end{array}%
\right.
\end{equation*}%
in $\mathcal{L}^{1}(\mathfrak{h})$ and $\mathcal{L}^{2}(\mathfrak{h})$,
respectively.
\end{theorem}

\begin{proof}
This assertion is proven in Section \ref{Section tech3} and refers to
Theorem \ref{Corollary existence}. Note that the uniqueness of the solution
is deduced from Theorem \ref{lemma existence 2 copy(1)}.
\end{proof}

 The Hilbert-Schmidt case given in Theorem \ref{lemma existence 2
copy(4)} is the important situation in view of the application of the flow
in quantum field theory, described in\ Section \ref{Diagonalization of
Quadratic}. This result can be generalized to the $L^{p}$--spaces $\mathcal{L%
}^{p}(\mathfrak{h})$, $p\in \lbrack 1,\infty ]$, constructed from Schatten
norms. Indeed, if $D_{0}\in \mathcal{L}^{2p}(\mathfrak{h})$ for some $p\in
\lbrack 1,\infty ]$, we expect the well-posedness of the operator-valued
flow by seeing it as a system of differential equations on the Banach space $%
\mathcal{L}^{p}\left( \mathfrak{h}\right) \times \mathcal{L}^{2p}\left( 
\mathfrak{h}\right) $. This is proven in Theorem \ref{Corollary existence
copy(1)} under the stronger condition that $\Upsilon _{0}\in \mathcal{B}(%
\mathfrak{h})$ is a bounded operator.  

\subsection{Ellipticity and Asymptotics of the Flow\label{Well--Posedness of
the Flow copy(1)}}

For any self-adjoint operator $\Upsilon _{0}\in \mathcal{B}\left( \mathfrak{h%
}\right) $ and every non-zero Hilbert-Schmidt operator $D_{0}=\pm
D_{0}^{\top }\in \mathcal{L}^{2}(\mathfrak{h})$, one can check that 
\begin{equation}
\mathrm{tr}\left( \Upsilon _{t}^{2}+4D_{t}D_{t}^{\ast }-\Upsilon
_{s}^{2}-4D_{s}D_{s}^{\ast }\right) =0\ ,\qquad s,t\in \mathbb{R}_{0}^{+}\ ,
\label{constant of motion eq1}
\end{equation}%
where $\Upsilon _{t}\doteq \Upsilon _{0}+\Delta _{t}$ for any $t\in \mathbb{R%
}_{0}^{+}$.  The proof of this statement is given by Theorem \ref%
{theorem constante of motion} (i), which is deliberately restricted to
bounded $\Upsilon _{0}\in \mathcal{B}\left( \mathfrak{h}\right) $ in order
to make clear the key arguments without too much technical issues. We do not
see a priori any obstruction to extend it in the unbounded case by following
the lines of arguments of Theorem \ref{theorem constante of motion} (ii).

Moreover, in the case that the initial operator $D_{0}=\pm D_{0}^{\top }\in 
\mathcal{B}\left( \mathfrak{h}\right) $\ is only bounded but satisfies $%
\Upsilon _{0}D_{0}=D_{0}\Upsilon _{0}^{\top }$ on the domain $\mathcal{D}%
(\Upsilon _{0}^{\top })\subseteq \mathfrak{h}$ of $\Upsilon _{0}^{\top }$
with 
\begin{equation}
\max \left\{ \Vert (\Upsilon _{0}+\mu \mathbf{1)}D_{0}(\Upsilon _{0}^{\top
}+\mu \mathbf{1)}^{-1}\Vert _{\mathrm{op}},\Vert (\Upsilon _{0}^{\top }+\mu 
\mathbf{1)}D_{0}^{\ast }(\Upsilon _{0}+\mu \mathbf{1)}^{-1}\Vert _{\mathrm{op%
}}\right\} \leq \mathrm{C}  \label{condition67}
\end{equation}
for some strictly positive constant $\mathrm{C}\in \mathbb{R}^{+}$, then,
for any $t,s\in \mathbb{R}_{0}^{+}$, 
\begin{equation}
\Upsilon _{t}D_{t}=D_{t}\Upsilon _{t}^{\top }\qquad \text{and}\qquad
\Upsilon _{t}^{2}+4D_{t}D_{t}^{\ast }=\Upsilon _{s}^{2}+4D_{s}D_{s}^{\ast }
\label{constant of motion eq2}
\end{equation}%
on the domains $\mathcal{D}(\Upsilon _{0}^{\top })=\mathcal{D}(\Upsilon
_{t}^{\top })$ and $\mathcal{D}(\Upsilon _{0}^{2})$, respectively. See
Theorem \ref{theorem constante of motion} (ii).

We can therefore see (\ref{constant of motion eq1}) or (\ref{constant of
motion eq2}) as a \emph{constant of motion} of the operator-valued flow of
Theorems \ref{lemma existence 2 copy(1)}--\ref{lemma existence 2 copy(4)}.
They are the first kind of ellipticity properties of the flow. To
easily see this, one can consider the system (\ref{flow equation-quadratic
delta}), or (\ref{flow equation-quadratic deltabis}), of ODEs in the
one-dimensional case, which is explicitly given in Section \ref{Flows on
Scalar Fields}. However, even if we prove (\ref{constant of motion eq1}) or (%
\ref{constant of motion eq2}) for unbounded $\Upsilon _{0}$, we would use
hypotheses that are too limited to be very useful to study the asymptotics
of the flow. In fact, because of the non-commutativity of operator spaces,
other ellipticity properties can be defined and studied, as it is done below
in Section \ref{Section tech5}.

We show here that the most natural ellipticity property is written in terms
of the operators 
\begin{equation}
\mathfrak{D}_{t}\doteq \Upsilon _{t}-\mu \mathbf{1}+4D_{t}\left( \Upsilon
_{t}^{\top }+\mu \mathbf{1}\right) ^{-1}D_{t}^{\ast }\ ,\qquad t\in \mathbb{R%
}_{0}^{+}\ ,  \label{D0}
\end{equation}%
with $\mu \in \mathbb{R}$ such that\footnote{%
Note that $\Upsilon _{0}\geq -\left( \mu -\varepsilon \right) \mathbf{1}$
iff $\Upsilon _{0}^{\top }\geq -\left( \mu -\varepsilon \right) \mathbf{1}$.}
$\Upsilon _{0}\geq -\left( \mu -\varepsilon \right) \mathbf{1}$ for some $%
\varepsilon \in \mathbb{R}^{+}$. Obviously, one cannot expect in general a
condition like $\mathfrak{D}_{t}=\mathrm{C}\mathbf{1}$ with $\mathrm{C}\in 
\mathbb{R}$ for all $t\in \mathbb{R}_{0}^{+}$, which would refer to a flow
on a fixed ellipse in terms of operators $X=\Upsilon _{t}$ and $Y=D_{t}$,
but we prove instead the following assertion:  

\begin{theorem}[Elliptic operator-valued flow]
\label{lemma constante of motion copy(6)}\mbox{}\newline
Take a self-adjoint operator $\Upsilon _{0}=\Upsilon _{0}^{\ast }$ and $%
D_{0}=\pm D_{0}^{\top }\in \mathcal{B}\left( \mathfrak{h}\right) $ ($%
D_{0}\neq 0$).

\begin{enumerate}
\item[\emph{(i)}] If $\Upsilon _{0}\in \mathcal{B}\left( \mathfrak{h}\right) 
$ then the mapping $t\mapsto \Vert \mathfrak{D}_{t}\Vert _{\mathrm{op}}$
from $\mathbb{R}_{0}^{+}$ to itself is decreasing.

\item[\emph{(ii)}] If $\Upsilon _{0}\geq -\left( \mu -\varepsilon \right) 
\mathbf{1}$ for some $\mu \in \mathbb{R}$ and $\varepsilon \in \mathbb{R}%
^{+} $ then the real-valued function defined by 
\begin{equation*}
\zeta \left( t\right) =\inf \sigma \left( \mathfrak{D}_{t}\right) \in \left[
-2\mu ,\infty \right) \ ,\qquad t\in \mathbb{R}_{0}^{+}\ ,
\end{equation*}%
is  upper  -semicontinuous\footnote{%
For bounded operators $\Upsilon _{0}\in \mathcal{B}\left( \mathfrak{h}%
\right) $, $\zeta $ is always continuous on $\mathbb{R}_{0}^{+}$. See Remark %
\ref{remark bounded2 copy(1)}.} at $t=0$ and continuous on $\mathbb{R}^{+}$
as well as decreasing with $\zeta \left( \mathbb{R}_{0}^{+}\right) \subseteq
\lbrack 0,\zeta \left( 0\right) ]$ whenever $\zeta \left( 0\right) \in 
\mathbb{R}_{0}^{+}$, while it is continuously increasing with $\zeta \left( 
\mathbb{R}_{0}^{+}\right) \subseteq \lbrack \zeta \left( 0\right) ,0)$
whenever $\zeta \left( 0\right) \in \mathbb{R}^{-}$.
\end{enumerate}
\end{theorem}

\begin{proof}
See Theorems \ref{theorem sympa0} and \ref{lemma constante of motion copy(3)}
(i).
\end{proof}

Note that Theorem \ref{lemma constante of motion copy(6)} gives only some of
the results proven in Section \ref{Section tech5}. For other important
results on the ellipticity properties of the flow, see Theorems \ref{lemma
constante of motion copy(3)} and \ref{Theorem sympa}. For instance, Theorem %
\ref{lemma constante of motion copy(6)} (i) needs bounded initial data $%
\Upsilon _{0}$. A generalization to unbounded $\Upsilon _{0}$ is given by
Theorem \ref{Theorem sympa} (ii) at the cost of additional assumptions,
namely $\mathfrak{D}_{0}\geq 0$ and $D_{0}\in \mathcal{L}^{2}(\mathfrak{h})$.

In fact, Theorem \ref{lemma constante of motion copy(6)} (i) is presented
here for the sake of simplicity. Indeed, this statement clearly shows the
general ellipticity property of the flow, since Theorem \ref{lemma constante
of motion copy(6)} (i) yields%
\begin{equation}
\zeta \left( t\right) \mathbf{1}\leq \Upsilon _{t}-\mu \mathbf{1}%
+4D_{t}\left( \Upsilon _{t}^{\top }+\mu \mathbf{1}\right) ^{-1}D_{t}^{\ast
}\leq \mathrm{C}\mathbf{1}\ ,\qquad t\in \mathbb{R}_{0}^{+}\ ,
\label{sdsdsdssdsd}
\end{equation}%
for some positive constant $\mathrm{C}\in \mathbb{R}_{0}^{+}$. In this case,
one could say that $X=\Upsilon _{t}$ is elliptically bounded (from below and
above) with respect to $Y=D_{t}$\textit{\ }for all times (up to the singular
case $\mu =0$)\textit{, }see Theorem \ref{lemma constante of motion copy(6)}%
. As already mentioned, a way to circumvent the necessary boundedness of $%
\Upsilon _{0}$ used in Theorem \ref{lemma constante of motion copy(6)} (i)
is given by Theorem \ref{Theorem sympa} (ii),  which is important to
study the convergence of the operator-valued flow, with possibly exponential
decay.  \ 

\begin{theorem}[Asymptotics of the operator-valued flow]
\label{lemma asymptotics1 copy(4)}\mbox{}\newline
Take a semibounded self-adjoint operator $\Upsilon _{0}=\Upsilon _{0}^{\ast
} $ and $\mu \in \mathbb{R}$ such that $\Upsilon _{0}\geq -\left( \mu
-\varepsilon \right) \mathbf{1}$ for some $\varepsilon \in \mathbb{R}^{+}$.
Let $D_{0}=\pm D_{0}^{\top }\in \mathcal{B}\left( \mathfrak{h}\right) $ ($%
D_{0}\neq 0$).

\begin{enumerate}
\item[\emph{(i)}] If $\Upsilon _{0}\in \mathcal{B}\left( \mathfrak{h}\right) 
$ then, as $t\rightarrow \infty $, $D$ and $\Delta $ converge respectively
in the strong operator topology to zero and in the weak operator topology to
an operator $\Delta _{\infty }\in \mathcal{B}\left( \mathfrak{h}\right) $
such that 
\begin{equation*}
\Upsilon _{\infty }\doteq \Upsilon _{0}+\Delta _{\infty }\geq (\mu
+\lim_{t\rightarrow \infty }\zeta \left( t\right) )\mathbf{1}\geq -\mu 
\mathbf{1}\ .
\end{equation*}

\item[\emph{(ii)}] If $\mu \in \mathbb{R}^{-}$ then, as $t\rightarrow \infty 
$, $D$ and $\Delta $ \emph{exponentially} converge in the norm topology
respectively to zero and an operator $\Delta _{\infty }\in \mathcal{B}(%
\mathfrak{h})$ such that 
\begin{equation*}
\Upsilon _{\infty }\doteq \Upsilon _{0}+\Delta _{\infty }\geq \left\vert \mu
\right\vert \mathbf{1}\ .
\end{equation*}

\item[\emph{(iii)}] If $\mu \neq 0$, $\mathfrak{D}_{0}\geq 0$ (see (\ref{D0}%
)) and $D_{0}\in \mathcal{L}^{2}(\mathfrak{h})$ then, as $t\rightarrow
\infty $, $D$ \emph{exponentially} converges in the Hilbert-Schmidt topology
(i.e., in $\mathcal{L}^{2}(\mathfrak{h})$) to zero, while $\Delta $ \emph{%
exponentially} converges in the trace norm topology (i.e., in $\mathcal{L}%
^{1}(\mathfrak{h})$) to an operator $\Delta _{\infty }\in \mathcal{L}^{1}(%
\mathfrak{h})$ satisfying 
\begin{equation*}
\Upsilon _{\infty }\doteq \Upsilon _{0}+\Delta _{\infty }\geq \left\vert \mu
\right\vert \mathbf{1}\ .
\end{equation*}
\end{enumerate}
\end{theorem}

\begin{proof}
Assertions (i), (ii) and (iii) correspond to Propositions \ref{limit prop1}, %
\ref{limit prop2} and \ref{limit prop3}, respectively.
\end{proof}

\begin{remark}
\label{lemma asymptotics1 copy(6)}\mbox{}\newline
It is clear that $\mathfrak{D}_{0}\geq 0$ does not hold true for
sufficiently large $\mu \in \mathbb{R}^{+}$, see (\ref{D0}). However, in
order to get a stronger exponential decay of the Hilbert-Schmidt norm $%
\left\Vert D_{t}\right\Vert _{2}$ and a more accurate estimate of the
spectral gap of $\Upsilon _{\infty }$ (see Theorem \ref{lemma asymptotics1
copy(4)} (iii)), if $\Upsilon _{\infty }$ is \emph{not} a positive operator
with spectral gap like in Theorem \ref{lemma asymptotics1 copy(4)} (ii), one
shall take the maximum value of $\mu \in \mathbb{R}^{+}$ such that $%
\mathfrak{D}_{0}\geq 0$ (provided such a positive $\mu $ exists).
\end{remark}

\begin{remark}
\label{lemma asymptotics1 copy(5)}\mbox{}\newline
Let $\alpha \in \mathbb{R}^{+}$ and assume that $\Upsilon _{0}\geq \alpha 
\mathbf{1}$. Then, upon choosing $\varepsilon \doteq \alpha $ and $\mu
\doteq \alpha /2$, observe that $\Upsilon _{0}\geq -\left( \mu -\varepsilon
\right) \mathbf{1=}(\alpha /2)\mathbf{1}$ and $\mathfrak{D}_{0}\geq (\alpha
/2)\mathbf{1}\geq 0$. As a result, if $D_{0}\in \mathcal{L}^{2}(\mathfrak{h}%
) $, Theorem \ref{lemma asymptotics1 copy(4)} (iii) can be applied to
positive operators $\Upsilon _{0}$ with spectral gap. In this case, note
that $\Upsilon _{\infty }\geq \alpha \mathbf{1}$, thanks to Theorem \ref%
{lemma asymptotics1 copy(4)} (ii).
\end{remark}

Using the constant of motion (\ref{constant of motion eq1}) and Theorem \ref%
{lemma asymptotics1 copy(4)} (iii), if $D_{0}^{\top }=\pm D_{0}$ and $%
t\rightarrow +\infty $, then observe that 
\begin{equation}
\mathrm{tr}\left( \Upsilon _{\infty }^{2}-\Upsilon
_{0}^{2}-4D_{0}D_{0}^{\ast }\right) =0\ ,  \label{constant of motion eq1bis}
\end{equation}%
where $\Upsilon _{\infty }\doteq \Upsilon _{0}+\Delta _{\infty }$ ,
at least when $\Upsilon _{0}$ is a bounded operator.  In the case where $%
\Upsilon _{0}D_{0}=D_{0}\Upsilon _{0}^{\top }$ on the domain $\mathcal{D}%
(\Upsilon _{0}^{\top })$ together with (\ref{condition67}), we can
invoke Theorem \ref{lemma asymptotics1 copy(4)} (ii)--(iii) together with
Lemma \ref{lemma square root} and get
\begin{equation}
\Upsilon _{\infty }=\sqrt{\Upsilon _{0}^{2}+4D_{0}D_{0}^{\ast }}
\label{constant of motion eq2bis}
\end{equation}%
on the domain $\mathcal{D}(\Upsilon _{0})$. In other words, $\Upsilon
_{\infty }$ can be fully characterized in this case.

Note that the condition $\mathfrak{D}_{0}\geq 0$ of Theorem \ref{lemma
asymptotics1 copy(4)} (iii), i.e., 
\begin{equation}
\Upsilon _{0}+4D_{0}\left( \Upsilon _{0}^{\top }+\mu \mathbf{1}\right)
^{-1}D_{0}^{\ast }\geq \mu \mathbf{1}\ ,  \label{conditions}
\end{equation}%
(see (\ref{D0})) together with $\Upsilon _{0}\geq -\left( \mu -\varepsilon
\right) \mathbf{1}$ for some $\varepsilon \in \mathbb{R}^{+}$ and $\mu \in 
\mathbb{R}$ can easily be satisfied. Take for instance a positive, possibly
unbounded, operator $\Upsilon _{0}$ with spectral gap, see Remark \ref{lemma
asymptotics1 copy(5)}. It is important however to observe that the
positivity of $\Upsilon _{0}$ is clearly not necessary to satisfy (\ref%
{conditions}). A sufficiently large spectral gap of the positive operator $%
D_{0}\left( \Upsilon _{0}^{\top }+\mu \mathbf{1}\right) ^{-1}D_{0}^{\ast }$
is sufficient to ensure (\ref{conditions}). For instance, Condition (\ref%
{conditions}) can obviously be deduced from the inequality 
\begin{equation*}
2D_{0}\left( \Upsilon _{0}^{\top }+\mu \mathbf{1}\right) ^{-1}D_{0}^{\ast
}\geq \mu \mathbf{1}
\end{equation*}%
when $\Upsilon _{0}\geq -\left( \mu -\varepsilon \right) \mathbf{1}$ for
some $\varepsilon \in \mathbb{R}^{+}$ and $\mu \in \mathbb{R}$. Since $%
\varepsilon \in \mathbb{R}^{+}$ can be taken arbitrarily small, this last
condition is for example always satisfied in the one-dimensional situation.
See Section \ref{Flows on Scalar Fields}. Note that it is remarkable in this
context that the limit operator $\Upsilon _{\infty }$ is \emph{always}
positive with a spectral gap, by Theorem \ref{lemma asymptotics1 copy(4)}
(iii), even if Condition (\ref{conditions}) includes situations where the
spectrum of the initial self-adjoint operator $\Upsilon _{0}$ carries
negative values.

If $\varphi \in \mathfrak{h}$ is an eigenvector of both operators $\Upsilon
_{0}$ and $D_{0}\left( \Upsilon _{0}^{\top }+\mu \mathbf{1}\right)
^{-1}D_{0}^{\ast }$, with respective eigenvalues $\lambda \in \mathbb{R}^{-}$
and $\gamma \in \mathbb{R}_{0}^{+}$ such that $\lambda +4\gamma \leq 0$,
then Condition (\ref{conditions}) does not hold true (because $\mu >-\lambda
>0$). In this case, one divides the flow into two parts: a flow on the space
generated by the eigenvector $\varphi $ and another one on its orthogonal
complement in $\mathfrak{h}$, on which (\ref{conditions}) can be verified.
Observe that the flow in the one--dimensional space can be reduced to a flow
on scalar fields that is explicitly known (see Section \ref{Flows on Scalar
Fields}).

Condition (\ref{conditions}) should only be viewed as a sufficient condition
to have a strong convergence of the elliptic flow. In particular, it is not
necessary for some convergence of the operator families $D$ and $\Delta $.
In the case $D_{0}\in \mathcal{L}^{2}(\mathfrak{h})$, one may also have
different assumptions leading to non-exponentially decaying Hilbert-Schmidt
norm $\left\Vert D_{t}\right\Vert _{2}$ while ensuring the existence of the
limit operator $\Delta _{\infty }$ in the trace norm topology. In addition,
one may ask the same question by removing the hypothesis $D_{0}\in \mathcal{L%
}^{2}(\mathfrak{h})$ and using either the weak/strong operator topology or
the operator norm, like in Theorem \ref{lemma asymptotics1 copy(4)} (i) and
(ii), respectively. For instance, Corollary \ref{Corollaire sympa} shows in
the bounded case $\Upsilon _{0},D_{0}\in \mathcal{B}\left( \mathfrak{h}%
\right) $ that 
\begin{equation*}
\int_{0}^{\infty }\left\langle \varphi ,D_{\tau }D_{\tau }^{\ast }\varphi
\right\rangle _{\mathfrak{h}}\mathrm{d}\tau <\infty \ ,\qquad \varphi \in 
\mathfrak{h}\ ,
\end{equation*}%
without the need of Condition (\ref{conditions}). This last result leads to
Theorem \ref{lemma asymptotics1 copy(4)} (i), which does not require (\ref%
{conditions}) but bounded operators $\Upsilon _{0},D_{0}\in \mathcal{B}%
\left( \mathfrak{h}\right) $. In fact, we need some integrability of the
operator-valued function $(D_{t}D_{t}^{\ast })_{t\geq 0}$, with for instance
a sufficiently strong polynomial decay (in some topology).

A limiting case should be the situation for which 
\begin{equation}
\Upsilon _{0}>0\text{\qquad and\qquad }\Upsilon _{0}+4D_{0}\left( \Upsilon
_{0}^{\top }\right) ^{-1}D_{0}^{\ast }\geq 0  \label{ghfhhg}
\end{equation}%
with $D_{0}\left( \Upsilon _{0}^{\top }\right) ^{-1}D_{0}^{\ast }$ being a
well-defined operator acting on $\mathfrak{h}$. Up to singular cases, these
assumptions may be \emph{necessary} for the convergence of the flow because
the existence of $\Upsilon _{\infty }$ seems to generally imply its
positivity, except in singular cases\footnote{%
One avoids the singular cases for which there is a $\psi \in \mathrm{ker}%
(D_{0})$ that is at the same time an eigenvector of $\Upsilon _{0}$ since,
in this situation, the flow is constant on the subspace generated by $\psi $.%
} . They correspond to take $\varepsilon ,\mu \rightarrow 0$ in the
conditions of Theorem \ref{lemma asymptotics1 copy(4)} with a possible limit
operator $\Upsilon _{\infty }\doteq \Upsilon _{0}+\Delta _{\infty }>0$
having a priori no spectral gap. By contrast, note that the assumptions of
Theorem \ref{lemma asymptotics1 copy(4)} (ii)--(iii) always lead to a
positive limit operator $\Upsilon _{\infty }$ with a spectral gap. The
conditions expressed by (\ref{ghfhhg}) look similar to the situation $\Omega
_{0}>0$, without spectral gap, studied in the hyperbolic case \cite%
{bach-bru-memo}. By analogy with the hyperbolic case, one may assume that $%
\left( \Upsilon _{0}\right) ^{-1/2}D_{0}$ is a Hilbert-Schmidt operator to
ensure the square-integrability of the Hilbert-Schmidt norm of $%
(D_{t})_{t\geq 0}$ and thus the existence of $\Delta _{\infty }$ as a
trace-class operator. The arguments to handle the non-spectral gap case in 
\cite{bach-bru-memo} are technically involved and long and we thus postpone
the analogous elliptical case for further studies.

To conclude, note that the Hilbert-Schmidt assumption $D_{0}\in \mathcal{L}%
^{2}(\mathfrak{h})$ is the most important case in view of the application of
the elliptic flow in quantum field theory. This is explained in the next
subsection.

\subsection{Application of the Flow in Quantum Field Theory\label%
{Diagonalization of Quadratic}}

 The Brockett-Wegner flow \cite{Brockett1,Wegner1,bach-bru} consists
in a (quadratically) non-linear first-order differential equation 
\begin{equation}
\forall t\in \mathbb{R}_{0}^{+}:\qquad \partial _{t}\mathrm{H}_{t}=\left[ 
\mathrm{H}_{t},\left[ \mathrm{H}_{t},\mathrm{N}\right] \right] \ ,\qquad 
\mathrm{H}_{t=0}\doteq \mathrm{H}_{0}=\mathrm{H}_{0}^{\ast }\ ,
\label{brocket flow}
\end{equation}%
for (possibly unbounded) operators all acting on a Hilbert space $\mathcal{F}
$, with $[A,B]\doteq AB-BA$ being the commutator between two operators $A$
and $B$ (provided it exists on some domain). Heuristically, one expects that
a solution $\left( \mathrm{H}_{t}\right) _{t\geq 0}$ to (\ref{brocket flow})
is given by 
\begin{equation}
\mathrm{H}_{t}=\mathrm{U}_{t,s}\mathrm{H}_{s}\mathrm{U}_{t,s}^{\ast }\
,\qquad s,t\in \mathbb{R}_{0}^{+}\ ,  \label{unitary equivalent}
\end{equation}%
where $(\mathrm{U}_{t,s})_{s,t\in \mathbb{R}_{0}^{+}}\subseteq \mathcal{B}%
\left( \mathcal{F}\right) $ is an evolution family of unitary operators
solving the non-autonomous (hyperbolic) evolution equation 
\begin{equation}
\forall s,t\in \mathbb{R}_{0}^{+}:\qquad \partial _{t}\mathrm{U}_{t,s}=-i%
\mathrm{G}_{t}\mathrm{U}_{t,s}\ ,\qquad \mathrm{U}_{s,s}\doteq \mathbf{1}\ ,
\label{non-autonomous evolution equation}
\end{equation}%
with self-adjoint generator $\mathrm{G}_{t}\doteq i\left[ \mathrm{N},\mathrm{%
H}_{t}\right] $. One could formally expect the convergence in some sense of $%
\mathrm{H}_{t}$ and $\mathrm{U}_{t,0}$ in the limit $t\rightarrow \infty $,
possibly leading to a unitarily equivalent Hamiltonian $\mathrm{H}_{\infty }=%
\mathrm{U}_{\infty ,0}\mathrm{\mathrm{H}}_{0}\mathrm{U}_{\infty ,0}^{\ast }$
satisfying $\mathrm{G}_{\infty }\doteq i\left[ \mathrm{H}_{\infty },\mathrm{N%
}\right] =0$.

In other words, Brockett-Wegner's method could formally be used to ($\mathrm{%
N}$--)diagonalize the initial self-adjoint operator $\mathrm{H}_{0}$. This
method is applied in \cite{QuadraticFermionic} to fermionic quadratic
Hamiltonians $\mathrm{H}_{0}$, $\mathrm{N}$ being the particle number
operator and $\mathcal{F}$ the fermionic Fock space. Doing so, one recovers
the non-autonomous system (\ref{flow equation-quadratic delta}) of ODEs
studied here. To understand more precisely this connection, let us first
define fermionic quadratic Hamiltonians.  

In quantum mechanics, one usually starts with a (one-particle) Hilbert space 
$\mathfrak{h}$ that is realized as a space $L^{2}(\mathcal{M})$ of
square-integrable, complex-valued functions on a measure space $(\mathcal{M},%
\mathfrak{a})$. A quantum system of $n\in \mathbb{N}$ fermions is then
described within the subspace $\wedge ^{n}\mathfrak{h}$ of totally
antisymmetric $n$--particle wave functions in $\mathfrak{h}^{\otimes n}$,
the $n$--fold tensor product of $\mathfrak{h}$.

In most of many-body quantum systems, the particle number is unknown. In
fact, in quantum statistical mechanics one usually studies physical
properties in the limit $n\rightarrow \infty $ of infinite particles, while
in quantum field theory one is often interested in time--dependent particle
numbers. To encode quantum many-body systems one regularly uses the
so-called Fock space which, for fermionic systems, refers to the Fock space 
\begin{equation*}
\mathcal{F}\doteq \bigoplus_{n=0}^{\infty }\wedge ^{n}\mathfrak{h}\ ,\qquad
\wedge ^{0}\mathfrak{h}\doteq \mathbb{C},
\end{equation*}%
the scalar product and the norm of which are respectively denoted by $%
\left\langle \cdot ,\cdot \right\rangle _{\mathcal{F}}$ and $\left\Vert
\cdot \right\Vert _{\mathcal{F}}$. Note that this scalar product is the sum
over $n\in \mathbb{N}$ of each canonical scalar product on the sector $%
\wedge ^{n}\mathfrak{h}$.

The use of the Fock space turned out to be very handy because of so--called%
\emph{\ creation/annihilation operators}: The annihilation operator $a\left(
\varphi \right) \in \mathcal{B}(\mathcal{F})$ of a fermion with wave
function $\varphi \in \mathfrak{h}$ is the (linear) operator uniquely
defined by the conditions $a\left( \varphi \right) (1,0,0,\ldots )=0 $ and%
\begin{equation*}
a\left( \varphi \right) \left( \psi _{1}\wedge \cdots \wedge \psi
_{n}\right) \doteq \frac{\sqrt{n}}{n!}\sum_{\pi \in \Pi _{n}}(-1)^{\mathrm{%
sgn}(\pi )}\left\langle \varphi ,\psi _{\pi \left( 1\right) }\right\rangle _{%
\mathfrak{h}}\psi _{\pi \left( 2\right) }\otimes \cdots \otimes \psi _{\pi
\left( n\right) }
\end{equation*}%
for any $n\in \mathbb{N}$ and $\psi _{1},\ldots ,\psi _{n}\in \mathfrak{h}$,
where $\Pi _{n}$ is the set of all permutations $\pi $ of $n$\ elements and $%
\mathrm{sgn}:\Pi _{n}\rightarrow \{-1,1\}$ denotes the signature of each
permutation, while $\psi _{1}\wedge \cdots \wedge \psi _{n}$ is the
orthogonal projection of $\psi _{1}\otimes \cdots \otimes \psi _{n}\in 
\mathfrak{h}^{\otimes n}$ onto the subspace of antisymmetric $n$--particle
wave functions: 
\begin{equation*}
\psi _{1}\wedge \cdots \wedge \psi _{n}\doteq \frac{1}{n!}\sum_{\pi \in \Pi
_{n}}(-1)^{\mathrm{sgn}(\pi )}\psi _{\pi \left( 1\right) }\otimes \cdots
\otimes \psi _{\pi \left( n\right) }\in \wedge ^{n}\mathfrak{h}\ .
\end{equation*}%
The creation operator of a fermion with wave function $\varphi \in \mathfrak{%
h}$ is the adjoint $a\left( \varphi \right) ^{\ast }$ of $a\left( \varphi
\right) $. Such operators are well-known to satisfy the Canonical
Anticommutation Relations (CAR): for all $\varphi _{1},\varphi _{2}\in 
\mathfrak{h}$,%
\begin{equation}
a(\varphi _{1})a(\varphi _{2})+a(\varphi _{2})a(\varphi _{1})=0,\quad
a(\varphi _{1})a(\varphi _{2})^{\ast }+a(\varphi _{2})^{\ast }a(\varphi
_{1})=\langle \varphi _{1},\varphi _{2}\rangle _{\mathfrak{h}}\mathbf{1}.
\label{CAR}
\end{equation}%
See \cite[p. 10]{BratteliRobinson}. The notation $\mathbf{1}$ used above is
the identity operator, acting here on the Fock space $\mathcal{F}$.

An important model in quantum field theory and quantum statistical mechanics
is given by fermionic quadratic operators, defined as follows: Take $%
E_{0}\in \mathbb{R}$, any Hilbert-Schmidt operator $D_{0}\in \mathcal{L}^{2}(%
\mathfrak{h})$ satisfying $D_{0}=-D_{0}^{\top }$ and a self-adjoint operator 
$\Upsilon _{0}=\Upsilon _{0}^{\ast }$ that is bounded from below on the
(separable) Hilbert space $\mathfrak{h}=L^{2}(\mathcal{M})$. Pick for
convenience some real orthonormal basis $\left\{ \varphi _{k}\right\}
_{k=1}^{\infty }$ in the dense domain $\mathcal{D}\left( \Upsilon
_{0}\right) \subseteq \mathfrak{h}$ of $\Upsilon _{0}$ and, for any $k\in 
\mathbb{N}$, let $a_{k}\doteq a\left( \varphi _{k}\right) $ be the
corresponding annihilation operator acting on the fermionic Fock space $%
\mathcal{F}$. We use also the notation 
\begin{equation*}
\left\{ X\right\} _{k,l}\doteq \left\langle \varphi _{k},X\varphi
_{l}\right\rangle \ ,\qquad k,l\in \mathbb{N}\ ,
\end{equation*}%
for operators $X$ acting on $\mathfrak{h}$ . Then, the fermionic quadratic
Hamiltonian $\mathrm{H}_{0}=\mathrm{H}_{0}^{\ast }$ acting on $\mathcal{F}$
can be defined by 
\begin{equation}
\mathrm{H}_{0}=\sum_{k,l\in \mathbb{N}}\{\Upsilon _{0}\}_{k,l}a_{k}^{\ast
}a_{l}+\{D_{0}\}_{k,l}a_{k}^{\ast }a_{l}^{\ast }+\{\bar{D}%
_{0}\}_{k,l}a_{l}a_{k}+E_{0}\mathbf{1}  \label{DefH0}
\end{equation}%
on its (dense) core 
\begin{equation}
\mathcal{D}_{0}\doteq \bigcup_{N\in \mathbb{N}}\left(
\bigoplus_{n=0}^{N}\left( \wedge ^{n}\mathcal{D}\left( \Upsilon _{0}\right)
\right) \right) \subseteq \mathcal{F}  \label{domain H0}
\end{equation}%
with $\wedge ^{0}\mathcal{D}\left( \Upsilon _{0}\right) \doteq \mathbb{C}$
while for $n\in \mathbb{N}$, $\wedge ^{n}\mathcal{D}\left( \Upsilon
_{0}\right) $ is the the subspace of totally antisymmetric $n$--particle
wave functions in $\mathcal{D}\left( \Upsilon _{0}\right) ^{\otimes n}$, the 
$n$--fold tensor product of $\mathcal{D}\left( \Upsilon _{0}\right)
^{\otimes n}\subseteq \mathfrak{h}^{\otimes n}$. Recall that $\mathbf{1}$ in
(\ref{DefH0}) is the identity operator on the Fock space $\mathcal{F}$.

For infinite-dimensional Hilbert spaces $\mathfrak{h}$ like $L^{2}(\mathbb{R}%
^{d})$ and $\ell ^{2}(\mathbb{Z}^{d})$, the self-adjointness of $\mathrm{H}%
_{0}$ is not obvious in general, but it is verified in \cite[Proposition 3.3]%
{QuadraticFermionic} using $\Upsilon _{0}=\Upsilon _{0}^{\ast }$ and $%
D_{0}\in \mathcal{L}^{2}(\mathfrak{h})$, which are both pivotal assumptions
here. Note that $\mathrm{H}_{0}$ does not depend upon the (possibly real)
orthonormal basis $\left\{ \varphi _{k}\right\} _{k=1}^{\infty }\subseteq 
\mathcal{D}\left( \Upsilon _{0}\right) \subseteq \mathfrak{h}$ chosen and
the condition $D_{0}=-D_{0}^{\top }$\ in (\ref{DefH0}) can be taken without
loss of generality because of (\ref{CAR}).

Hamiltonians like (\ref{DefH0}) appear very often in theoretical physics and
can for instance encode in some sense interaction effects. In fact,
interparticle interactions usually play a pivotal role in physical
phenomena, but they make the\ physical system extremely difficult to study
in a mathematically rigorous way and effective models coming from different
approximations and  Ans\"{a}zte  are normally used and regularly
lead to fermionic quadratic Hamiltonians, like in the BCS theory of
superfluidity \cite{BCS1,BCS2,BCS3} or in the (generalized) Hartree-Fock
theory \cite[ Sections 5.7 and 6.10]{Bru-Pedra-livre}. In these effective
models, $D_{0}$ usually accounts for interaction effects, whereas $\Upsilon
_{0}$ is the Hamiltonian describing a free fermion in the one-particle
Hilbert space $\mathfrak{h}$.

Fermionic quadratic Hamiltonians, as defined by (\ref{DefH0})--(\ref{domain
H0}), are generally unbounded in spite of the boundedness of fermionic
creation and annihilation operators \cite[Proposition 5.2.2]%
{BratteliRobinson}. However, when $D_{0}=0$, $\mathrm{H}_{0}$ is only the
so-called second quantization of the operator $\Upsilon _{0}$ acting on $%
\mathfrak{h}$ (see, e.g., \cite[Section 5.2.1]{BratteliRobinson}) and hence,
its spectral properties can be deduced from the spectral properties of $%
\Upsilon _{0}$. Hamiltonians of this form are named $\mathrm{N}$\emph{%
--diagonal} since they commute with the particle number operator $\mathrm{N}$%
, which can be defined by (\ref{DefH0})--(\ref{domain H0}) for $%
D_{0}=E_{0}=0 $ and $\Upsilon _{0}=\mathbf{1}$.

In this context, it is very useful to find a unitary operator $\mathrm{U}$
acting on the Fock space such that $\mathrm{UH}_{0}\mathrm{U}^{\ast }$
commutes with the particle number operator $\mathrm{N}$, because in this
case $\mathrm{UH}_{0}\mathrm{U}^{\ast }$ would be the second quantization of
some operator $\Upsilon _{\infty }$, the off-$\mathrm{N}$\emph{--}diagonal
part being zero in this situation. (I.e., $\mathrm{UH}_{0}\mathrm{U}^{\ast }$
would equal (\ref{DefH0}) with $\Upsilon _{\infty }=\Upsilon _{0}$ and $%
D_{0}=0$.) Mathematical results on the $\mathrm{N}$--diagonalization of
general quadratic operators like (\ref{DefH0}) have been obtained in \cite%
{Friedrichs,Berezin,Kato-Mugibayashi}, but since the sixties no new
mathematical result on the fermionic case has been obtained\footnote{%
Note that Araki presented in 1968 \cite{Araki} a very general method for the
\textquotedblleft $\mathrm{N}$--diagonalization\textquotedblright\ of
bilinear Hamiltonians, but it does not solve the issue addressed here. For
more explanations, see \cite{QuadraticFermionic}.},  although the
research on the bosonic case was recently much more active, see in
particular \cite{bach-bru-memo,Solovej-Nam}, driven by the mathematical
justifications of effective models.

The non-linear operator-valued elliptic flow studied in this paper for
Hilbert-Schmidt initial data $D_{0}$ allows one to solve this problem under
much more general conditions than before.  The link between the $%
\mathrm{N}$--diagonalization of fermionic quadratic Hamiltonians and the
elliptic flow studied here is constructed from the Brockett-Wegner flow (\ref%
{brocket flow}) when $\mathrm{H}_{0}$ and $\mathrm{N}$ are respectively the
fermionic quadratic Hamiltonian and the particle number operator.

The existence of an operator family solution to (\ref{brocket flow}) is very
questionable. Recall indeed  that fermionic quadratic Hamiltonians,
including the particle number operator $\mathrm{N}$, as defined by (\ref%
{DefH0})--(\ref{domain H0}), are generally unbounded for infinite
dimensional Hilbert spaces $\mathfrak{h}$. Thus, the well-posedness of the
differential equation (\ref{brocket flow}) is a priori utterly unclear in
the present case. This is in fact a general and very serious issue of the
Brockett-Wegner flow for which very limited results in mathematics are
proven (see only \cite{bach-bru-memo,bach-bru} for unbounded operators),
although the Brockett-Wegner flow has been largely used (in a formal way) in
theoretical physics \cite{Kehrein}.

We get around this problem in \cite{QuadraticFermionic} by observing that a
solution $\left( \mathrm{H}_{t}\right) _{t\geq 0}$ to (\ref{brocket flow})
should also be a family of quadratic Hamiltonians of the form 
\begin{equation}
\mathrm{H}_{t}\doteq \sum_{k,l\in \mathbb{N}}\left\{ \Upsilon _{t}\right\}
_{k,l}a_{k}^{\ast }a_{l}+\left\{ D_{t}\right\} _{k,l}a_{k}^{\ast
}a_{l}^{\ast }+\left\{ \bar{D}_{t}\right\} _{k,l}a_{l}a_{k}+\left(
E_{0}-8\int_{0}^{t}\left\Vert D_{\tau }\right\Vert _{2}^{2}\mathrm{d}\tau
\right) \mathbf{1}  \label{DefH0bis}
\end{equation}%
with $\Upsilon _{t}\doteq \Upsilon _{0}+\Delta _{t}$ for each $t\in \mathbb{R%
}_{0}^{+}$ and where $\Delta $ and $D$ are solution to the elliptic
operator-valued flow (\ref{flow equation-quadratic deltabis}). In a similar
same way, the infinitesimal generator in (\ref{non-autonomous evolution
equation}) should be of the form 
\begin{equation}
\mathrm{G}_{t}\doteq i\left[ \mathrm{N},\mathrm{H}_{t}\right]
=2i\sum_{k,l\in \mathbb{N}}\left\{ D_{t}\right\} _{k,l}a_{k}^{\ast
}a_{l}^{\ast }+\left\{ \bar{D}_{t}\right\} _{k,l}a_{k}a_{l}\ ,\qquad t\in 
\mathbb{R}_{0}^{+}\ .  \label{generator}
\end{equation}%
In other words, the Brockett-Wegner flow (\ref{brocket flow}) for operators
acting on the Fock space $\mathcal{F}$ can be replaced by the elliptic flow (%
\ref{flow equation-quadratic deltabis}) for operators acting on the
one-particle Hilbert space $\mathfrak{h}$.

Using the continuous functions $(\Delta ,D)\in C(\mathbb{R}_{0}^{+};\mathcal{%
L}^{2}(\mathfrak{h})^{2})$ of Theorem \ref{lemma existence 2 copy(4)}
together with Equations (\ref{DefH0})--(\ref{domain H0}) and the theory of
non-autonomous evolution equations \cite[Section VII.1]{bach-bru-memo} (see (%
\ref{non-autonomous evolution equation})), one can then construct the
families $\left( \mathrm{H}_{t}\right) _{t\geq 0}$, $\left( \mathrm{G}%
_{t}\right) _{t\geq 0}$ and $(\mathrm{U}_{t,s})_{s,t\in \mathbb{R}_{0}^{+}}$
of operators acting on the fermionic Fock space $\mathcal{F}$ and verifies
Equality (\ref{unitary equivalent}). This is done in the paper \cite%
{QuadraticFermionic}. Note that the Hilbert-Schmidt property of the family $%
D\equiv (D_{t})_{t\geq 0}$ is crucial here to be able to make sense of $%
\left( \mathrm{H}_{t}\right) _{t\geq 0}$ and $\left( \mathrm{G}_{t}\right)
_{t\geq 0}$ as self-adjoint operators, by (\ref{DefH0bis})--(\ref{generator}%
) and \cite[Proposition 3.3]{QuadraticFermionic}. Then, using Theorem \ref%
{lemma asymptotics1 copy(4)} (iii), we show in \cite{QuadraticFermionic} the
convergence in some sense of $\mathrm{H}_{t}$ and $\mathrm{U}_{t,0}$ in the
limit $t\rightarrow \infty $, leading to a unitarily equivalent Hamiltonian 
\begin{equation*}
\mathrm{H}_{\infty }=\mathrm{U}_{\infty ,0}\mathrm{\mathrm{H}}_{0}\mathrm{U}%
_{\infty ,0}^{\ast }
\end{equation*}%
satisfying $\left[ \mathrm{H}_{\infty },\mathrm{N}\right] =0$. In other
words, one obtains the following theorem:

\begin{theorem}[Diagonalization of quadratic Hamiltonians -- \protect\cite%
{QuadraticFermionic}]
\label{thm3}\mbox{}\newline
Let $E_{0}\in \mathbb{R}$. Take $D_{0}=-D_{0}^{\top }\in \mathcal{L}^{2}(%
\mathfrak{h})$ and $\Upsilon _{0}=\Upsilon _{0}^{\ast }$, both acting on $%
\mathfrak{h}$. Assume that 
\begin{equation*}
\Upsilon _{0}\geq -\left( \mu -\varepsilon \right) \mathbf{1}\qquad \text{and%
}\qquad \Upsilon _{0}+4D_{0}\left( \Upsilon _{0}^{\top }+\mu \mathbf{1}%
\right) ^{-1}D_{0}^{\ast }\geq \mu \mathbf{1}
\end{equation*}%
for some $\mu \in \mathbb{R}\backslash \{0\}$ and $\varepsilon \in \mathbb{R}%
^{+}$. Then, there is a unitary transformation\footnote{%
It is the strong limit $t\rightarrow +\infty $ of the family $\left( \mathrm{%
U}_{t,0}\right) _{t\geq 0}$ discussed above.} $\mathrm{U}$ such that 
\begin{equation*}
\mathrm{U\mathrm{H}}_{0}\mathrm{U}^{\ast }=\sum_{k,l\in \mathbb{N}%
}\{\Upsilon _{\infty }\}_{k,l}a_{k}^{\ast }a_{l}+\left(
E_{0}-8\int_{0}^{\infty }\left\Vert D_{\tau }\right\Vert _{2}^{2}\mathrm{d}%
\tau \right) \mathbf{1}
\end{equation*}%
with the operator family $D$ and the Hamiltonian $\Upsilon _{\infty }\doteq
\Upsilon _{0}+\Delta _{\infty }$ being defined via Theorems \ref{lemma
existence 2 copy(4)} and \ref{lemma asymptotics1 copy(4)}. In particular, $%
\mathrm{U\mathrm{H}}_{0}\mathrm{U}^{\ast }$ is $\mathrm{N}$--diagonal.
\end{theorem}

This new result represents a significant improvement. Compare indeed the
previously known result proven in \cite{Berezin,Kato-Mugibayashi} with
Theorem \ref{thm3}. For instance, observe that no gap and Hilbert-Schmidt
conditions, like $\Upsilon _{0}\geq \alpha \mathbf{1}$ with $\alpha \in 
\mathbb{R}^{+}$ and $[\Upsilon _{0},D_{0}]\in \mathcal{L}^{2}(\mathfrak{h})$%
, are used in Theorem \ref{thm3}, in contrast with \cite[Theorem 8.2]%
{Berezin} and \cite[Theorem 3]{Kato-Mugibayashi}. In some cases, we can even
be able to fully characterize the spectrum of\ the fermionic quadratic
Hamiltonian by using (\ref{constant of motion eq2bis}), which is also a new
outcome. See also Equation (\ref{constant of motion eq1bis}). Note finally
that the lower semiboundedness of $\Upsilon _{0}$ in Theorem \ref{thm3} is
also a natural condition, see discussions in \cite[Section 2.6]%
{QuadraticFermionic}.

This theorem refers to \cite[Theorem 2.4]{QuadraticFermionic}. Its proof is
nontrivial and rather long, see \cite[Section 3]{QuadraticFermionic}. Note
finally that, in \cite{QuadraticFermionic}, we also strengthen the
mathematical foundations of quadratic Hamiltonians as self-adjoint operators
acting on fermionic Fock spaces.

\section{Technical Results on the Flow\label{Technical Results}}

The case $D_{0}=0$ is trivial since it leads to a constant flow and it is
thus always omitted in this section.

\subsection{Local Solution to the Flow -- Strong Operator Topology\label%
{Section tech1}}

The strategy for the local existence of a solution to (\ref{flow
equation-quadratic deltabis}) is the same as the one used to study the
hyperbolic case (\ref{flow equation-quadratic}), see \cite[Section V.1]%
{bach-bru-memo}. We do it via an auxiliary non-autonomous parabolic
evolution equation, which corresponds to use the two-parameter operator
family $(W_{t,s})_{t\geq s}$ defined for all $s,t\in \mathbb{R}_{0}^{+}$
with $t\geq s$ by 
\begin{eqnarray}
W_{t,s} &\equiv &W_{t,s}\left( \Delta \right) \doteq \mathrm{e}^{-2\left(
t-s\right) \Upsilon _{0}}  \label{definition of W} \\
&&+\sum_{n=1}^{\infty }\left( -2\right) ^{n}\int_{s}^{t}\mathrm{d}\tau
_{1}\cdots \int_{s}^{\tau _{n-1}}\mathrm{d}\tau _{n}\mathrm{e}^{-2\left(
t-\tau _{1}\right) \Upsilon _{0}}\left( \underset{j=1}{\overset{n}{\prod }}%
\Delta _{\tau _{j}}\mathrm{e}^{-2\left( \tau _{j}-\tau _{j+1}\right)
\Upsilon _{0}}\right) \ ,  \notag
\end{eqnarray}%
where $\tau _{n+1}\doteq s$ inside the product, given any strongly
continuous family $\left( \Delta _{t}\right) _{t\geq 0}$ and semibounded
operator $\Upsilon _{0}\geq -\mu \mathbf{1}$ with $\mu \in \mathbb{R}$. See
for instance \cite[Lemma 33]{bach-bru-memo} applied to the positive operator 
$\Upsilon _{0}+\mu \mathbf{1}$ and $\left( -\Delta _{t}\right) _{t\geq 0}$.
In particular, it satisfies the cocycle property\ $W_{t,x}W_{x,s}=W_{t,s}$
for any $s,x,t\in \mathbb{R}_{0}^{+}$ satisfying $t\geq x\geq s$, while $%
W_{t,s}$ is jointly strongly continuous in $s$ and $t$. Similar properties
also hold for $\left( W_{t,s}^{\ast }\right) _{t\geq s}$, $\left(
W_{t,s}^{\top }\right) _{t\geq s}$ and $\left( W_{t,s}^{\top }\right)
_{t\geq s}^{\ast }$. Additionally, for any $s,t\in \mathbb{R}_{0}^{+}$ with $%
t\geq s$, 
\begin{equation}
\Vert W_{t,s}\Vert _{\mathrm{op}}=\Vert W_{t,s}^{\top }\Vert _{\mathrm{op}%
}\leq \mathrm{e}^{2\left( r\left( t\right) +\mu \right) \left( t-s\right) }\
,  \label{upper bound for W0}
\end{equation}%
with%
\begin{equation*}
r\left( t\right) \doteq \sup_{\alpha \in \left[ 0,t\right] }\Vert \Delta
_{t}\Vert _{\mathrm{op}}=\sup_{\alpha \in \left[ 0,t\right] }\Vert \Delta
_{t}^{\top }\Vert _{\mathrm{op}}
\end{equation*}%
and $\Upsilon _{0}\geq -\mu \mathbf{1}$ for some $\mu \in \mathbb{R}$. Then,
we study the existence of a solution $\Delta $ to the initial value problem 
\begin{equation}
\forall t\in \lbrack 0,T]:\qquad \partial _{t}\Delta
_{t}=16W_{t,0}D_{0}W_{t,0}^{\top }\left( W_{t,0}^{\top }\right) ^{\ast
}D_{0}^{\ast }W_{t,0}^{\ast }\ ,\quad \Delta _{0}=0\ ,
\label{initial value problemlocal}
\end{equation}%
for sufficiently small $T\in \mathbb{R}^{+}$. This is done via a contraction
mapping principle.

To this end, for any fixed $T\in \mathbb{R}^{+}\cup \{\infty \}$, we use the
shorter notation $\mathfrak{C}$ for the Banach space of strongly continuous
mappings $t\mapsto \Delta _{t}$\ from $\left[ 0,T\right] \subseteq \mathbb{R}%
_{0}^{+}$ to the set $\mathcal{B}\left( \mathfrak{h}\right) $ of bounded
operators acting on the separable, complex Hilbert space $\mathfrak{h}$. Its
norm is denoted as usual by 
\begin{equation}
\left\Vert \Delta \right\Vert _{\infty }\doteq \sup_{t\in \left[ 0,T\right]
}\left\Vert \Delta _{t}\right\Vert _{\mathrm{op}}\ ,\qquad \Delta \in 
\mathfrak{C}\ ,  \label{norm for contraction mapping principle}
\end{equation}%
with $\left[ 0,\infty \right] \equiv \mathbb{R}_{0}^{+}$. Also, $\mathbf{B}%
_{r}\left( X\right) \subseteq \mathfrak{C}$ denotes the open ball of radius $%
r>0$ centered at $X\doteq \left( X_{t}\right) _{t\in \left[ 0,T\right] }\in 
\mathfrak{C}$. Below we take $\Delta \in \mathbf{B}_{r}\left( 0\right) $ for
some finite constants $r,T>0$. We are now in a position to prove the
existence of a solution $\Delta \in \mathfrak{C}$ to the initial value
problem (\ref{initial value problemlocal}) for sufficiently small $T\in 
\mathbb{R}^{+}$, referring here to the condition 
\begin{equation}
0\leq T\leq T_{0}\doteq \min \left\{ \frac{\ln 2}{2\sup \left\{ \mu
,1\right\} },\frac{1}{2^{15/2}\Vert D_{0}\Vert _{\mathrm{op}}}\right\} \ .
\label{local time}
\end{equation}%
This is given by the following lemma:\ 

\begin{lemma}[Local solution to (\protect\ref{initial value problemlocal})]
\label{lemma existence 2 copy(2)}\mbox{}\newline
Assume $\Upsilon _{0}=\Upsilon _{0}^{\ast }\geq -\mu \mathbf{1}$ with $\mu
\in \mathbb{R}$ and $D_{0}\in \mathcal{B}\left( \mathfrak{h}\right) $ ($%
D_{0}\neq 0$). Let $T\in (0,T_{0}]$. Then there is a unique strongly
continuous operator family $\Delta \in \mathfrak{C}$ that is a strong
solution in $\mathfrak{h}$ to the initial value problem (\ref{initial value
problemlocal}). Furthermore, for any $t\in \lbrack 0,T]$, $\Delta _{t}$ is
self-adjoint and locally Lipschitz continuous in the norm topology.
\end{lemma}

\begin{proof}
The proof is essentially the same as the one of \cite[Lemma 34]%
{bach-bru-memo}. For the reader's convenience we shortly reproduce it in
this special case. Assume without loss of generality that $\Upsilon
_{0}=\Upsilon _{0}^{\ast }\geq -\mu \mathbf{1}$ with $\mu >0$ and $D_{0}\in 
\mathcal{B}\left( \mathfrak{h}\right) $ ($D_{0}\neq 0$). Fix the constants $%
r,T>0$. For any $\Delta \in \mathfrak{C}$ and times $s,t\in \lbrack 0,T]$
with $t\geq s$, we use the notation $W_{t,s}=W_{t,s}^{\Delta }$ for the
two-parameter operator family defined by (\ref{definition of W}). For any $%
\Delta \in \mathfrak{C}$, $W_{t,s}^{\Delta }$ is jointly strongly continuous
in $s$ and $t$ and, for any $s,t\in \lbrack 0,T]$ with $t\geq s$, 
\begin{equation}
\Vert W_{t,s}^{\Delta }\Vert _{\mathrm{op}}=\Vert (W_{t,s}^{\Delta })^{\top
}\Vert _{\mathrm{op}}\leq \mathrm{e}^{2\left( \Vert \Delta \Vert _{\infty
}+\mu \right) \left( t-s\right) }\ ,  \label{upper bound for W}
\end{equation}%
thanks to (\ref{upper bound for W0}) and (\ref{norm for contraction mapping
principle}). As a consequence, given $D_{0}\in \mathcal{B}\left( \mathfrak{h}%
\right) $ ($D_{0}\neq 0$), we can define the mapping $\mathfrak{I}$ from $%
\mathfrak{C}$ to itself by 
\begin{equation}
\mathfrak{I}(\Delta )_{t}\doteq 16\int_{0}^{t}W_{\tau ,0}^{\Delta
}D_{0}\left( W_{\tau ,0}^{\Delta }\right) ^{\top }\left( \left( W_{\tau
,0}^{\Delta }\right) ^{\top }\right) ^{\ast }D_{0}^{\ast }\left( W_{\tau
,0}^{\Delta }\right) ^{\ast }\mathrm{d}\tau  \label{defIfrak}
\end{equation}%
for all $t\in \left[ 0,T\right] $ and $\Delta \in \mathfrak{C}$. If $r>0$
and $\Delta =\Delta ^{\ast }\in \mathbf{B}_{r}\left( 0\right) $ then we get
from (\ref{upper bound for W}) and (\ref{defIfrak}) that 
\begin{equation}
\left\Vert \mathfrak{I}\left( \Delta \right) \right\Vert _{\infty }\leq 
\frac{2}{r+\mu }\left( \mathrm{e}^{8\left( r+\mu \right) T}-1\right)
\left\Vert D_{0}\right\Vert _{\mathrm{op}}^{2}\ .  \label{toto1}
\end{equation}%
Now, let $\Delta ^{\left( 1\right) },\Delta ^{\left( 2\right) }\in \mathbf{B}%
_{r}\left( 0\right) $. Again from (\ref{upper bound for W}) and (\ref%
{defIfrak}), 
\begin{equation}
\left\Vert \mathfrak{I}(\Delta ^{\left( 1\right) })-\mathfrak{I}(\Delta
^{\left( 2\right) })\right\Vert _{\infty }\leq \frac{32}{3\left( r+\mu
\right) }\left( \mathrm{e}^{6\left( r+\mu \right) T}-1\right) \left\Vert
D_{0}\right\Vert _{\mathrm{op}}^{2}\left\Vert \Lambda \right\Vert _{\infty }
\label{condition new 0}
\end{equation}%
with $\Lambda _{t}\doteq W_{t,0}^{\Delta ^{\left( 1\right)
}}-W_{t,0}^{\Delta ^{\left( 2\right) }}$ for any $t\in \lbrack 0,T]$. The
family $\left( W_{t,s}^{\Delta }\right) _{t\geq s}$ is the (unique) solution
to the integral equation $W=\mathcal{T}\left( W\right) $, with%
\begin{equation}
\left[ \mathcal{T}\left( W\right) \right] _{t,s}=\mathrm{e}^{-2\left(
t-s\right) \Upsilon _{0}}-2\int_{s}^{t}\mathrm{e}^{-2\left( t-\tau \right)
\Upsilon _{0}}\Delta _{\tau }W_{\tau ,s}\mathrm{d}\tau \ ,  \label{ghghghgh}
\end{equation}%
which follows from a standard contraction mapping principle argument. As a
consequence, for any $t\in \lbrack 0,T]$, 
\begin{equation*}
\Lambda _{t}\doteq W_{t,0}^{\Delta ^{\left( 1\right) }}-W_{t,0}^{\Delta
^{\left( 2\right) }}=2\int_{0}^{t}\mathrm{e}^{-2\left( t-\tau \right)
\Upsilon _{0}}\left\{ (\Delta _{\tau }^{(2)}-\Delta _{\tau }^{(1)})W_{\tau
,0}^{\Delta ^{\left( 1\right) }}-\Delta _{\tau }^{(2)}\Lambda _{\tau
}\right\} \mathrm{d}\tau \ ,
\end{equation*}%
which, together with (\ref{upper bound for W}), in turn implies that 
\begin{equation*}
\left\Vert \Lambda \right\Vert _{\infty }\leq \mathrm{e}^{2T\mu }\left(
2rT\left\Vert \Lambda \right\Vert _{\infty }+\frac{1}{r}\left( \mathrm{e}%
^{2rT}-1\right) \left\Vert \Delta ^{\left( 1\right) }-\Delta ^{\left(
2\right) }\right\Vert _{\infty }\right) \ .
\end{equation*}%
When $2rT\mathrm{e}^{2T\mu }<1$, this leads to the inequality 
\begin{equation*}
\left\Vert \Lambda \right\Vert _{\infty }\leq \frac{\mathrm{e}^{2T\mu
}\left( \mathrm{e}^{2rT}-1\right) }{r\left( 1-2rT\mathrm{e}^{2T\mu }\right) }%
\left\Vert \Delta ^{\left( 1\right) }-\Delta ^{\left( 2\right) }\right\Vert
_{\infty }\ .
\end{equation*}%
Inserting this in (\ref{condition new 0}), we get 
\begin{equation}
\left\Vert \mathfrak{I}(\Delta ^{\left( 1\right) })-\mathfrak{I}(\Delta
^{\left( 2\right) })\right\Vert _{\infty }\leq \frac{32\mathrm{e}^{2T\mu
}\left( \mathrm{e}^{6\left( r+\mu \right) T}-1\right) \left( \mathrm{e}%
^{2rT}-1\right) }{3\left( r+\mu \right) r\left( 1-2rT\mathrm{e}^{2T\mu
}\right) }\left\Vert D_{0}\right\Vert _{\mathrm{op}}^{2}\left\Vert \Delta
^{\left( 1\right) }-\Delta ^{\left( 2\right) }\right\Vert _{\infty }
\label{toto2}
\end{equation}%
for any $\Delta ^{\left( 1\right) },\Delta ^{\left( 2\right) }\in \mathbf{B}%
_{r}\left( 0\right) $, provided that $2rT\mathrm{e}^{2T\mu }<1$. Therefore,
upon choosing%
\begin{equation}
r\doteq 16\sqrt{2}\left\Vert D_{0}\right\Vert _{\mathrm{op}}\qquad \text{and}%
\qquad T\leq T_{0}\ ,  \label{flow equation T}
\end{equation}%
we observe that $2rT\mathrm{e}^{2T\mu }<1$ and 
\begin{equation}
\left\Vert \mathfrak{I}\left( \Delta \right) \right\Vert _{\infty }\leq 
\frac{2}{16\sqrt{2}}\left( 16\mathrm{e}-1\right) \left\Vert D_{0}\right\Vert
_{\mathrm{op}}<r  \label{flow equation condition (i)}
\end{equation}%
by (\ref{toto1}), while we infer from (\ref{toto2}) that 
\begin{equation}
\left\Vert \mathfrak{I}(\Delta ^{\left( 1\right) })-\mathfrak{I}(\Delta
^{\left( 2\right) })\right\Vert _{\infty }<\frac{1}{2}\left\Vert \Delta
^{\left( 1\right) }-\Delta ^{\left( 2\right) }\right\Vert _{\infty }\ .
\label{estimatecontraction}
\end{equation}%
In other words, the mapping 
\begin{equation*}
\mathfrak{I}:\overline{\mathbf{B}_{r}\left( 0\right) }\rightarrow \mathbf{B}%
_{r}\left( 0\right) \subset \overline{\mathbf{B}_{r}\left( 0\right) }
\end{equation*}%
is a contraction. Consequently, the contraction mapping principle on the
closed subspace defined by $\overline{\mathbf{B}_{r}\left( 0\right) }%
\subseteq \mathfrak{C}$, which is equipped with the norm $\left\Vert \cdot
\right\Vert _{\infty }$, yields a unique fixed point $\Delta =\mathfrak{I}%
\left( \Delta \right) $ with $\Delta =\Delta ^{\ast }\in \overline{\mathbf{B}%
_{r}\left( 0\right) }$, for $t\in \left[ 0,T\right] $ (keeping in mind (\ref%
{flow equation T})). Moreover, since $\left\Vert \Delta \right\Vert _{\infty
}\leq r$, by (\ref{upper bound for W}), it is clear that $\Delta \in 
\mathfrak{C}$ is Lipschitz continuous in the norm topology for any $t\in %
\left[ 0,T\right] $.
\end{proof}

Thanks to Lemma \ref{lemma existence 2 copy(2)} we can construct an
operator-valued function $D:[0,T]\rightarrow \mathcal{B}(\mathfrak{h})$
defined by 
\begin{equation}
D_{t}\doteq W_{t,0}D_{0}W_{t,0}^{\top }\ ,\qquad t\in \left[ 0,T\right] \ ,
\label{D_explicit_solutionbis}
\end{equation}%
where $W_{t,s}=W_{t,s}^{\Delta }$ refers to the two-parameter operator
family defined by (\ref{definition of W}) with $\Delta \in \mathfrak{C}$
being the strong solution in $\mathfrak{h}$ to the initial value problem (%
\ref{initial value problemlocal}), as given by Lemma \ref{lemma existence 2
copy(2)} for any $T\in (0,T_{0}]$. Note that the mapping $t\mapsto D_{t}$
from $[0,T]$ to $\mathcal{B}(\mathfrak{h})$ is continuous for the strong
(operator) topology in $\mathcal{B}(\mathfrak{h})$, i.e., $D\in \mathfrak{C}$%
. Recall the notation (\ref{sdssds}), that is, 
\begin{equation*}
\Upsilon _{t}\doteq \Upsilon _{0}+\Delta _{t}\ ,
\end{equation*}%
whenever the solution $\Delta \in \mathfrak{C}$ to (\ref{initial value
problemlocal}) exists.

\begin{lemma}[Existence of $D$ for small times]
\label{lemma existence 2}\mbox{}\newline
Assume $\Upsilon _{0}=\Upsilon _{0}^{\ast }\geq -\mu \mathbf{1}$ with $\mu
\in \mathbb{R}$ and $D_{0}\in \mathcal{B}\left( \mathfrak{h}\right) $ ($%
D_{0}\neq 0$). Let $T\in (0,T_{0}]$. Then, $D\in \mathfrak{C}$ defined by (%
\ref{D_explicit_solutionbis}) satisfies 
\begin{equation}
D_{t}=D_{0}-2\int_{0}^{t}\left( \Upsilon _{\tau }D_{\tau }+D_{\tau }\Upsilon
_{\tau }^{\top }\right) \mathrm{d}\tau \ ,\qquad t\in \left[ 0,T\right] \ .
\label{sdsdsd0}
\end{equation}%
Moreover, $D_{t>0}\mathfrak{h}\subseteq \mathcal{D}(\Upsilon _{0})$ and $D\ $%
is the unique strong solution on the domain $\mathcal{D}(\Upsilon _{0}^{\top
})$ to 
\begin{equation}
\partial _{t}D_{t}=-2\left( \Upsilon _{t}D_{t}+D_{t}\Upsilon _{t}^{\top
}\right) \ ,\qquad t\in \left( 0,T\right] \ .  \label{sdsdsd}
\end{equation}%
It is also locally Lipschitz continuous in the norm topology outside $t=0$.
\end{lemma}

\begin{proof}
The assertions are consequences of properties of the two-parameter operator
family defined by (\ref{definition of W}) with $\Delta \in \mathfrak{C}$
given by Lemma \ref{lemma existence 2 copy(2)}. The corresponding arguments
are relatively simple adaptations of \cite[Lemmata 33 and 35, Corollary 37]%
{bach-bru-memo}. Note in particular that the operator $\Upsilon _{t}$ has
domain $\mathcal{D}\left( \Upsilon _{t}\right) =\mathcal{D}\left( \Upsilon
_{0}\right) $ and is (Lipschitz) continuous in the norm topology for $t\in %
\left[ 0,T\right] $ with $T\leq T_{0}$, $T_{0}$ being defined by (\ref{local
time}). Because $\Upsilon _{t}=\Upsilon _{t}^{\ast }\geq \Upsilon _{0}\geq
-\mu \mathbf{1}$, $-2\Upsilon _{t}$ is semibounded from above (i.e. its
resolvent set contains the half-plane in $\mathbb{C}$ for which the real
part is greater than $2\mu $), and hence, it is the generator of a bounded
analytic semigroup $\left( \mathrm{e}^{-2\alpha \Upsilon _{t}}\right)
_{\alpha \geq 0}$ for any fixed $t\in \left[ 0,T_{0}\right] $. See, e.g., 
\cite[Chapter II, Definition 4.5 and Theorem 4.6]{EngelNagel}. Consequently,
the Cauchy problem 
\begin{equation}
\left\{ 
\begin{array}{llllll}
\forall s\in \lbrack 0,T),\ t\in (s,T] & : & \partial _{t}W_{t,s}=-2\Upsilon
_{t}W_{t,s} & , & W_{s,s}\doteq \mathbf{1} & , \\ 
\forall t\in (0,T],\ s\in \lbrack 0,t] & : & \partial
_{s}W_{t,s}=2W_{t,s}\Upsilon _{s} & , & W_{t,t}\doteq \mathbf{1} & ,%
\end{array}%
\right.  \label{eq diff 4}
\end{equation}%
is a parabolic evolution equation. Existence and uniqueness of its solution
are standard in this case, see, e.g., \cite[p. 407, Theorem 2.2]{Schnaubelt1}
or \cite[Chap. 5, Theorem 6.1]{Pazy}. Its solution is nothing else than the
evolution family $(W_{t,s})_{t\geq s\geq 0}$ defined by (\ref{definition of
W}) for $s\in \lbrack 0,T]\ $and $t\in \lbrack s,T]$. Compare for instance (%
\ref{eq diff 4}) with (\ref{ghghghgh}). It is important to observe that the
derivatives with respect to $t$ and $s$ in (\ref{eq diff 4}) are in the
strong sense in $\mathfrak{h}$ and $\mathcal{D}\left( \Upsilon _{0}\right) $%
, respectively. It means in particular that for any $s\in \lbrack 0,T]\ $and 
$t\in (s,T]$, $W_{t,s}\mathfrak{h}\subseteq \mathcal{D}\left( \Upsilon
_{0}\right) $. In fact, by a straightforward adaptation of the proof of \cite%
[Lemma 35]{bach-bru-memo}, for any positive times $s,t\geq 0$ such that $t>s$%
,$\left( \Upsilon _{t}W_{t,s}\right) _{t,s\geq 0}\subseteq \mathcal{B}\left( 
\mathfrak{h}\right) $ is not only a family of bounded operators but it is
also a strongly continuous one in $t$ (provided $t>s$), see below Equation (%
\ref{hhhhh}). It remains to discuss the uniqueness of the solution to (\ref%
{sdsdsd}), which is proven exactly like in \cite[Corollary 37]{bach-bru-memo}%
. We omit the details.
\end{proof}

In other words, Lemmata \ref{lemma existence 2 copy(2)}--\ref{lemma
existence 2} give a unique solution to the elliptic flow (\ref{flow
equation-quadratic deltabis}) for any semibounded operator $\Upsilon _{0}$
acting on the Hilbert space $\mathfrak{h}$ and each bounded operator $%
D_{0}\in \mathcal{B}(\mathfrak{h})$. We now extend this solution to all
times.

\subsection{Global Solution to the Flow -- Strong Operator Topology\label%
{Section tech2}}

By Lemma \ref{lemma existence 2 copy(2)}, there is a unique solution $\Delta
\in \mathfrak{C}$ to the initial value problem (\ref{initial value
problemlocal}) for any $T\in (0,T_{0}]$ and it is thus natural to define the
(possibly infinite) maximal time for which such a solution exists, that is, 
\begin{equation}
T_{\max }\doteq \sup \left\{ T\geq 0:\exists \mathrm{\mathrm{\ }}\text{a
solution}\mathrm{\ }(\Delta _{t})_{t\in \lbrack 0,T]}\in \mathfrak{C}\ \text{%
to\ (\ref{initial value problemlocal})}\right\} \in (0,\infty ]\ .
\label{Tmax}
\end{equation}%
Obviously, $T_{\max }\geq T_{0}>0$, thanks to Lemma \ref{lemma existence 2
copy(2)}. Our aim is to prove that $T_{\max }=\infty $. But, before doing
that, we first extend Lemmata \ref{lemma existence 2 copy(2)}--\ref{lemma
existence 2} to all positive times $t<T_{\max }$:

\begin{proposition}[Well-posedness and uniqueness of the solution to the flow%
]
\label{lemma existence 2 copy(3)}\mbox{}\newline
Assume $\Upsilon _{0}=\Upsilon _{0}^{\ast }\geq -\mu \mathbf{1}$ with $\mu
\in \mathbb{R}$ and $D_{0}\in \mathcal{B}\left( \mathfrak{h}\right) $ ($%
D_{0}\neq 0$). Then, the assertions of Lemmata \ref{lemma existence 2
copy(2)}--\ref{lemma existence 2} hold true for all $T\in (0,T_{\max })$.
\end{proposition}

\begin{proof}
Following the proofs of Lemmata \ref{lemma existence 2 copy(2)}--\ref{lemma
existence 2}, one easily checks all the corresponding assertions for all $%
T\in (0,T_{\max })$, except the uniqueness of the strong solution in $%
\mathfrak{h}$ to the initial value problem (\ref{initial value problemlocal}%
) since no contraction mapping principle is used here. This is proven in two
steps. The first one simplifies the estimates for the second step, while
being pivotal to extend the flow to all times in Theorem \ref{section
extension gap}:\medskip

\noindent \underline{Step 1:} By Equation (\ref{eq diff 4}) extended to $%
T=T_{\max }$ (cf. \cite[Lemma 35 (ii)]{bach-bru-memo}) and $\Upsilon
_{0}=\Upsilon _{0}^{\ast }\geq -\mu \mathbf{1}$, we compute that 
\begin{equation}
\partial _{t}\left\{ W_{t,s}^{\ast }\mathrm{e}^{-4\mu \left( t-s\right)
}W_{t,s}\right\} =-4W_{t,s}^{\ast }\left( \left( \Upsilon _{0}+\mu \mathbf{1}%
\right) +\Delta _{t}\right) W_{t,s}\leq 0\ ,  \label{petit inequalitybis}
\end{equation}%
for all $s\in \lbrack 0,T_{\max })$ and $t\in (s,T_{\max })$. Integrating
this we obtain 
\begin{equation*}
\left( W_{t,s}\mathrm{e}^{-2\mu \left( t-s\right) }\right) ^{\ast }W_{t,s}%
\mathrm{e}^{-2\mu \left( t-s\right) }\leq \mathbf{1}
\end{equation*}%
for $s\in \lbrack 0,T_{\max })$ and $t\in (s,T_{\max })$. Therefore, 
\begin{equation}
\left\Vert W_{t,s}\right\Vert _{\mathrm{op}}=\left\Vert W_{t,s}^{\top
}\right\Vert _{\mathrm{op}}\leq \mathrm{e}^{2\mu \left( t-s\right) }\
,\qquad s\in \lbrack 0,T_{\max }),\ t\in \lbrack s,T_{\max })\ .
\label{petit inequalitybisbis}
\end{equation}%
By Equation (\ref{D_explicit_solutionbis}), if follows that 
\begin{equation}
\left\Vert D_{t}\right\Vert _{\mathrm{op}}\leq \mathrm{e}^{4\mu \left(
t-s\right) }\left\Vert D_{s}\right\Vert _{\mathrm{op}}\leq \mathrm{e}^{4\mu
t}\left\Vert D_{0}\right\Vert _{\mathrm{op}}\ ,\qquad s\in \lbrack 0,T_{\max
}),\ t\in \lbrack s,T_{\max })\ .  \label{sdsdsdsdsdsdsdsdsdsdsd}
\end{equation}%
This last bound is in fact crucial to extend below the flow to all times.
\medskip

\noindent \underline{Step 2:} Assume without loss of generality that $\mu >0$%
. For any $T\in (0,T_{\max })$, let $\Delta ^{\left( 1\right) },\Delta
^{\left( 2\right) }\in \mathfrak{C}$ be two (non-zero) solutions to the
initial value problem (\ref{initial value problemlocal}), which also define
via (\ref{definition of W}) two evolution operators $(W_{t,s}^{\Delta
^{\left( 1\right) }})_{t\geq s}$ and $(W_{t,s}^{\Delta ^{\left( 2\right)
}})_{t\geq s}$. Observe from (\ref{sdsdsdsdsdsdsdsdsdsdsd}) that%
\begin{equation*}
r=\sup \left\{ \Vert \Delta ^{\left( 1\right) }\Vert _{\infty },\Vert \Delta
^{\left( 2\right) }\Vert _{\infty }\right\} <\frac{2\left( \mathrm{e}^{8\mu
T}-1\right) }{\mu }\left\Vert D_{0}\right\Vert _{\mathrm{op}}^{2}<\infty \ .
\end{equation*}%
Define 
\begin{equation*}
\delta =\min \left\{ \frac{\ln 2}{2\sup \left\{ \mu ,1\right\} },\frac{\mu 
\mathrm{e}^{-8\mu T}}{600\left\Vert D_{0}\right\Vert _{\mathrm{op}}^{2}}%
,T\right\} >0\ .
\end{equation*}%
Then, in the way we get (\ref{toto2}), by using (\ref{petit inequalitybisbis}%
) and (\ref{sdsdsdsdsdsdsdsdsdsdsd}) we obtain that 
\begin{equation*}
\underset{t\in \left[ 0,\delta \right] }{\sup }\Vert \Delta ^{\left(
1\right) }-\Delta ^{\left( 2\right) }\Vert _{\mathrm{op}}<\frac{1}{2}%
\underset{t\in \left[ 0,\delta \right] }{\sup }\Vert \Delta ^{\left(
1\right) }-\Delta ^{\left( 2\right) }\Vert _{\mathrm{op}}\ ,
\end{equation*}%
i.e., $\Delta _{t}^{\left( 1\right) }=\Delta _{t}^{\left( 2\right) }$ for
all $t\in \lbrack 0,\delta ]$. Now, we can shift the starting point from $%
t=0 $ to any fixed $x\in (0,T]$ to deduce that $\Delta _{t}^{\left( 1\right)
}=\Delta _{t}^{\left( 2\right) }$ for any $t\in \left[ x,x+\delta \right]
\cap \left[ 0,T\right] $. In other words, by recursively using these
arguments, there is a unique solution $\Delta _{t}$ in $\mathfrak{C}$ to the
initial value problem\ (\ref{initial value problemlocal}) for any $T\in
(0,T_{\max })$.
\end{proof}

We are now in a position to prove the existence of a global solution to the
operator flow:

\begin{theorem}[Global solution to the flow]
\label{section extension gap}\mbox{}\newline
Assume $\Upsilon _{0}=\Upsilon _{0}^{\ast }\geq -\mu \mathbf{1}$ with $\mu
\in \mathbb{R}$ and $D_{0}\in \mathcal{B}\left( \mathfrak{h}\right) $ ($%
D_{0}\neq 0$). Then, $T_{\max }=\infty $ and the assertions of Lemmata \ref%
{lemma existence 2 copy(2)}--\ref{lemma existence 2} hold true for $%
T\rightarrow \infty $, i.e., on $\mathbb{R}_{0}^{+}$. If additionally $%
D_{0}^{\top }=\pm D_{0}$, then $D_{t}^{\top }=\pm D_{t}$ for any $t\in 
\mathbb{R}_{0}^{+}$.
\end{theorem}

\begin{proof}
To obtain the assertion, we only need to prove $T_{\max }=\infty $, thanks
to Proposition \ref{lemma existence 2 copy(3)}. Note in this context that $%
D_{0}^{\top }=\pm D_{0}$ clearly implies $D_{t}^{\top }=\pm D_{t}$, since 
\begin{equation*}
\left( D_{t}\right) ^{\top }=\left( W_{t,0}D_{0}W_{t,0}^{\top }\right)
^{\top }=\pm W_{t,0}D_{0}W_{t,0}^{\top }=\pm D_{t}\ .
\end{equation*}%
To prove that $T_{\max }=\infty $, we use the so-called \emph{blow-up
alternative} expressed as follows:\ If $T_{\max }<\infty $ then 
\begin{equation}
\underset{t\rightarrow T_{\max }^{-}}{\lim }\left\Vert D_{t}\right\Vert _{%
\mathrm{op}}=\infty \ .  \label{blowup}
\end{equation}%
If this property holds true, the theorem follows because of Inequality (\ref%
{sdsdsdsdsdsdsdsdsdsdsd}). Therefore, we need to prove (\ref{blowup})
whenever $T_{\max }<\infty $. This done by contradiction: Let $\Upsilon
_{0}=\Upsilon _{0}^{\ast }\geq -\mu \mathbf{1}$ with $\mu \in \mathbb{R}$
and $D_{0}\in \mathcal{B}\left( \mathfrak{h}\right) $ ($D_{0}\neq 0$).
Assume the finiteness of $T_{\max }<\infty $ and the existence of a sequence 
$(t_{n})_{n\in \mathbb{N}}\subseteq \lbrack 0,T_{\max })$ converging to $%
T_{\max }$ such that 
\begin{equation}
\kappa \doteq \underset{n\in \mathbb{N}}{\sup }\left\Vert
D_{t_{n}}\right\Vert _{\mathrm{op}}^{2}<\infty \ .
\label{toto3bisassumption}
\end{equation}%
Let 
\begin{equation*}
\delta _{\kappa }\doteq \min \left\{ \frac{\ln 2}{2\sup \left\{ \mu
,1\right\} },\frac{1}{2^{15/2}\kappa }\right\} >0.
\end{equation*}%
Using essentially the same proof as the one of Lemma \ref{lemma existence 2
copy(2)} with the new starting point $t_{n}$ for each fixed $n\in \mathbb{N}$%
, there exists a unique $(\tilde{\Delta}_{t})_{t\in \lbrack
t_{n},t_{n}+\delta _{\kappa }]}\in \mathfrak{C}$ that is a strong solution
in $\mathfrak{h}$ to the initial value problem 
\begin{equation}
\forall t\in \lbrack t_{n},t_{n}+\delta _{\kappa }]:\qquad \partial _{t}%
\tilde{\Delta}_{t}=16W_{t,t_{n}}D_{t_{n}}W_{t,t_{n}}^{\top }\left(
W_{t,t_{n}}^{\top }\right) ^{\ast }D_{t_{n}}^{\ast }W_{t,t_{n}}^{\ast }\
,\quad \tilde{\Delta}_{t=t_{n}}\doteq \Delta _{t_{n}}\ .
\label{initial problem2}
\end{equation}%
Using the cocycle property as well as (\ref{D_explicit_solutionbis}), one
can thus check that $(\Delta _{t})_{t\in \lbrack t_{n},t_{n}+\delta _{\kappa
}]\cap \lbrack 0,T_{\max })}$ is also a strong solution in $\mathfrak{h}$ to
the initial value problem (\ref{initial problem2}). Hence, $\tilde{\Delta}%
_{t}=\Delta _{t}$ for any $t\in \lbrack t_{n},t_{n}+\delta _{\kappa }]\cap
\lbrack 0,T_{\max })$ and we obtain a solution to (\ref{initial value
problemlocal}) for $T=t_{n}+\delta _{\kappa }$, i.e., for all times in the
interval $[0,t_{n}+\delta _{\kappa }]$. This leads to a contradiction since $%
\delta _{\kappa }>0$ does not depend on $n\in \mathbb{N}$ and $(t_{n})_{n\in 
\mathbb{N}}$ converges to $T_{\max }<\infty $. We thus conclude that either $%
T_{\max }=\infty $ or $\kappa =\infty $.
\end{proof}

\begin{corollary}[Existence and uniqueness of the solution to the flow]
\label{coro uniqueness}\mbox{}\newline
Assume $\Upsilon _{0}=\Upsilon _{0}^{\ast }\geq -\mu \mathbf{1}$ with $\mu
\in \mathbb{R}$ and $D_{0}\in \mathcal{B}\left( \mathfrak{h}\right) $ ($%
D_{0}\neq 0$). Then, there exists a unique solution $\Delta \equiv (\Delta
_{t})_{t\geq 0},D\equiv (D_{t})_{t\geq 0}$ of strongly continuous mappings
on $\mathcal{B}\left( \mathfrak{h}\right) $ to the initial value problem (%
\ref{flow equation-quadratic deltabis}).
\end{corollary}

\begin{proof}
The existence of a solution to (\ref{flow equation-quadratic delta}) is
given by Theorem \ref{section extension gap}. Then, the uniqueness is proven
in the following way: Take a pair $(\tilde{\Delta},\tilde{D})$ of strongly
continuous mappings on $\mathcal{B}\left( \mathfrak{h}\right) $ which is
solution to the initial value problem (\ref{flow equation-quadratic deltabis}%
). Using the two-parameter operator family $((W_{t,s}(\tilde{\Delta}%
))_{t\geq s}$ defined by (\ref{definition of W}) and arguments similar to 
\cite[Corollary 37]{bach-bru-memo}, one deduces that (\ref%
{D_explicit_solutionbis}) holds true for $\tilde{D}$, meaning that $\tilde{%
\Delta}$ solves (\ref{initial value problemlocal}) for all $T\in \mathbb{R}%
^{+}$, like the first operator family $\Delta $. By uniqueness of the
solution to (\ref{initial value problemlocal}) (see Theorem \ref{section
extension gap}), $\Delta =\tilde{\Delta}$, which in turn implies $\tilde{D}%
=D $, thanks to (\ref{D_explicit_solutionbis}).
\end{proof}

\subsection{Extension to the Schatten Topology\label{Section tech3}}

From now on, $\Delta$ and $D$ are always given by Theorem \ref{section
extension gap}:

\begin{itemize}
\item $\Delta $ is the strong solution in $\mathfrak{h}$ to the initial
value problem (\ref{initial value problemlocal}). See Lemma \ref{lemma
existence 2 copy(2)}, which holds true even for $T\rightarrow \infty $,
i.e., on $\mathbb{R}_{0}^{+}$.

\item The operator-valued function $D:[0,T]\rightarrow \mathcal{B}(\mathfrak{%
h})$ is defined by (\ref{D_explicit_solutionbis}). See Lemma \ref{lemma
existence 2}, which holds true even for $T\rightarrow \infty $, i.e., on $%
\mathbb{R}_{0}^{+}$.
\end{itemize}

\noindent Recall that $C\left( \mathbb{R}_{0}^{+};\mathcal{X}\right) $
denotes the space of continuous functions from $\mathbb{R}_{0}^{+}$ to a
Banach space $\mathcal{X}$.

 For the reader's convenience, let us start by reminding the H\"{o}%
lder inequality for Schatten norms: Using the usual notation $\mathcal{B}(%
\mathfrak{h})\equiv \mathcal{L}^{\infty }(\mathfrak{h})$ and $\left\Vert
\cdot \right\Vert _{\mathrm{op}}\equiv \left\Vert \cdot \right\Vert _{\infty
}$ as well as the convention $1/\infty =0$, for any $r,p,q\in \lbrack
1,\infty ]$ satisfying $1/p+1/q=1/r$,%
\begin{equation*}
\left\Vert XY\right\Vert _{r}\leq \left\Vert X\right\Vert _{p}\left\Vert
Y\right\Vert _{q},\qquad X\in \mathcal{L}^{p}\left( \mathfrak{h}\right) ,\
Y\in \mathcal{L}^{q}\left( \mathfrak{h}\right) .
\end{equation*}%
Now, if $D_{0}\in \mathcal{L}^{2p}(\mathfrak{h})$ for $p\in \lbrack 1,\infty
)$ then we deduce from (\ref{D_explicit_solutionbis}) and (\ref{petit
inequalitybisbis}) together with the H\"{o}lder inequality that $D_{t}\in 
\mathcal{L}^{2p}(\mathfrak{h})$ for all $t\in \mathbb{R}_{0}^{+}$ with
Schatten $2p$--norm bounded as follows: 
\begin{equation}
\left\Vert D_{t}\right\Vert _{2p}\leq \mathrm{e}^{4\mu \left( t-s\right)
}\left\Vert D_{s}\right\Vert _{2p}\leq \mathrm{e}^{4\mu t}\left\Vert
D_{0}\right\Vert _{2p}\ ,\qquad s\in \mathbb{R}_{0}^{+},\ t\in \lbrack
s,\infty )\ .  \label{petit inequalitybisbis2}
\end{equation}%
See also Theorem \ref{section extension gap}. In fact, one gets the
following properties:

\begin{lemma}[Continuity in the Schatten topology]
\label{Lemma HS1 copy(1)}\mbox{}\newline
Let $p\in \lbrack 1,\infty ]$. Assume $\Upsilon _{0}=\Upsilon _{0}^{\ast
}\geq -\mu \mathbf{1}$ with $\mu \in \mathbb{R}$ and $D_{0}\in \mathcal{L}%
^{2p}(\mathfrak{h})$ ($D_{0}\neq 0$).

\begin{enumerate}
\item[\emph{(i)}] $\Delta $\ is locally Lipschitz continuous in $\mathcal{L}%
^{p}(\mathfrak{h})$ on $\mathbb{R}_{0}^{+}$.

\item[\emph{(ii)}] $D$ is locally Lipschitz continuous in $\mathcal{L}^{2p}(%
\mathfrak{h})$ on $\mathbb{R}^{+}$.
\end{enumerate}
\end{lemma}

\begin{proof}
By the triangle and H\"{o}lder inequalities, for all $s\in \mathbb{R}_{0}^{+}
$ and $t\in \left[ s,+\infty \right) $, 
\begin{equation*}
\left\Vert \Delta _{t}-\Delta _{s}\right\Vert _{p}\leq 16\left( t-s\right)
\sup_{\tau \in \left[ s,t\right] }\left\Vert D_{\tau }\right\Vert _{2p}^{2}\
,
\end{equation*}%
which proves Assertion (i), thanks to (\ref{petit inequalitybisbis2}).
Additionally, by (\ref{D_explicit_solutionbis}) together with the triangle
and H\"{o}lder inequalities,%
\begin{equation*}
\left\Vert D_{t}-D_{s}\right\Vert _{2p}\leq \left( \left\Vert \left(
W_{t,0}-W_{s,0}\right) \right\Vert _{\mathrm{op}}\left\Vert
W_{s,0}\right\Vert _{\mathrm{op}}+\left\Vert W_{t,0}\right\Vert _{\mathrm{op}%
}\left\Vert \left( W_{t,0}-W_{s,0}\right) \right\Vert _{\mathrm{op}}\right)
\left\Vert D_{0}\right\Vert _{2p}\ .
\end{equation*}
By using the proofs of \cite[Lemmata 33 and 35]{bach-bru-memo} to the
evolution family $(W_{t,0})_{t\geq 0}$ defined by (\ref{definition of W}),
one shows that, for any $\delta \in \mathbb{R}^{+}$, the mapping $t\mapsto
W_{t,0}$ is Lipschitz continuous on $[\delta ,\infty )$ in the norm
topology. Using the last inequality and (\ref{petit inequalitybisbis}), we
thus deduce Assertion (ii). As compared with \cite[Theorem 45 (i)]%
{bach-bru-memo} note finally that we only have a local Lipschitz continuity
in (i)--(ii), because of the upper bounds (\ref{petit inequalitybisbis}) and
(\ref{petit inequalitybisbis2}). The global Lipschitz continuity of $\Delta $
and $D$ holds true at least in the case $\mu \leq 0$.
\end{proof}

Because of this last lemma, one may wonder whether the flow proven in the
strong (operator) topology could be established in the Schatten topology,
provided $D_{0}\in \mathcal{L}^{2p}(\mathfrak{h})$ for some $p\in \lbrack
1,\infty ]$. We explain this by starting with the more important case of the
Hilbert-Schmidt topology, i.e., for $p=1$.  

\begin{lemma}[Hilbert-Schmidt continuity]
\label{Lemma HS1}\mbox{}\newline
Assume $\Upsilon _{0}=\Upsilon _{0}^{\ast }\geq -\mu \mathbf{1}$ with $\mu
\in \mathbb{R}$ and $D_{0}\in \mathcal{L}^{2}(\mathfrak{h})$ ($D_{0}\neq 0$%
). Then, $D\in C(\mathbb{R}_{0}^{+};\mathcal{L}^{2}(\mathfrak{h}))$ and the
operator families $(\Upsilon _{t}D_{t})_{t\in \mathbb{R}^{+}}$ and $%
(D_{t}\Upsilon _{t}^{\top })_{t\in \mathbb{R}^{+}}$ are continuous in $%
\mathcal{L}^{2}(\mathfrak{h})$, where $\Upsilon _{t}\doteq \Upsilon
_{0}+\Delta _{t}$ for any $t\in \mathbb{R}_{0}^{+}$.
\end{lemma}

\begin{proof}
The proof is  neither trivial nor short , but it is identical to
the one of \cite[Theorem 45]{bach-bru-memo}. Detailed arguments are thus
omitted. In fact, one basically uses Lebesgue's dominated convergence
theorem, cyclicity of the trace and a direct adaptation of the proofs of 
\cite[Lemmata 33 and 35]{bach-bru-memo} to the evolution family $%
(W_{t,s})_{t\geq s\geq 0}$ defined by (\ref{definition of W}). For instance, 
$(\Upsilon _{t}D_{t})_{t\in \mathbb{R}^{+}}$ belongs to $\mathcal{L}^{2}(%
\mathfrak{h})$ as a consequence of the following fact: For any $s\in \mathbb{%
R}_{0}^{+}$ and $t\in (s,\infty )$, 
\begin{equation}
\left\Vert \Upsilon _{t}W_{t,s}\right\Vert _{\mathrm{op}}\leq
C_{1}+C_{2}\left( t-s\right) ^{-1}  \label{hhhhh}
\end{equation}%
with $C_{1},C_{2}<\infty $. Compare this upper bound with \cite[Lemma 35 (i)]%
{bach-bru-memo}. Since $(\Upsilon _{t}W_{t,s})_{t>s\geq 0}$ is additionally
strongly continuous in $t>s$ (cf. \cite[Lemma 35 (i)]{bach-bru-memo}), one
uses Lebesgue's dominated convergence theorem and $D_{0}\in \mathcal{L}^{2}(%
\mathfrak{h})$ to deduce that $(\Upsilon _{t}D_{t})_{t\in \mathbb{R}^{+}}$
is continuous in $\mathcal{L}^{2}(\mathfrak{h})$. All the other assertions
are proven in a similar way.
\end{proof}

As a consequence, the flow holds true in the Hilbert-Schmidt topology:

\begin{theorem}[Well--posedness of the flow on $\mathcal{L}^{2}\left( 
\mathfrak{h}\right) $]
\label{Corollary existence}\mbox{}\newline
Assume $\Upsilon _{0}=\Upsilon _{0}^{\ast }\geq -\mu \mathbf{1}$ with $\mu
\in \mathbb{R}$ and $D_{0}\in \mathcal{L}^{2}(\mathfrak{h})$ ($D_{0}\neq 0$%
). Then, $(\Delta ,D)\in C(\mathbb{R}_{0}^{+};\mathcal{L}^{1}(%
\mathfrak{h})\times \mathcal{L}^{2}(\mathfrak{h}))$ is the unique
solution to the system of differential equations 
\begin{equation}
\left\{ 
\begin{array}{lllll}
\forall t\in \mathbb{R}_{0}^{+} & : & \partial _{t}X_{t}=16Y_{t}Y_{t}^{\ast }
& , & X_{0}=0\ , \\ 
\forall t\in \mathbb{R}^{+} & : & \partial _{t}Y_{t}=-2\left( \left(
X_{t}+\Upsilon _{0}\right) Y_{t}+Y_{t}\left( X_{t}+\Upsilon _{0}\right)
^{\top }\right) & , & Y_{0}=D_{0}\ ,%
\end{array}%
\right.  \label{toto dif eq}
\end{equation}%
in $\mathcal{L}^{1}(\mathfrak{h})$ and $\mathcal{L}^{2}(\mathfrak{h})$,
respectively.
\end{theorem}

\begin{proof}
Note first that $(\Delta ,D)\in C(\mathbb{R}_{0}^{+};\mathcal{L}^{1}(%
\mathfrak{h})\times \mathcal{L}^{2}(\mathfrak{h}))$ directly results from
Lemmata \ref{Lemma HS1 copy(1)} and \ref{Lemma HS1}. Next, we infer from (%
\ref{petit inequalitybisbis2}) and Theorem \ref{section extension gap} (cf. (%
\ref{initial value problemlocal}) and (\ref{sdsdsd})) together with  the triangle and H\"{o}lder inequalities that, for any $t\in \mathbb{R}%
_{0}^{+}$, $\delta \in \mathbb{R}^{+}$ and every $p\in \lbrack 1,\infty ]$, 
\begin{equation}
\left\Vert \delta ^{-1}\left\{ \Delta _{t+\delta }-\Delta _{t}\right\}
-16D_{t}D_{t}^{\ast }\right\Vert _{p}\leq 16\left( \mathrm{e}^{4\mu \delta
}+1\right) \delta ^{-1}\int_{t}^{t+\delta }\left\Vert D_{t}-D_{\tau
}\right\Vert _{2p}\left\Vert D_{t}\right\Vert _{2p}\mathrm{d}\tau
\label{strengthening 1}
\end{equation}%
and, using the notation $\Upsilon _{t}\doteq \Upsilon _{0}+\Delta _{t}$, 
\begin{eqnarray}
&&\left\Vert \delta ^{-1}\left\{ D_{t+\delta }-D_{t}\right\} +2\Upsilon
_{t}D_{t}+2D_{t}\Upsilon _{t}^{\top }\right\Vert _{2p}
\label{strengthening 2} \\
&\leq &2\delta ^{-1}\int_{t}^{t+\delta }\left\{ \left\Vert \Upsilon
_{t}D_{t}-\Upsilon _{\tau }D_{\tau }\right\Vert _{2p}+\left\Vert
D_{t}\Upsilon _{t}^{\top }-D_{\tau }\Upsilon _{\tau }^{\top }\right\Vert
_{2p}\right\} \ \mathrm{d}\tau \ .  \notag
\end{eqnarray}%
Mutadis mutandis for sufficiently small $\delta <0$ such that $t+\delta
\geq 0$. From (\ref{strengthening 1})--(\ref{strengthening 2}) at $p=1$
together with the Lebesgue differentiation theorem whose hypothesis are
fulfilled by virtue of Lemma \ref{Lemma HS1}, we infer that $(\Delta ,D)$ is
a solution to the system of differential equations (\ref{toto dif eq}).
Finally, if $(X,Y)\in C(\mathbb{R}_{0}^{+};\mathcal{L}^{2}\left( \mathfrak{h}%
\right) ^{2})$, then $t\mapsto X_{t}$ is a strongly continuous mapping\ from 
$\mathbb{R}_{0}^{+}$ to $\mathcal{B}\left( \mathfrak{h}\right) $ and $Y\in
C\left( \mathbb{R}_{0}^{+};\mathcal{B}(\mathfrak{h})\right) $. If $(X,Y)$
additionally solves (\ref{toto dif eq}), then $X\ $is a strong solution in $%
\mathfrak{h}$ to the initial value problem (\ref{initial value problemlocal}%
) for $T\rightarrow \infty $, i.e., on $\mathbb{R}_{0}^{+}$, while $Y\ $is
the unique strong solution on the domain $\mathcal{D}(\Upsilon _{0}^{\top })$
to (\ref{sdsdsd}), again for $T\rightarrow \infty $. By Theorem \ref{section
extension gap}, we deduce that $X=\Delta $ and $Y=D$. (Uniqueness of the
family $(\Delta ,D)$ can also be deduced from Theorem \ref{section extension
gap} and \cite[Lemma 99]{bach-bru-memo}, which can be extended to the lower
semibounded case.)
\end{proof}

 If the initial data $D_{0}$ belongs to $\mathcal{L}^{2}(\mathfrak{h})$%
, the last theorem shows that the operator-valued flow (\ref{flow
equation-quadratic delta}) becomes an ODEs on the Hilbert space $\mathcal{L}%
^{2}(\mathfrak{h})$ of Hilbert-Schmidt operators. One can ask whether this
observation can be generalized to any $L^{p}$--spaces $\mathcal{L}^{p}(%
\mathfrak{h})$, $p\in \lbrack 1,\infty )$, constructed from Schatten norms.
See in particular Lemma \ref{Lemma HS1 copy(1)}. We give a partial answer to
this question via the following assertion:

\begin{theorem}[Well--posedness of the flow on $\mathcal{L}^{p}\left( 
\mathfrak{h}\right) \times \mathcal{L}^{2p}\left( \mathfrak{h}\right) $]
\label{Corollary existence copy(1)}\mbox{}\newline
Let $p\in \lbrack 1,\infty ]$ and $\Upsilon _{0}\in \mathcal{B}(\mathfrak{h}%
) $. Assume $D_{0}\in \mathcal{L}^{2p}(\mathfrak{h})$ ($D_{0}\neq 0$). Then, 
$(\Delta ,D)\in C(\mathbb{R}_{0}^{+};\mathcal{L}^{p}(\mathfrak{h})\times 
\mathcal{L}^{2p}(\mathfrak{h}))$ is the unique solution to the system of
differential equations 
\begin{equation*}
\left\{ 
\begin{array}{lllll}
\forall t\in \mathbb{R}_{0}^{+} & : & \partial _{t}X_{t}=16Y_{t}Y_{t}^{\ast }
& , & X_{0}=0\ , \\ 
\forall t\in \mathbb{R}_{0}^{+} & : & \partial _{t}Y_{t}=-2\left( \left(
X_{t}+\Upsilon _{0}\right) Y_{t}+Y_{t}\left( X_{t}+\Upsilon _{0}\right)
^{\top }\right) & , & Y_{0}=D_{0}\ ,%
\end{array}%
\right.
\end{equation*}%
in $\mathcal{L}^{p}(\mathfrak{h})$ and $\mathcal{L}^{2p}(\mathfrak{h})$,
respectively.
\end{theorem}

\begin{proof}
If $\Upsilon _{0}\in \mathcal{B}(\mathfrak{h})$, observe that the mapping $%
t\mapsto W_{t,0}$ defined by (\ref{definition of W}) become continuous on $%
\mathbb{R}_{0}^{+}$ in the norm topology. For more details, see below
Equations (\ref{eq-9})--(\ref{smooth funcions}). Then, by combining (\ref%
{D_explicit_solutionbis}) with the H\"{o}lder inequality, we deduce that $%
D\in C(\mathbb{R}_{0}^{+};\mathcal{L}^{2p}(\mathfrak{h}))$. In the same way,
the operator families $(\Upsilon _{t}D_{t})_{t\in \mathbb{R}_{0}^{+}}$ and $%
(D_{t}\Upsilon _{t}^{\top })_{t\in \mathbb{R}_{0}^{+}}$ are continuous in $%
\mathcal{L}^{2p}(\mathfrak{h})$, thanks to Lemma \ref{Lemma HS1 copy(1)}
(i). (Use also that $\Vert X\Vert _{p}\geq \Vert X\Vert _{\mathrm{op}}$ for
any $p\in \lbrack 1,\infty ]$). Therefore, we can use (\ref{strengthening 1}%
)--(\ref{strengthening 2}) and similar inequalities for sufficiently small $%
\delta <0$ such that $t+\delta \geq 0$ in order to deduce that $(\Delta ,D)$
is a solution to the system of differential equations (\ref{toto dif eq}).
The uniqueness of the solution is proven like in Theorem \ref{Corollary
existence}.
\end{proof}

To generalize Theorem \ref{Corollary existence copy(1)} for lower
semibounded $\Upsilon _{0}$, as in Theorem \ref{Corollary existence} for $%
p=1 $, one needs to extend Lemma \ref{Lemma HS1} to the Schatten (norm)
topology. One can follow the same line of arguments, used here for the
Hilbert-Schmidt topology. This can only be possible when $D_{0}\in \mathcal{L%
}^{2q}(\mathfrak{h})$ ($D_{0}\neq 0$) for $q=2^{p-1}$ with $p\in \mathbb{N}$%
. Even if it seems to work perfectly, it leads to cumbersome induction
arguments, all using the properties of the trace. We refrain from doing so,
as the assumption $D_{0}\in \mathcal{L}^{2q}(\mathfrak{h})$ for $q=2^{p-1}$
with $p\in \mathbb{N}$ is very special whereas the Hilbert-Schmidt case is
our main interest here, in view of the applications in quantum field theory.

\subsection{Constants of Motion\label{Section tech4}}

Again, $\Delta $ and $D$ are always given by Theorem \ref{section extension
gap}. In this section we study the constant of motion of the flow.
Note that the unbounded case $\Upsilon _{0}\notin \mathcal{B}\left( 
\mathfrak{h}\right) $ yields domain issues that obscure the general ideas
and the arguments. So, we first deliberately restrict our first proofs to
bounded $\Upsilon _{0}\in \mathcal{B}\left( \mathfrak{h}\right) $ in order
to make clear the key arguments without too much technical estimates.  

If $\Upsilon _{0}=\Upsilon _{0}^{\ast }\in \mathcal{B}\left( \mathfrak{h}%
\right) $ and $D_{0}\in \mathcal{B}\left( \mathfrak{h}\right) $ ($D_{0}\neq
0 $) then the evolution operator $(W_{t,s})_{t\geq s}$ defined by (\ref%
{definition of W}), with $\Delta :\mathbb{R}_{0}^{+}\rightarrow \mathcal{B}%
\left( \mathfrak{h}\right) $ being taken from Theorem \ref{section extension
gap}, can be written as a usual Dyson series: For any $s,t\in \mathbb{R}%
_{0}^{+}$, 
\begin{equation}
W_{t,s}=\mathbf{1}+\sum_{n=1}^{\infty }\left( -2\right) ^{n}\int_{s}^{t}%
\mathrm{d}\tau _{1}\cdots \int_{s}^{\tau _{n-1}}\mathrm{d}\tau
_{n}\;\Upsilon _{\tau _{1}}\cdots \Upsilon _{\tau _{n}}  \label{eq-9}
\end{equation}%
with $\Upsilon _{t}\doteq \Upsilon _{0}+\Delta _{t}$ for any $t\in \mathbb{R}%
^{+}$, keeping in mind that $\Delta _{t}$ is continuous in the (operator)
norm topology. In particular, it is norm-continuous for all times, including 
$t=0$, and satisfies the following non-autonomous evolution equations 
\begin{equation}
\left\{ 
\begin{array}{llllll}
\forall s,t\in \mathbb{R}_{0}^{+} & : & \partial _{t}W_{t,s}=-2\Upsilon
_{t}W_{t,s} & , & W_{s,s}\doteq \mathbf{1} & . \\ 
\forall s,t\in \mathbb{R}_{0}^{+} & : & \partial
_{s}W_{t,s}=2W_{t,s}\Upsilon _{s} & , & W_{t,t}\doteq \mathbf{1} & .%
\end{array}%
\right.  \label{eq 10}
\end{equation}%
The derivatives with respect to $t$ and $s$ are done in the Banach space $%
\mathcal{B}\left( \mathfrak{h}\right) $, i.e., in the norm topology. The
operator-valued function $D:\mathbb{R}_{0}^{+}\rightarrow \mathcal{B}(%
\mathfrak{h})$ defined by (\ref{D_explicit_solutionbis}) (for $T\rightarrow
\infty $) then satisfies 
\begin{equation}
\partial _{t}D_{t}=-2\left( \Upsilon _{t}D_{t}+D_{t}\Upsilon _{t}^{\top
}\right) \ ,\qquad t\in \mathbb{R}_{0}^{+}\ ,  \label{derivative D t0}
\end{equation}%
including $t=0$, in $\mathcal{B}\left( \mathfrak{h}\right) $. It is in
particular locally Lipschitz continuous in the norm topology for all times.
Therefore, using a bootstrap argument and Leibniz rule for 
differentiable  bounded-operator families (as is done in Lemma \ref{lemma
constante of motion}), one deduces in this case that 
\begin{equation}
\left( D_{t}\right) _{t\in \mathbb{R}_{0}^{+}},\left( \Delta _{t}\right)
_{t\in \mathbb{R}_{0}^{+}}\in C^{\infty }\left( \mathbb{R}_{0}^{+};\mathcal{B%
}\left( \mathfrak{h}\right) \right) \ .  \label{smooth funcions}
\end{equation}%
Using this technical (albeit non necessary) simplification and the operators 
\begin{equation}
K_{t}\doteq \Upsilon _{t}D_{t}-D_{t}\Upsilon _{t}^{\top }\ ,\qquad t\in 
\mathbb{R}_{0}^{+}\ ,  \label{commutator type}
\end{equation}%
one arrives at two interesting lemmata:

\begin{lemma}[Towards a constant of motion -- I]
\label{lemma constante of motion}\mbox{}\newline
Assume $\Upsilon _{0}=\Upsilon _{0}^{\ast }\in \mathcal{B}\left( \mathfrak{h}%
\right) $ and $D_{0}\in \mathcal{B}\left( \mathfrak{h}\right) $ ($D_{0}\neq
0 $). Let $\Upsilon _{t}\doteq \Upsilon _{0}+\Delta _{t}$ for each $t\in 
\mathbb{R}_{0}^{+}$. For any $s\in \mathbb{R}_{0}^{+}$ and $t\in \lbrack
s,\infty )$,%
\begin{equation*}
\Upsilon _{t}^{2}+4D_{t}D_{t}^{\ast }=\Upsilon _{s}^{2}+4D_{s}D_{s}^{\ast
}+8\int_{s}^{t}\left( K_{\tau }D_{\tau }^{\ast }+D_{\tau }K_{\tau }^{\ast
}\right) \mathrm{d}\tau \ .
\end{equation*}
\end{lemma}

\begin{proof}
Using the equality 
\begin{equation*}
\left( \Upsilon _{t}^{\top }\right) ^{\ast }=\overline{\Upsilon }%
_{t}=\Upsilon _{t}^{\top },\qquad t\in \mathbb{R}_{0}^{+}\ ,
\end{equation*}%
we compute from Theorem \ref{section extension gap} and (\ref{derivative D
t0}) that, for any $t\in \mathbb{R}_{0}^{+}$, 
\begin{equation}
\partial _{t}\left\{ D_{t}D_{t}^{\ast }\right\} =-2\left( \Upsilon
_{t}D_{t}D_{t}^{\ast }+D_{t}D_{t}^{\ast }\Upsilon _{t}\right)
-4D_{t}\Upsilon _{t}^{\top }D_{t}^{\ast }  \label{eq sup 0}
\end{equation}%
as well as 
\begin{equation}
\partial _{t}\left\{ \Upsilon _{t}^{2}+4D_{t}D_{t}^{\ast }\right\} =8\left(
\Upsilon _{t}D_{t}-D_{t}\Upsilon _{t}^{\top }\right) D_{t}^{\ast
}+8D_{t}\left( D_{t}^{\ast }\Upsilon _{t}-\Upsilon _{t}^{\top }D_{t}^{\ast
}\right) \ ,  \label{eq sup 1}
\end{equation}%
both in the Banach space $\mathcal{B}\left( \mathfrak{h}\right) $. To
compute these derivatives, note that one verifies Leibniz rule for 
differentiable    bounded-operator families. This is easily checked. For
instance, using the limit $\delta \rightarrow 0^{+}$ of the
inequality 
\begin{eqnarray}
\left\Vert \delta ^{-1}\left( \Upsilon _{t\pm \delta }^{2}-\Upsilon
_{t}^{2}\right) -\left( \partial _{t}\Upsilon _{t}\right) \Upsilon
_{t}-\Upsilon _{t}\left( \partial _{t}\Upsilon _{t}\right) \right\Vert _{%
\mathrm{op}} &\leq &\left\Vert \Upsilon _{t\pm \delta }\right\Vert _{\mathrm{%
op}}\left\Vert \delta ^{-1}\left( \Upsilon _{t\pm \delta }-\Upsilon
_{t}\right) -\partial _{t}\Upsilon _{t}\right\Vert _{\mathrm{op}}  \notag \\
&&+\left\Vert \Upsilon _{t}\right\Vert _{\mathrm{op}}\left\Vert \delta
^{-1}\left( \Upsilon _{t\pm \delta }-\Upsilon _{t}\right) -\partial
_{t}\Upsilon _{t}\right\Vert _{\mathrm{op}}  \notag \\
&&+\left\Vert \partial _{t}\Upsilon _{t}\right\Vert _{\mathrm{op}}\left\Vert
\Upsilon _{t\pm \delta }-\Upsilon _{t}\right\Vert _{\mathrm{op}}
\label{sup23}
\end{eqnarray}%
for any $\delta \in \mathbb{R}\backslash \{0\}$ and $t\in \mathbb{R}_{0}^{+}$
such that $t\pm \delta \geq 0$, we deduce that 
\begin{equation*}
\partial _{t}\Upsilon _{t}^{2}=\left( \partial _{t}\Upsilon _{t}\right)
\Upsilon _{t}+\Upsilon _{t}\left( \partial _{t}\Upsilon _{t}\right)
\end{equation*}%
in $\mathcal{B}\left( \mathfrak{h}\right) $. The lemma then follows from (%
\ref{eq sup 1}). 
\end{proof}

The operator family $(K_{t})_{t\in \mathbb{R}_{0}^{+}}$ defined by (\ref%
{commutator type}) and appearing in Lemma \ref{lemma constante of motion} is
a kind of commutator. Interestingly, as is observed in \cite[Lemma 53]%
{bach-bru-memo} for the hyperbolic case, this family can be explicitly
written in a very similar way to the operator-valued function $D$:

\begin{lemma}[Towards a constant of motion -- II]
\label{lemma constante of motion copy(7)}\mbox{}\newline
Assume $\Upsilon _{0}=\Upsilon _{0}^{\ast }\in \mathcal{B}\left( \mathfrak{h}%
\right) $ and $D_{0}\in \mathcal{B}\left( \mathfrak{h}\right) $ ($D_{0}\neq
0 $). Let $\Upsilon _{t}\doteq \Upsilon _{0}+\Delta _{t}$ for each $t\in 
\mathbb{R}_{0}^{+}$. If $D_{0}^{\top }=\pm D_{0}$ then 
\begin{equation*}
K_{t}\doteq \Upsilon _{t}D_{t}-D_{t}\Upsilon _{t}^{\top
}=W_{t,s}K_{s}W_{t,s}^{\top }\ ,\qquad t,s\in \mathbb{R}_{0}^{+}\ .
\end{equation*}
\end{lemma}

\begin{proof}
Assume $D_{0}^{\top }=\pm D_{0}$ and $\Upsilon _{0}\in \mathcal{B}\left( 
\mathfrak{h}\right) $, which avoids technical complications associated with
domain issues. By Theorem \ref{section extension gap}, i.e., Lemmata \ref%
{lemma existence 2 copy(2)}--\ref{lemma existence 2} for $T\rightarrow
\infty $, and Leibniz rule for  differentiable  bounded-operator
families, one checks that%
\begin{equation}
\partial _{t}K_{t}=-2\left( \Upsilon _{t}K_{t}+K_{t}\Upsilon _{t}^{\top
}\right) \ ,\qquad t\in \mathbb{R}_{0}^{+}\ ,  \label{K derivative3}
\end{equation}%
in the Banach space $\mathcal{B}\left( \mathfrak{h}\right) $. Note that we
also use above (\ref{derivative D t0}) to include $t=0$ as well as the
equality $D_{t}^{\top }=\pm D_{t}$ which holds true for any $t\in \mathbb{R}%
_{0}^{+}$, provided $D_{0}^{\top }=\pm D_{0}$. By combining (\ref{K
derivative3}) with (\ref{eq 10}) it follows that, for any $t,s\in \mathbb{R}%
_{0}^{+}$, 
\begin{equation*}
\partial _{s}\{W_{t,s}K_{s}W_{t,s}^{\top }\}=0\ ,
\end{equation*}%
in $\mathcal{B}\left( \mathfrak{h}\right) $. This implies the assertion.
\end{proof}

 The extension of Lemmata \ref{lemma constante of motion}--\ref{lemma
constante of motion copy(7)} to lower semibounded operators $\Upsilon _{0}$ 
is cumbersome to do it concretely, in detail. Therefore, we only extend
Lemma \ref{lemma constante of motion copy(7)} to the semibounded case, which
is the more interesting result.

To this end, we use the operators 
\begin{equation}
\mathfrak{d}_{t}\doteq \left( \Upsilon _{0}+\mu \mathbf{1}\right) \Delta
_{t}\left( \Upsilon _{0}+\mu \mathbf{1}\right) ^{-1}\quad \mathrm{and}\quad 
\boldsymbol{V}_{t,s}\doteq \left( \Upsilon _{0}+\mu \mathbf{1}\right)
W_{t,s}\left( \Upsilon _{0}+\mu \mathbf{1}\right) ^{-1}\ ,
\label{evoluation operators2}
\end{equation}%
for any $t\in \mathbb{R}_{0}^{+}$ and $s\in \lbrack 0,t]$. If $\Upsilon
_{0}\geq -\left( \mu -\varepsilon \right) \mathbf{1}$ for some $\mu \in 
\mathbb{R}$ and $\varepsilon \in \mathbb{R}^{+}$ and 
\begin{equation}
\max \left\{ \Vert (\Upsilon _{0}+\mu \mathbf{1)}D_{0}(\Upsilon _{0}^{\top
}+\mu \mathbf{1)}^{-1}\Vert _{\mathrm{op}},\Vert (\Upsilon _{0}^{\top }+\mu 
\mathbf{1)}D_{0}^{\ast }(\Upsilon _{0}+\mu \mathbf{1)}^{-1}\Vert _{\mathrm{op%
}}\right\} \leq \mathrm{C}  \label{conditionplus}
\end{equation}%
for some strictly positive constant $\mathrm{C}\in \mathbb{R}^{+}$, then the
operators of (\ref{evoluation operators2}) are well-defined and satisfy the
following properties:

\begin{lemma}[$\boldsymbol{V}_{t,s}$ as an evolution operator]
\label{lemma constante of motion copy(8)}\mbox{ }\newline
Take a self-adjoint operator $\Upsilon _{0}=\Upsilon _{0}^{\ast }$ and $\mu
\in \mathbb{R}$ such that $\Upsilon _{0}\geq -\left( \mu -\varepsilon
\right) \mathbf{1}$ for some $\varepsilon \in \mathbb{R}^{+}$. Let $%
D_{0}=\pm D_{0}^{\top }\in \mathcal{B}\left( \mathfrak{h}\right) $ ($%
D_{0}\neq 0$). If (\ref{conditionplus}) holds true then one has:

\begin{enumerate}
\item[\emph{(i)}] $(\mathfrak{d}_{t})_{t\geq 0}\in C(\mathbb{R}_{0}^{+};%
\mathcal{B}(\mathfrak{h}))$, while $(\boldsymbol{V}_{t,s})_{t\geq s\geq
0}\subseteq \mathcal{B}(\mathfrak{h})$.

\item[\emph{(ii)}] $\boldsymbol{V}_{t,s}$ satisfies the cocycle property\ $%
\boldsymbol{V}_{t,x}\boldsymbol{V}_{x,s}=\boldsymbol{V}_{t,s}$ for any $%
s,x,t\in \mathbb{R}_{0}^{+}$ so that $t\geq x\geq s$.

\item[\emph{(iii)}] $(\boldsymbol{V}_{t,s})_{t\geq s\geq 0}$ is jointly
strongly continuous in $s$ and $t$.

\item[\emph{(iv)}] The evolution family $(\boldsymbol{V}_{t,s})_{t\geq s\geq
0}$ is the solution to the non-autonomous evolution equations%
\begin{equation}
\left\{ 
\begin{array}{llllll}
\forall s\in \mathbb{R}_{0}^{+},\ t\in (s,\infty ) & : & \partial _{t}%
\boldsymbol{V}_{t,s}=-2\left( \Upsilon _{0}+\mathfrak{\Delta }_{t}\right) 
\boldsymbol{V}_{t,s} & , & \boldsymbol{V}_{s,s}:=\mathbf{1} & . \\ 
\forall t\in \mathbb{R}^{+},\ s\in \lbrack 0,t] & : & \partial _{s}%
\boldsymbol{V}_{t,s}=2\boldsymbol{V}_{t,s}\left( \Upsilon _{0}+\mathfrak{%
\Delta }_{s}\right) & , & \boldsymbol{V}_{t,t}:=\mathbf{1} & .%
\end{array}%
\right.
\end{equation}%
The derivatives with respect to $t$ and $s$ are in the strong sense in $%
\mathfrak{h}$ and $\mathcal{D}\left( \Upsilon _{0}\right) $, respectively.
\end{enumerate}

Similar properties as (i)--(iv) also hold for the mappings 
\begin{equation}
t\mapsto (\Upsilon _{0}^{\sharp }+\mu \mathbf{1)}\Delta _{t}^{\sharp
}(\Upsilon _{0}^{\sharp }+\mu \mathbf{1)}^{-1}\ ,\quad \left( s,t\right)
\mapsto \boldsymbol{V}_{t,s,\sharp }\doteq (\Upsilon _{0}^{\sharp }+\mu 
\mathbf{1)}W_{t,s}^{\sharp }(\Upsilon _{0}^{\sharp }+\mu \mathbf{1)}^{-1}\ ,
\label{fglkh}
\end{equation}%
where $\sharp =\top ,\ast $. \newline
\end{lemma}

\begin{proof}
The proof is done in several steps. Note that the corresponding arguments
for the other mappings (\ref{fglkh}) are basically the same and they are
largely omitted. \medskip

\noindent \underline{Step 1:} Recall that the proof of Lemma \ref{lemma
existence 2 copy(2)} shows that $\Delta \in \mathfrak{C}$ is the unique
strongly continuous operator family $\Delta \in \mathfrak{C}$ that is a
strong solution in $\mathfrak{h}$ to the equation $\Delta =\mathfrak{I}%
(\Delta )$, where $\mathfrak{I}$ is the mapping from $\mathfrak{C}$ to
itself defined by (\ref{defIfrak}) for some $T\in (0,T_{0}]$. The proof is
done via the contraction mapping principle. In particular, one shows the
existence of the fixed point by successive applications of the mapping $%
\mathfrak{I\circ \cdots \circ I}(X)$ on some initial strongly continuous
operator family $X=X^{\ast }\in \mathfrak{C}$, like $X=0$. Assume that $X\in 
\mathfrak{C}$ satisfies the bounds%
\begin{equation*}
\max \left\{ \Vert (\Upsilon _{0}+\mu \mathbf{1)}X(\Upsilon _{0}+\mu \mathbf{%
1)}^{-1}\Vert _{\infty },\Vert (\Upsilon _{0}^{\top }+\mu \mathbf{1)}X^{\top
}(\Upsilon _{0}^{\top }+\mu \mathbf{1)}^{-1}\Vert _{\infty }\right\} \leq 
\mathrm{B}\ ,
\end{equation*}%
for some $\mathrm{B}\in \mathbb{R}^{+}$, with the mapping 
\begin{equation*}
t\mapsto (\Upsilon _{0}+\mu \mathbf{1)}X_{t}(\Upsilon _{0}+\mu \mathbf{1)}%
^{-1}\qquad \text{and}\qquad t\mapsto (\Upsilon _{0}^{\top }+\mu \mathbf{1)}%
X_{t}^{\top }(\Upsilon _{0}^{\top }+\mu \mathbf{1)}^{-1}
\end{equation*}%
from $[0,T]$ to $\mathcal{B}(\mathfrak{h})$ being strongly continuous. Then
we deduce from (\ref{definition of W}) and the triangle inequality that, for
any $s,t\in \lbrack 0,T]$,%
\begin{equation}
\max_{\sharp =\top ,\ast ,\varnothing }\Vert (\Upsilon _{0}^{\sharp }+\mu 
\mathbf{1)}(W_{t,s}(X\mathbf{))}^{\sharp }(\Upsilon _{0}^{\sharp }+\mu 
\mathbf{1)}^{-1}\Vert _{\mathrm{op}}\leq \mathrm{e}^{2\left( \mathrm{B}+\mu
\right) \left( t-s\right) }\ ,  \label{eqsup1}
\end{equation}%
similar to\ Inequality (\ref{upper bound for W0}). If additionnally (\ref%
{conditionplus}) holds true, then we deduce from Equations (\ref{defIfrak}),
(\ref{eqsup1}) and the triangle inequality that 
\begin{equation*}
\left\Vert (\Upsilon _{0}^{\sharp }+\mu \mathbf{1})\mathfrak{I}(X)^{\sharp
}(\Upsilon _{0}^{\sharp }+\mu \mathbf{1})^{-1}\right\Vert _{\infty }\leq 16%
\mathrm{C}^{2}\int_{0}^{T}\mathrm{e}^{8\left( \mathrm{B}+\mu \right) \tau }%
\mathrm{d}\tau \ ,
\end{equation*}%
where $\sharp =\top ,\varnothing $. One also easily checks that the mappings 
\begin{equation*}
t\mapsto (\Upsilon _{0}+\mu \mathbf{1)}\mathfrak{I}(X)_{t}(\Upsilon _{0}+\mu 
\mathbf{1)}^{-1}\qquad \text{and}\qquad t\mapsto (\Upsilon _{0}^{\top }+\mu 
\mathbf{1)}\mathfrak{I}(X)_{t}^{\top }(\Upsilon _{0}^{\top }+\mu \mathbf{1)}%
^{-1}
\end{equation*}%
from $[0,T]$ to $\mathcal{B}(\mathfrak{h})$ are both strongly continuous,
being even Lipschitz continuous in $\mathcal{B}(\mathfrak{h})$ on $[0,T]$.
In particular, if we take a time $T$ satisfying (\ref{local time}) and%
\footnote{%
If $\mathrm{B}=-\mu $ then (\ref{sdsdssdsdsdsddsd}) has to be understood as $%
T<\mathrm{B}/(16\mathrm{C}^{2})$, which corresponds to the limit $%
\lim_{x\rightarrow 0^{+}}x^{-1}\ln (1+x)=1$ in (\ref{sdsdssdsdsdsddsd}).} 
\begin{equation}
T<\frac{1}{8\left( \mathrm{B}+\mu \right) }\ln \left( \frac{\mathrm{B}}{2%
\mathrm{C}^{2}}\left( \mathrm{B}+\mu \right) +1\right) ,
\label{sdsdssdsdsdsddsd}
\end{equation}%
then 
\begin{equation*}
\max \left\{ \Vert (\Upsilon _{0}+\mu \mathbf{1)}\mathfrak{I}(X)(\Upsilon
_{0}+\mu \mathbf{1)}^{-1}\Vert _{\infty },\Vert (\Upsilon _{0}^{\top }+\mu 
\mathbf{1)}\mathfrak{I}(X)^{\top }(\Upsilon _{0}^{\top }+\mu \mathbf{1)}%
^{-1}\Vert _{\infty }\right\} \leq \mathrm{B}\ .
\end{equation*}%
We deduce that the fixed point $\Delta \in \mathfrak{C}$ must satisfy 
\begin{equation*}
\max \left\{ \Vert (\Upsilon _{0}+\mu \mathbf{1)}\Delta (\Upsilon _{0}+\mu 
\mathbf{1)}^{-1}\Vert _{\infty },\Vert (\Upsilon _{0}^{\top }+\mu \mathbf{1)}%
\Delta ^{\top }(\Upsilon _{0}^{\top }+\mu \mathbf{1)}^{-1}\Vert _{\infty
}\right\} \leq \mathrm{B}
\end{equation*}%
with both mappings 
\begin{equation*}
t\mapsto (\Upsilon _{0}+\mu \mathbf{1)}\Delta _{t}(\Upsilon _{0}+\mu \mathbf{%
1)}^{-1}\qquad \text{and}\qquad t\mapsto (\Upsilon _{0}^{\top }+\mu \mathbf{%
1)}\Delta _{t}^{\top }(\Upsilon _{0}^{\top }+\mu \mathbf{1)}^{-1}
\end{equation*}%
from $[0,T]$ to $\mathcal{B}(\mathfrak{h})$ being not only strongly
continuous, but even Lipschitz continuous in $\mathcal{B}(\mathfrak{h})$ (on 
$[0,T]$). As in \cite[Lemma 33]{bach-bru-memo}, one thus verifies that, for
any $s,x,t\in \lbrack 0,T]$ so that $t\geq x\geq s$, $\boldsymbol{V}%
_{t,s}\in \mathcal{B}(\mathfrak{h})$ and $\boldsymbol{V}_{t,x}\boldsymbol{V}%
_{x,s}=\boldsymbol{V}_{t,s}$, while it is jointly strongly continuous in $s$
and $t$, provided the maximum time $T\in \mathbb{R}^{+}$ satisfies (\ref%
{local time}) and (\ref{sdsdssdsdsdsddsd}). \medskip

\noindent \underline{Step 2:} Fix once and for all $T\in \mathbb{R}^{+}$
such that (\ref{local time}) and (\ref{sdsdssdsdsdsddsd}) hold true. Take $%
\eta \in (0,T)$. Recall that 
\begin{equation}
\Upsilon _{t}=\Upsilon _{0}+\Delta _{\eta }+16\int_{\eta }^{t}D_{\tau
}D_{\tau }^{\ast }\mathrm{d}\tau \ ,\qquad t\in \lbrack \eta ,\infty )\ ,
\label{fghfghh}
\end{equation}%
and define the operator 
\begin{equation}
\mathfrak{d}_{\eta ,t}\doteq \left( \Upsilon _{0}+\mu \mathbf{1}\right)
\left( 16\int_{\eta }^{t}D_{\tau }D_{\tau }^{\ast }\mathrm{d}\tau \right)
\left( \Upsilon _{0}+\mu \mathbf{1}\right) ^{-1},\qquad t\in \lbrack \eta
,\infty )\ .  \label{evolution operato bis}
\end{equation}%
Since $D_{t>0}\mathfrak{h}\subseteq \mathcal{D}(\Upsilon _{0})$, we
interchange the operator $(\Upsilon _{0}+\mu \mathbf{1})$ and the integral
and use (\ref{fghfghh}) to arrive at the equality 
\begin{equation*}
\left( \Upsilon _{0}+\mu \mathbf{1}\right) \int_{\eta }^{t}D_{\tau }D_{\tau
}^{\ast }\mathrm{d}\tau =\int_{\eta }^{t}\Upsilon _{\tau }D_{\tau }D_{\tau
}^{\ast }\mathrm{d}\tau +\mu \int_{\eta }^{t}D_{\tau }D_{\tau }^{\ast }%
\mathrm{d}\tau -16\int_{\eta }^{t}\mathrm{d}\tau _{1}\int_{0}^{\tau _{1}}%
\mathrm{d}\tau _{2}D_{\tau _{2}}D_{\tau _{2}}^{\ast }D_{\tau _{1}}D_{\tau
_{1}}^{\ast }.
\end{equation*}%
Using (\ref{D_explicit_solutionbis}) (with $T\rightarrow \infty $), (\ref%
{hhhhh}) and Theorem \ref{section extension gap}, it follows that the
mapping $t\mapsto \mathfrak{d}_{\eta ,t}$ is locally Lipschitz continuous
from $[\eta ,\infty )$ to $\mathcal{B}\left( \mathfrak{h}\right) $ for any
fixed $\eta \in (0,T)$. Since 
\begin{equation*}
\mathfrak{d}_{t}=\mathfrak{d}_{\eta }+\mathfrak{d}_{\eta ,t}\ ,\qquad t\in
\lbrack \eta ,\infty )\ ,
\end{equation*}%
using Step 1, we thus deduce that $(\mathfrak{d}_{t})_{t\geq 0}\in C(\mathbb{%
R}_{0}^{+};\mathcal{B}(\mathfrak{h}))$, which is even locally Lipschitz
continuous.\medskip

\noindent \underline{Step 3:} For any $\eta \in (0,T)$, observe that $(%
\boldsymbol{V}_{t,s})_{t\geq s\geq \eta }\subseteq \mathcal{B}(\mathfrak{h})$%
, thanks to (\ref{definition of W}), (\ref{D_explicit_solutionbis}), (\ref%
{hhhhh}), (\ref{evoluation operators2}) and the hypothesis (\ref%
{conditionplus}). Combined with Step 1, it implies that $(\boldsymbol{V}%
_{t,s})_{t\geq s\geq 0}\subseteq \mathcal{B}(\mathfrak{h})$. One also
directly verifies the cocycle property of $(\boldsymbol{V}_{t,s})_{t\geq
s\geq 0}$, since the operator family $\left( W_{t,s}\right) _{t\geq s}$
satisfies the cocycle property. Moreover, for any $t\in \mathbb{R}_{0}^{+}$
and $s\in \lbrack 0,t]$, 
\begin{eqnarray*}
\boldsymbol{V}_{t,s} &=&\mathrm{e}^{-2\left( t-s\right) \Upsilon _{0}} \\
&&+\sum_{n=1}^{\infty }\left( -2\right) ^{n}\int_{s}^{t}\mathrm{d}\tau
_{1}\cdots \int_{s}^{\tau _{n-1}}\mathrm{d}\tau _{n}\mathrm{e}^{-2\left(
t-\tau _{1}\right) \Upsilon _{0}}\left( \underset{j=1}{\overset{n}{\prod }}%
\mathfrak{d}_{\tau _{j}}\mathrm{e}^{-2\left( \tau _{j}-\tau _{j+1}\right)
\Upsilon _{0}}\right) \ ,
\end{eqnarray*}%
because of (\ref{definition of W}). Above, we swap the operator $(\Upsilon
_{0}+\mu \mathbf{1})$ and the sum or the (Riemann) integrals. This is
justified by the fact that $(\mathfrak{d}_{t})_{t\geq 0}\in C(\mathbb{R}%
_{0}^{+};\mathcal{B}(\mathfrak{h}))$, thanks to Step 2. We omit the details.
In particular, for any $t\in \mathbb{R}_{0}^{+}$ and $s\in \lbrack 0,t]$, 
\begin{equation*}
\boldsymbol{V}_{t,s}=\mathrm{e}^{-2\left( t-s\right) \Upsilon
_{0}}-2\int_{s}^{t}\mathrm{e}^{-2\left( t-\tau \right) \Upsilon _{0}}%
\mathfrak{d}_{\tau }\boldsymbol{V}_{\tau ,s}\mathrm{d}\tau \ .
\end{equation*}
As a consequence, one can straightforwardly adapt the proofs of \cite[%
Lemmata 33 and 35]{bach-bru-memo} to show Assertions (ii)--(iv).
\end{proof}

Lemma \ref{lemma constante of motion copy(8)} can be used to obtain the
following proposition, which is the extension of Lemma \ref{lemma constante
of motion copy(7)} to lower semibounded operators $\Upsilon _{0}$:

\begin{proposition}[Towards a constant of motion -- Unbounded case]
\label{lemma constante of motion copy(2)}\mbox{}\newline
Take a self-adjoint operator $\Upsilon _{0}=\Upsilon _{0}^{\ast }$ and $\mu
\in \mathbb{R}$ such that $\Upsilon _{0}\geq -\left( \mu -\varepsilon
\right) \mathbf{1}$ for some $\varepsilon \in \mathbb{R}^{+}$. Let $%
D_{0}=\pm D_{0}^{\top }\in \mathcal{B}\left( \mathfrak{h}\right) $ ($%
D_{0}\neq 0$). Assume (\ref{conditionplus}). Then, 
\begin{equation*}
K_{t}(\Upsilon _{0}^{\top }+\mu \mathbf{1)}^{-1}=W_{t,s}K_{s}W_{t,s}^{\top
}(\Upsilon _{0}^{\top }+\mu \mathbf{1)}^{-1}\ ,\qquad t,s\in \mathbb{R}%
_{0}^{+}\ .
\end{equation*}
\end{proposition}

\begin{proof}
For vectors $\varphi \in \mathcal{D}(\Upsilon _{0}^{\top })$ and $\psi \in 
\mathcal{D}(\Upsilon _{0})$, and strictly positive times $t\in \mathbb{R}%
^{+} $, we compute from Theorem \ref{section extension gap} that 
\begin{equation*}
\partial _{t}\left\langle \psi ,K_{t}\varphi \right\rangle _{\mathfrak{h}%
}=-2\left\langle \psi ,\left( \Upsilon _{t}K_{t}+K_{t}\Upsilon _{t}^{\top
}\right) \varphi \right\rangle _{\mathfrak{h}}\ .
\end{equation*}%
Using this together with (\ref{eq diff 4}), $W_{t,s}^{\ast }\mathfrak{h}%
\subseteq \mathcal{D}\left( \Upsilon _{0}\right) $ and $W_{t,s}^{\top }%
\mathfrak{h}\subseteq \mathcal{D}\left( \Upsilon _{0}^{\top }\right) $ when $%
t>s$, we obtain that, for any $\varphi \in \mathcal{D}(\Upsilon _{0}^{\top
}) $, $\psi \in \mathcal{D}(\Upsilon _{0})$, $t\in \mathbb{R}^{+}$ and $s\in
(0,t)$, 
\begin{equation*}
\partial _{s}\left\langle \psi ,W_{t,s}K_{s}W_{t,s}^{\top }\varphi
\right\rangle _{\mathfrak{h}}=\partial _{s}\left\langle W_{t,s}^{\ast }\psi
,K_{s}W_{t,s}^{\top }\varphi \right\rangle _{\mathfrak{h}}=0\ ,
\end{equation*}%
which in turn implies that 
\begin{equation*}
\left\langle \psi ,W_{t,x}K_{x}W_{t,x}^{\top }\varphi \right\rangle _{%
\mathfrak{h}}=\left\langle \psi ,W_{t,s}K_{s}W_{t,s}^{\top }\varphi
\right\rangle _{\mathfrak{h}}\ ,\qquad x\in (s,t)\ .
\end{equation*}%
If $t>x>s>0$, then, similar to (\ref{hhhhh}), $\Upsilon _{t}^{\top
}W_{t,x}^{\top }\in \mathcal{B}(\mathfrak{h})$ and the operators $%
W_{t,x}K_{x}W_{t,x}^{\top }$ and $W_{t,s}K_{s}W_{t,s}^{\top }$ are both
bounded. By density of $\mathcal{D}(\Upsilon _{0}^{\top })$ and $\mathcal{D}%
(\Upsilon _{0})$ in $\mathfrak{h}$, the last equality yields%
\begin{equation}
W_{t,x}K_{x}W_{t,x}^{\top }=W_{t,s}K_{s}W_{t,s}^{\top }\ ,\qquad t\in 
\mathbb{R}^{+},\ s\in (0,t),\ x\in (s,t)\ .  \label{weqewqe}
\end{equation}%
It remains to do the limits $x\rightarrow t^{-}$ and $s\rightarrow 0^{+}$ in
(\ref{weqewqe}) to deduce the assertion. This is done by using Theorem \ref%
{section extension gap} and Lemma \ref{lemma constante of motion copy(8)}
together with (\ref{D_explicit_solutionbis}) (with $T\rightarrow \infty $).
For instance, use that, for any $\varphi \in \mathfrak{h}$, $s\in \mathbb{R}%
_{0}^{+}$ and $\delta \in \mathbb{R}^{+}$ such that $s\pm \delta \geq 0$,%
\begin{multline*}
\left\Vert \left( \Upsilon _{s\pm \delta }D_{s\pm \delta }-\Upsilon
_{s}D_{s}\right) \left( \Upsilon _{0}^{\top }+\mu \mathbf{1}\right)
^{-1}\varphi \right\Vert _{\mathfrak{h}} \\
\leq \left\Vert \Upsilon _{s\pm \delta }\left( \Upsilon _{0}+\mu \mathbf{1}%
\right) ^{-1}\right\Vert _{\mathrm{op}}\left\Vert \left( \Upsilon _{0}+\mu 
\mathbf{1}\right) \left( D_{s\pm \delta }-D_{s}\right) \left( \Upsilon
_{0}^{\top }+\mu \mathbf{1}\right) ^{-1}\varphi \right\Vert _{\mathfrak{h}}
\\
+\left\Vert \Upsilon _{s\pm \delta }-\Upsilon _{s}\right\Vert _{\mathrm{op}%
}\left\Vert D_{s}\left( \Upsilon _{0}^{\top }+\mu \mathbf{1}\right)
^{-1}\right\Vert _{\mathrm{op}}
\end{multline*}%
combined (having (\ref{fglkh}) in mind) with 
\begin{eqnarray}
&&\left\Vert \left( \Upsilon _{0}+\mu \mathbf{1}\right) \left( D_{s\pm
\delta }-D_{s}\right) \left( \Upsilon _{0}^{\top }+\mu \mathbf{1}\right)
^{-1}\varphi \right\Vert _{\mathfrak{h}}  \notag \\
&=&\left\Vert \left( \Upsilon _{0}+\mu \mathbf{1}\right) \left( W_{s\pm
\delta ,0}D_{0}W_{s\pm \delta ,0}^{\top }-W_{s,0}D_{0}W_{s,0}^{\top }\right)
\left( \Upsilon _{0}^{\top }+\mu \mathbf{1}\right) ^{-1}\varphi \right\Vert
_{\mathfrak{h}}  \notag \\
&\leq &\left\Vert \boldsymbol{V}_{s\pm \delta ,0}\right\Vert _{\mathrm{op}%
}\left\Vert \left( \Upsilon _{0}+\mu \mathbf{1}\right) D_{0}\left( \Upsilon
_{0}+\mu \mathbf{1}\right) ^{-1}\right\Vert _{\mathrm{op}}\left\Vert \left( 
\boldsymbol{V}_{s\pm \delta ,0,\top }-\boldsymbol{V}_{s,0,\top }\right)
\varphi \right\Vert _{\mathfrak{h}}  \notag \\
&&+\left\Vert \left( \boldsymbol{V}_{s\pm \delta ,0}-\boldsymbol{V}%
_{s,0}\right) \left( \Upsilon _{0}+\mu \mathbf{1}\right) D_{0}W_{s,0}^{\top
}\left( \Upsilon _{0}^{\top }+\mu \mathbf{1}\right) ^{-1}\varphi \right\Vert
_{\mathfrak{h}},  \label{sdfsfsddfsdfsdfsdfsfs}
\end{eqnarray}%
thanks to (\ref{D_explicit_solutionbis}) and the triangle inequality. Then
using the last two estimates together with Theorem \ref{section extension
gap}, Equation (\ref{conditionplus}) and Lemma \ref{lemma constante of
motion copy(8)}, one arrives at the following strong limit:\ 
\begin{equation*}
\lim_{t\rightarrow s}\left( \Upsilon _{t}D_{t}\right) \left( \Upsilon
_{0}^{\top }+\mu \mathbf{1}\right) ^{-1}=\Upsilon _{s}D_{s}\left( \Upsilon
_{0}^{\top }+\mu \mathbf{1}\right) ^{-1}.
\end{equation*}%
All the other limits (in the strong topology) can be done exactly in the
same way. One shows in particular the continuity of the family $%
(K_{t})_{t\geq 0}$ in the strong sense in $\mathcal{D}(\Upsilon _{0}^{\top
}) $.
\end{proof}

Lemmata \ref{lemma constante of motion} and \ref{lemma constante of motion
copy(8)}, and Proposition \ref{lemma constante of motion copy(2)} can then
be used to obtain constants of motion on the operator flow in the
Hilbert-Schmidt or \textquotedblleft commutative\textquotedblright\ cases:

\begin{theorem}[Constants of motion]
\label{theorem constante of motion}\mbox{}\newline
Take a self-adjoint operator $\Upsilon _{0}=\Upsilon _{0}^{\ast }$ and $\mu
\in \mathbb{R}$ such that $\Upsilon _{0}\geq -\left( \mu -\varepsilon
\right) \mathbf{1}$ for some $\varepsilon \in \mathbb{R}^{+}$. Let $\Upsilon
_{t}\doteq \Upsilon _{0}+\Delta _{t}$ for each $t\in \mathbb{R}_{0}^{+}$.

\begin{enumerate}
\item[\emph{(i)}]  Assume $D_{0}^{\top }=\pm D_{0}\in \mathcal{L}^{2}(%
\mathfrak{h})$ and $\Upsilon _{0}=\Upsilon _{0}^{\ast }\in \mathcal{B}\left( 
\mathfrak{h}\right) $. Then, for any $t,s\in \mathbb{R}_{0}^{+}$, 
\begin{equation*}
\mathrm{tr}\left( \Upsilon _{t}^{2}+4D_{t}D_{t}^{\ast }-\Upsilon
_{s}^{2}-4D_{s}D_{s}^{\ast }\right) =0\ .
\end{equation*}

\item[\emph{(ii)}] Assume that $D_{0}^{\top }=\pm D_{0}\in \mathcal{B}\left( 
\mathfrak{h}\right) $ and $\Upsilon _{0}D_{0}=D_{0}\Upsilon _{0}^{\top }$.
Then, for any $t,s\in \mathbb{R}_{0}^{+}$, $\Upsilon _{t}D_{t}=D_{t}\Upsilon
_{t}^{\top }$ and 
\begin{equation}
\left( \Upsilon _{t}^{2}+4D_{t}D_{t}^{\ast }-\Upsilon
_{s}^{2}-4D_{s}D_{s}^{\ast }\right) \left( \Upsilon _{0}^{\top }+\mu \right)
^{-1}=0\ .  \label{constmotionexplicit}
\end{equation}%
 
\end{enumerate}
\end{theorem}

\begin{proof}
Observe first that $D_{0}^{\top }=\pm D_{0}\in \mathcal{B}\left( \mathfrak{h}%
\right) $ yields $D_{t}^{\top }=\pm D_{t}$ for any $t\in \mathbb{R}_{0}^{+}$%
, thanks to Theorem \ref{section extension gap}. To prove Assertion (i), one
uses the cyclicity of the trace combined with $\mathrm{tr}\left( X^{\top
}\right) =\mathrm{tr}\left( \mathcal{C}X^{\ast }\mathcal{C}\right) =\mathrm{%
tr}\left( X\right) $ and $D_{t}=\pm D_{t}^{\top }\in \mathcal{L}^{2}\left( 
\mathfrak{h}\right) $ to deduce that 
\begin{equation}
\mathrm{tr}\left( \Upsilon _{t}D_{t}D_{t}^{\ast }\right) =\mathrm{tr}\left(
D_{t}^{\ast }\Upsilon _{t}D_{t}\right) =\mathrm{tr}\left( D_{t}\Upsilon
_{t}^{\top }D_{t}^{\ast }\right) =\mathrm{tr}\left( D_{t}D_{t}^{\ast
}\Upsilon _{t}\right)  \label{inequality trivial}
\end{equation}%
for any $t\in \mathbb{R}_{0}^{+}$. Then, together with Lemma \ref{lemma
constante of motion} and Tonelli's theorem, Assertion (i) follows. 
To prove Assertion (ii), observe that, according to Proposition \ref{lemma
constante of motion copy(2)}, if $K_{0}=0$ then $K_{t}=0$, i.e., $\Upsilon
_{t}D_{t}=D_{t}\Upsilon _{t}^{\top }$, for any $t\in \mathbb{R}_{0}^{+}$.
Now, like in the bounded case (Lemma \ref{lemma constante of motion}), we
compute from Theorem \ref{section extension gap} that, for any $t\in \mathbb{%
R}^{+}$, 
\begin{equation*}
\partial _{t}\left\{ \left( \Upsilon _{t}^{2}+4D_{t}D_{t}^{\ast }\right)
(\Upsilon _{0}+\mu \mathbf{1)}^{-1}\right\} =8\left( K_{t}D_{t}^{\ast
}+D_{t}K_{t}^{\ast }\right) (\Upsilon _{0}+\mu \mathbf{1)}^{-1}=0\ ,
\end{equation*}%
in the strong sense in $\mathfrak{h}$, which implies 
\begin{equation}
\left( \Upsilon _{t}^{2}+4D_{t}D_{t}^{\ast }-\Upsilon
_{s}^{2}-4D_{s}D_{s}^{\ast }\right) \left( \Upsilon _{0}^{\top }+\mu \right)
^{-1}=0  \label{suplimit}
\end{equation}%
for any $t,s\in \mathbb{R}^{+}$. To arrive at this result, we use the
evolution families studied in Lemma \ref{lemma constante of motion copy(8)}
and the inequality 
\begin{eqnarray}
&&\left\Vert \left( \delta ^{-1}\left( \Upsilon _{t\pm \delta }^{2}-\Upsilon
_{t}^{2}\right) -\left( \partial _{t}\Upsilon _{t}\right) \Upsilon
_{t}-\Upsilon _{t}\left( \partial _{t}\Upsilon _{t}\right) \right) \left(
\Upsilon _{0}+\mu \mathbf{1}\right) ^{-1}\varphi \right\Vert _{\mathfrak{h}}
\notag \\
&\leq &\left\Vert \Upsilon _{t\pm \delta }\left( \Upsilon _{0}+\mu \mathbf{1}%
\right) ^{-1}\right\Vert _{\mathrm{op}}\left\Vert \left( \Upsilon _{0}+\mu 
\mathbf{1}\right) \left( \delta ^{-1}\left( \Upsilon _{t\pm \delta
}-\Upsilon _{t}\right) -\partial _{t}\Upsilon _{t}\right) \left( \Upsilon
_{0}+\mu \mathbf{1}\right) ^{-1}\varphi \right\Vert  \notag \\
&&+\left\Vert \delta ^{-1}\left( \Upsilon _{t\pm \delta }-\Upsilon
_{t}\right) -\partial _{t}\Upsilon _{t}\right\Vert _{\mathrm{op}}\left\Vert
\Upsilon _{t}\left( \Upsilon _{0}+\mu \mathbf{1}\right) ^{-1}\right\Vert _{%
\mathrm{op}}  \notag \\
&&+\left\Vert \Upsilon _{t\pm \delta }-\Upsilon _{t}\right\Vert _{\mathrm{op}%
}\left\Vert \partial _{t}\Upsilon _{t}\right\Vert _{\mathrm{op}}
\label{yuyuuiy}
\end{eqnarray}%
for any $\varphi \in \mathfrak{h}$ and $t,\delta \in \mathbb{R}^{+}$ such
that $t\pm \delta \geq 0$. Compare this last inequality with (\ref{sup23}).
Note in particular that, for any $t,\delta \in \mathbb{R}^{+}$ such that $%
t\pm \delta \geq 0$, 
\begin{eqnarray*}
&&\left( \Upsilon _{0}+\mu \mathbf{1}\right) \left( \delta ^{-1}\left(
\Upsilon _{t\pm \delta }-\Upsilon _{t}\right) -\partial _{t}\Upsilon
_{t}\right) \left( \Upsilon _{0}+\mu \mathbf{1}\right) ^{-1} \\
&=&\left( \Upsilon _{0}+\mu \mathbf{1}\right) \left( \delta
^{-1}\int_{t}^{t\pm \delta }D_{\tau }D_{\tau }^{\ast }\mathrm{d}\tau
-D_{t}D_{t}^{\ast }\right) \left( \Upsilon _{0}+\mu \mathbf{1}\right) ^{-1}
\end{eqnarray*}%
which, for any $\varphi \in \mathfrak{h}$, leads to the inequality 
\begin{eqnarray*}
&&\left\Vert \left( \Upsilon _{0}+\mu \mathbf{1}\right) \left( \delta
^{-1}\left( \Upsilon _{t\pm \delta }-\Upsilon _{t}\right) -\partial
_{t}\Upsilon _{t}\right) \left( \Upsilon _{0}+\mu \mathbf{1}\right)
^{-1}\varphi \right\Vert _{\mathfrak{h}} \\
&=&\max_{\tau \in \left[ t\wedge \left( t\pm \delta \right) ,t\vee \left(
t\pm \delta \right) \right] }\left\Vert \left( \Upsilon _{0}+\mu \mathbf{1}%
\right) \left( D_{\tau }D_{\tau }^{\ast }-D_{t}D_{t}^{\ast }\right) \left(
\Upsilon _{0}+\mu \mathbf{1}\right) ^{-1}\varphi \right\Vert _{\mathfrak{h}%
}\ .
\end{eqnarray*}%
By (\ref{sdfsfsddfsdfsdfsdfsfs}) and Lemma \ref{lemma constante of motion
copy(8)}, it follows that 
\begin{equation*}
\lim_{\delta \rightarrow 0^{+}}\left( \Upsilon _{0}+\mu \mathbf{1}\right)
\left( \delta ^{-1}\left( \Upsilon _{t\pm \delta }-\Upsilon _{t}\right)
-\partial _{t}\Upsilon _{t}\right) \left( \Upsilon _{0}+\mu \mathbf{1}%
\right) ^{-1}=0\ ,
\end{equation*}%
in the strong topology, which, combined with (\ref{yuyuuiy}) and Theorem \ref%
{section extension gap} in turn implies Leibniz rule: 
\begin{equation*}
\lim_{\delta \rightarrow 0^{+}}\left( \delta ^{-1}\left( \Upsilon _{t\pm
\delta }^{2}-\Upsilon _{t}^{2}\right) \left( \Upsilon _{0}+\mu \mathbf{1}%
\right) ^{-1}=\left( \partial _{t}\Upsilon _{t}\right) \Upsilon
_{t}+\Upsilon _{t}\left( \partial _{t}\Upsilon _{t}\right) \right) \left(
\Upsilon _{0}+\mu \mathbf{1}\right) ^{-1}\ ,
\end{equation*}%
again in the strong topology. It remains to do the limit $s\rightarrow 0^{+}$
in (\ref{suplimit}) to deduce the assertion. By\ Theorem \ref{section
extension gap}, $D_{s}D_{s}^{\ast }$ strongly converges to $D_{0}D_{0}^{\ast
}$. Meanwhile, the convergence of the operator $\Upsilon _{s}^{2}$ to $%
\Upsilon _{0}^{2}$, as $s\rightarrow 0^{+}$, follows from the estimate[ 
\begin{equation*}
\left\Vert \left( \Upsilon _{s}^{2}-\Upsilon _{0}^{2}\right) \left( \Upsilon
_{0}+\mu \mathbf{1}\right) ^{-1}\right\Vert _{\mathrm{op}}\leq \left\Vert 
\mathfrak{d}_{s}\right\Vert _{\mathrm{op}}+\left\Vert \Delta _{s}\right\Vert
_{\mathrm{op}}+\left\Vert \Delta _{s}\right\Vert _{\mathrm{op}}^{2}\ ,\qquad
s\in \mathbb{R}_{0}^{+}\ ,
\end{equation*}%
together with Theorem \ref{section extension gap} and Lemma \ref{lemma
constante of motion copy(8)} (i).  
\end{proof}

Theorem \ref{theorem constante of motion} gives strong constraints on the
operator flow. In particular, Theorem \ref{theorem constante of motion} (ii)
allows us to explicit compute the operator $\Upsilon _{t}^{2}$\ in the limit 
$t\rightarrow \infty $ under the reasonable assumption that $%
D_{t}D_{t}^{\ast }$\ vanishes in the appropriate topology.

\begin{remark}[Extension to semibounded operators]
\label{theorem constante of motion copy(1)}\mbox{}\newline
 Lemma \ref{lemma constante of motion} and Theorem \ref{theorem
constante of motion} (i) could be extended to all self-adjoint operators $%
\Upsilon _{0}$ bounded from below, using Lemma \ref{lemma constante of
motion copy(8)}. See also \cite[Section V.2]{bach-bru-memo}. 
\end{remark}

\subsection{Elliptic Properties of the Flow\label{Section tech5}}

The constants of motion given by Theorem \ref{theorem constante of motion}
can be viewed as an ellipticity property of the operator flow, as one can
see in the one-dimensional case. See (\ref{constant motion scalar})--(\ref%
{constant motion scalar2}). The conditions of Theorem \ref{theorem constante
of motion} are however quite restrictive, even if we consider the assertion
for operators $\Upsilon _{0}$ that are only bounded from below, see Remark %
\ref{theorem constante of motion copy(1)}. Indeed, Theorem \ref{theorem
constante of motion} (i) needs Hilbert-Schmidt initial data $D_{0}=\pm
D_{0}^{\top }\in \mathcal{L}^{2}(\mathfrak{h})$, while Theorem \ref{theorem
constante of motion} (ii) requires $\Upsilon _{0}D_{0}=D_{0}\Upsilon
_{0}^{\top }$ with bounded initial data $D_{0}^{\top }=\pm D_{0}\in \mathcal{%
B}\left( \mathfrak{h}\right) $. In this section we study the elliptic
properties of the flow without such restrictions, more precisely for lower
semibounded operators $\Upsilon _{0}$ and bounded (possibly
non-Hilbert-Schmidt) $D_{0}\in \mathcal{B}\left( \mathfrak{h}\right) $.
Different and more general arguments than those used to prove Theorem \ref%
{theorem constante of motion} are necessary.

In all the section, $\Delta $ and $D$ are always given by Theorem \ref%
{section extension gap}. We consider the new operators 
\begin{equation}
\mathfrak{D}_{t}\doteq \Upsilon _{t}-\mu \mathbf{1}+4\mathfrak{B}%
_{t}\,,\qquad t\in \mathbb{R}_{0}^{+}\ ,  \label{def D frac}
\end{equation}%
with domain $\mathcal{D}(\mathfrak{D}_{t})=\mathcal{D}(\Upsilon _{0})$ for
every $t\in \mathbb{R}_{0}^{+}$, where $\Upsilon _{t}\doteq \Upsilon
_{0}+\Delta _{t}$ for each $t\in \mathbb{R}_{0}^{+}$ and 
\begin{equation}
\mathfrak{B}_{t}\doteq D_{t}\left( \Upsilon _{t}^{\top }+\mu \mathbf{1}%
\right) ^{-1}D_{t}^{\ast }\in \mathcal{B}\left( \mathfrak{h}\right)
\,,\qquad t\in \mathbb{R}_{0}^{+}\ ,  \label{def D frac2}
\end{equation}%
with $\mu \in \mathbb{R}$ and $\varepsilon \in \mathbb{R}^{+}$ such that $%
\Upsilon _{0}^{\top }\geq -\left( \mu -\varepsilon \right) \mathbf{1}$,
which is equivalent\footnote{%
Recall that $X\geq 0$ iff $X^{\top }\geq 0$.} to $\Upsilon _{0}\geq -\left(
\mu -\varepsilon \right) \mathbf{1}$. Under such a condition, $\mathfrak{B}%
_{t}$ is well-defined as a bounded operator for any $t\in \mathbb{R}_{0}^{+}$%
, because 
\begin{equation}
\Vert \mathfrak{B}_{t}\Vert _{\mathrm{op}}\leq \varepsilon ^{-1}\left\Vert
D_{t}\right\Vert _{\mathrm{op}}^{2}<\infty \ ,\qquad t\in \mathbb{R}%
_{0}^{+}\ .  \label{upper bound toto0}
\end{equation}%
The operator families $\mathfrak{B}\equiv (\mathfrak{B}_{t})_{t\geq 0}$ and $%
\mathfrak{D}\equiv (\mathfrak{D}_{t})_{t\geq 0}$ are strongly 
differentiable , at least for strictly positive times $t>0$, like $%
D\equiv (D_{t})_{t\geq 0}$ (cf. Theorem \ref{section extension gap}).

\begin{lemma}[Differentially of $\mathfrak{B}$ and $\mathfrak{D}$]
\label{lemma constante of motion copy(5)}\mbox{}\newline
Take a self-adjoint operator $\Upsilon _{0}=\Upsilon _{0}^{\ast }$ and $\mu
\in \mathbb{R}$ such that $\Upsilon _{0}\geq -\left( \mu -\varepsilon
\right) \mathbf{1}$ for some $\varepsilon \in \mathbb{R}^{+}$. Let $%
D_{0}=\pm D_{0}^{\top }\in \mathcal{B}\left( \mathfrak{h}\right) $ ($%
D_{0}\neq 0$). Then $\mathfrak{B}\in \mathfrak{C}$ with $(\mathfrak{B}%
_{t})_{t>0}\in C\left( \mathbb{R}^{+};\mathcal{B}(\mathfrak{h})\right) $
while $(\mathfrak{D}_{t})_{t>0}$ is a strongly continuous family of
unbounded operators with domain $\mathcal{D}(\Upsilon _{0})$. Additionally,
for any $t\in \mathbb{R}^{+}$ and in the strong sense on the domain $%
\mathcal{D}\left( \Upsilon _{0}\right) $, 
\begin{eqnarray}
\partial _{t}\mathfrak{B}_{t} &=&-2\left( \mathfrak{D}_{t}\mathfrak{B}_{t}+%
\mathfrak{B}_{t}\mathfrak{D}_{t}\right) -4D_{t}D_{t}^{\ast }\ ,
\label{sdsdsd12} \\
\partial _{t}\mathfrak{D}_{t} &=&-8\left( \mathfrak{D}_{t}\mathfrak{B}_{t}+%
\mathfrak{B}_{t}\mathfrak{D}_{t}\right) \ .  \label{sdsdsd1}
\end{eqnarray}
\end{lemma}

\begin{proof}
The assumption $D_{0}^{\top }=\pm D_{0}$   yields $%
D_{t}^{\top }=\pm D_{t}$ for any $t\in \mathbb{R}_{0}^{+}$, thanks to
Theorem \ref{section extension gap}. Recall also that $\Upsilon _{0}^{\top
}\geq -\left( \mu -\varepsilon \right) \mathbf{1}$ is a consequence of $%
\Upsilon _{0}\geq -\left( \mu -\varepsilon \right) \mathbf{1}$ for some $%
\varepsilon \in \mathbb{R}^{+}$ and $\mu \in \mathbb{R}$. First, define 
\begin{equation*}
X\equiv \left( X_{t}\right) _{t\geq 0}\doteq \left( \left( \Upsilon
_{t}^{\top }+\mu \mathbf{1}\right) ^{-1}\right) _{t\geq 0}\subseteq \mathcal{%
B}(\mathfrak{h})\ .
\end{equation*}%
By Theorem \ref{section extension gap}, the operator family $\Delta $ is
(locally Lipschitz) continuous in the norm topology and we infer from the
triangle inequality and the resolvent identity that 
\begin{equation*}
\left\Vert X_{t+\delta }-X_{t}\right\Vert _{\mathrm{op}}\leq \varepsilon
^{-2}\left\Vert \Delta _{t+\delta }-\Delta _{t}\right\Vert _{\mathrm{op}%
}\,\qquad t\in \mathbb{R}_{0}^{+}\ ,
\end{equation*}%
from which we deduce that 
\begin{equation}
X\in C\left( \mathbb{R}_{0}^{+};\mathcal{B}(\mathfrak{h})\right) \ .
\label{derivative2}
\end{equation}%
Note from Theorem \ref{section extension gap} that the operator family $D$
is (locally Lipschitz) continuous in the norm topology only outside $t=0$
where it is simply strongly continuous. It follows that%
\begin{equation*}
\left( \mathfrak{B}_{t}\right) _{t>0}\in C\left( \mathbb{R}^{+};\mathcal{B}(%
\mathfrak{h})\right) \ ,
\end{equation*}%
which, together with Theorem \ref{section extension gap}, in turn implies
that 
\begin{equation*}
(\mathfrak{D}_{t}-\Upsilon _{0})_{t>0}\in C\left( \mathbb{R}^{+};\mathcal{B}(%
\mathfrak{h})\right) \ .
\end{equation*}%
In particular, $(\mathfrak{D}_{t})_{t>0}$ is a strongly continuous family of
(possibly unbounded) operators on the domain $\mathcal{D}(\Upsilon _{0})$.
Additionally, remark from the triangle inequality and the resolvent identity
as well as Theorem \ref{section extension gap} and the fact that $X^{\ast }=%
\overline{X^{\top }}=\overline{X}^{\top }$ and $D_{t}=\pm D_{t}^{\top }$
that, for any $t\in \mathbb{R}^{+}$ and $\delta \in \mathbb{R}\backslash
\{0\}$ such that $\left( t+\delta \right) \in \mathbb{R}_{0}^{+}$, 
\begin{eqnarray*}
\left\Vert \delta ^{-1}\left( X_{t+\delta }-X_{t}\right) +16X_{t}D_{t}^{\ast
}D_{t}X_{t}\right\Vert _{\mathrm{op}} &\leq &16\varepsilon ^{-2}\max_{\tau
\in \left[ t\wedge \left( t+\delta \right) ,t\vee \left( t+\delta \right) %
\right] }\left\Vert D_{t}D_{t}^{\ast }-D_{\tau }D_{\tau }^{\ast }\right\Vert
_{\mathrm{op}} \\
&&+16^{2}\varepsilon ^{-3}\delta \max_{\tau \in \left[ t\wedge \left(
t+\delta \right) ,t\vee \left( t+\delta \right) \right] }\left\Vert D_{\tau
}D_{\tau }^{\ast }\right\Vert _{\mathrm{op}}^{2},
\end{eqnarray*}%
where the notation $s\wedge t$\ and $s\vee t$ stand respectively for the
minimum and the maximum of real numbers $s,t\in \mathbb{R}$. Recall that the
family $(D_{t})_{t>0}$ is continuous in the norm topology, see Theorem \ref%
{section extension gap}. As a consequence, by taking the limit $\delta
\rightarrow 0^{+}$ in the above inequality, for any fixed $t\in \mathbb{R}%
^{+}$, we obtain the following right-derivative 
\begin{equation*}
\partial _{t}^{+}X_{t}=-16X_{t}D_{t}D_{t}^{\ast }X_{t}\ ,
\end{equation*}%
in the norm topology, i.e., in the Banach space $\mathcal{B}\left( \mathfrak{%
h}\right) $. In the same way, by taking the limit $\delta \rightarrow 0^{-}$%
, we obtain the left-derivative. The above right-derivative on $t=0$ holds
true only in the strong sense on $\mathfrak{h}$, since the operator family $%
D $ is only strongly continuous on $t=0$. Using this last information, the
first derivative of the lemma is justified by combining Theorem \ref{section
extension gap} with the upper bound%
\begin{eqnarray}
&&\left\Vert \left( \delta ^{-1}(\mathfrak{B}_{t+\delta }-\mathfrak{B}%
_{t})-\partial _{t}\{D_{t}\}X_{t}D_{t}^{\ast }-D_{t}\partial
_{t}\{X_{t}\}D_{t}^{\ast }-D_{t}X_{t}\partial _{t}\{D_{t}^{\ast }\}\right)
\varphi \right\Vert _{\mathfrak{h}}  \notag \\
&\leq &\varepsilon ^{-1}\left\Vert D_{t+\delta }\right\Vert _{\mathrm{op}%
}\left\Vert \left( \delta ^{-1}(D_{t+\delta }^{\ast }-D_{t}^{\ast
})-\partial _{t}\{D_{t}^{\ast }\}\right) \varphi \right\Vert _{\mathfrak{h}%
}+\varepsilon ^{-1}\left\Vert D_{t+\delta }-D_{t}\right\Vert _{\mathrm{op}%
}\left\Vert \partial _{t}\{D_{t}^{\ast }\}\varphi \right\Vert _{\mathfrak{h}}
\notag \\
&&+\left\Vert D_{t}\right\Vert _{\mathrm{op}}\left\Vert X_{t+\delta
}-X_{t}\right\Vert _{\mathrm{op}}\left\Vert \partial _{t}\{D_{t}^{\ast
}\}\varphi \right\Vert _{\mathfrak{h}}+\left\Vert D_{t+\delta }\right\Vert _{%
\mathrm{op}}\left\Vert \left( \delta ^{-1}(X_{t+\delta }-X_{t})-\partial
_{t}\{X_{t}\}\right) D_{t}^{\ast }\varphi \right\Vert _{\mathfrak{h}}  \notag
\\
&&+\left\Vert D_{t+\delta }-D_{t}\right\Vert _{\mathrm{op}}\left\Vert
\partial _{t}\{X_{t}\}\right\Vert _{\mathrm{op}}\left\Vert D_{t}^{\ast
}\right\Vert _{\mathrm{op}}+\left\Vert \left( \delta ^{-1}(D_{t+\delta
}-D_{t})-\partial _{t}\{D_{t}\}\right) X_{t}D_{t}^{\ast }\varphi \right\Vert
_{\mathfrak{h}}\ ,  \label{Leibniz rule}
\end{eqnarray}%
for any $\varphi \in \mathcal{D}(\Upsilon _{0})$, $t\in \mathbb{R}^{+}$ and $%
\delta \in \mathbb{R}\backslash \{0\}$ such that $\left( t+\delta \right)
\in \mathbb{R}_{0}^{+}$. Note that $X_{t}D_{t}^{\ast }\mathfrak{h}\subseteq 
\mathcal{D}\left( \Upsilon _{0}\right) $. In order to deduce the last
derivative, it thus suffices to combine Equation (\ref{sdsdsd12}) with
Theorem \ref{section extension gap}.
\end{proof}

\begin{remark}[Bounded case]
\label{remark bounded1}\mbox{}\newline
For bounded operators $\Upsilon _{0}\in \mathcal{B}\left( \mathfrak{h}%
\right) $, $\mathfrak{B},\mathfrak{D}\in C(\mathbb{R}_{0}^{+};\mathcal{B}(%
\mathfrak{h}))$ and (\ref{sdsdsd12})--(\ref{sdsdsd1}) can be extended to $%
s=0 $.
\end{remark}

Similar to (\ref{D_explicit_solutionbis}), the derivatives computed in\
Lemma \ref{lemma constante of motion copy(5)} allow us to compute the
operator families $\mathfrak{D}$ and $\mathfrak{B}$ via the evolution
operators 
\begin{equation}
\mathfrak{V}_{t,s}\doteq \mathbf{1}+\sum_{n=1}^{\infty }\left( -8\right)
^{n}\int_{s}^{t}\mathrm{d}\tau _{1}\cdots \int_{s}^{\tau _{n-1}}\mathrm{d}%
\tau _{n}\;\mathfrak{B}_{\tau _{1}}\cdots \mathfrak{B}_{\tau _{n}}
\label{def D frac3}
\end{equation}%
and 
\begin{eqnarray}
\mathfrak{U}_{t,s} &\doteq &\mathrm{e}^{-2\left( t-s\right) \Upsilon
_{0}}+\sum_{n=1}^{\infty }\left( -2\right) ^{n}\int_{s}^{t}\mathrm{d}\tau
_{1}\cdots  \label{def D frac4} \\
&&\qquad \cdots \int_{s}^{\tau _{n-1}}\mathrm{d}\tau _{n}\mathrm{e}%
^{-2\left( t-\tau _{1}\right) \Upsilon _{0}}\left( \underset{j=1}{\overset{n}%
{\prod }}\left( \Delta _{\tau _{j}}-\mu \mathbf{1+}4\mathfrak{B}_{\tau
_{j}}\right) \mathrm{e}^{-2\left( \tau _{j}-\tau _{j+1}\right) \Upsilon
_{0}}\right) \ ,  \notag
\end{eqnarray}%
for any $s\in \mathbb{R}_{0}^{+}$ and $t\in \left[ s,\infty \right) $, where 
$\tau _{n+1}\doteq s$. Compare this last definition with (\ref{definition of
W}). By Proposition \ref{lemma constante of motion copy(5)} and \cite[Lemma
33]{bach-bru-memo}, these evolution operators are well-defined in $\mathcal{B%
}(\mathfrak{h})$ and strongly continuous in the parameters $s,t$. In fact,
 they  are the (unique) solutions to the non-autonomous evolution
equations 
\begin{equation}
\partial _{t}\mathfrak{V}_{t,s}=-8\mathfrak{B}_{t}\mathfrak{V}_{t,s}\qquad 
\text{and}\qquad \partial _{t}\mathfrak{U}_{t,s}=-2\mathfrak{D}_{t}\mathfrak{%
U}_{t,s}  \label{sdsdsd3sdsdsd3}
\end{equation}%
in the strong sense in $\mathfrak{h}$ for any $s\in \mathbb{R}_{0}^{+}$ and $%
t\in \left( s,\infty \right) $, with $\mathfrak{V}_{s,s}\doteq \mathfrak{U}%
_{s,s}\doteq \mathbf{1}$. Additionally, for any $s\in \mathbb{R}_{0}^{+}$
and $t\in \left( s,\infty \right) $, 
\begin{equation}
\partial _{s}\mathfrak{V}_{t,s}=8\mathfrak{V}_{t,s}\mathfrak{B}_{s}\qquad 
\text{and}\qquad \partial _{s}\mathfrak{U}_{t,s}=2\mathfrak{U}_{t,s}%
\mathfrak{D}_{s}  \label{sdsdsd3}
\end{equation}%
the derivatives being in the strong sense in $\mathfrak{h}$ and $\mathcal{D}%
\left( \Upsilon _{0}\right) $, respectively. Note that $\mathfrak{V}_{t,x}%
\mathfrak{V}_{x,s}=\mathfrak{V}_{t,s}$ and $\mathfrak{U}_{t,x}\mathfrak{U}%
_{x,s}=\mathfrak{U}_{t,s}$ for any $s,x,t\in \mathbb{R}_{0}^{+}$ with $s\leq
x\leq t$. One can also check that, for any $s\in \mathbb{R}_{0}^{+}$ and $%
t\in \left( s,\infty \right) $, 
\begin{equation}
\partial _{t}\mathfrak{V}_{t,s}^{\ast }=-8\mathfrak{V}_{t,s}^{\ast }%
\mathfrak{B}_{t}\qquad \text{and}\qquad \partial _{t}\mathfrak{U}%
_{t,s}^{\ast }=-2\mathfrak{U}_{t,s}^{\ast }\mathfrak{D}_{t}
\label{sdsdsd3sdsdsd3+1}
\end{equation}%
in the strong sense in $\mathfrak{h}$ and $\mathcal{D}\left( \Upsilon
_{0}\right) $, respectively, as well as 
\begin{equation}
\partial _{s}\mathfrak{V}_{t,s}^{\ast }=8\mathfrak{B}_{s}\mathfrak{V}%
_{t,s}^{\ast }\qquad \text{and}\qquad \partial _{s}\mathfrak{U}_{t,s}^{\ast
}=2\mathfrak{D}_{s}\mathfrak{U}_{t,s}^{\ast }  \label{sdsdsd4}
\end{equation}%
the derivatives being both in the strong sense in $\mathfrak{h}$.

If $s=t\in \mathbb{R}_{0}^{+}$, then the partial derivatives of $%
(t,s)\mapsto \mathfrak{V}_{t,s}$ stated in (\ref{sdsdsd3sdsdsd3})--(\ref%
{sdsdsd4}) also hold true. See for instance \cite[Chap. 5, Theorem 5.2]{Pazy}%
, which gives a detailed proof of basic properties of $(\mathfrak{V}%
_{t,s})_{t,s\in \mathbb{R}_{0}^{+}}$, also well-defined for $s\geq t$. See
below Equation (\ref{DFDFDDF}). Additionally, the derivative family $\left(
\partial _{s}\mathfrak{U}_{t,s}\right) _{t\geq s>0}$ given in (\ref{sdsdsd3}%
) also hold true for $s=t\in \mathbb{R}^{+}$, as it is shown in \cite[Lemma
35]{bach-bru-memo} on an analogous situation.

We are now in a position to give another explicit expression of the operator
families $\mathfrak{D}$ and $\mathfrak{B}$ by invoking Lemma \ref{lemma
constante of motion copy(5)}. We have indeed the following statement:

\begin{proposition}[$\mathfrak{B}$ and $\mathfrak{D}$ as function of initial
data]
\label{lemma constante of motion copy(1)}\mbox{}\newline
Take a self-adjoint operator $\Upsilon _{0}=\Upsilon _{0}^{\ast }$ and $\mu
\in \mathbb{R}$ such that $\Upsilon _{0}\geq -\left( \mu -\varepsilon
\right) \mathbf{1}$ for some $\varepsilon \in \mathbb{R}^{+}$. Let $%
D_{0}=\pm D_{0}^{\top }\in \mathcal{B}\left( \mathfrak{h}\right) $ ($%
D_{0}\neq 0$). Then, for any $s\in \mathbb{R}^{+}$ and $t\in \left[ s,\infty
\right) $, 
\begin{equation*}
\mathfrak{D}_{t}=\mathfrak{V}_{t,s}\mathfrak{D}_{s}\mathfrak{V}_{t,s}^{\ast
}\quad \text{and}\quad \mathfrak{B}_{t}=\mathfrak{U}_{t,s}\mathfrak{B}_{s}%
\mathfrak{U}_{t,s}^{\ast }-4\int_{s}^{t}\mathfrak{U}_{t,\tau }D_{\tau
}D_{\tau }^{\ast }\mathfrak{U}_{t,\tau }^{\ast }\mathrm{d}\tau \ .
\end{equation*}
\end{proposition}

\begin{proof}
Assume for the moment that $\Upsilon _{0}\in \mathcal{B}(\mathfrak{h})$.
Then, (\ref{sdsdsd1}), (\ref{sdsdsd3}) and (\ref{sdsdsd4}) hold true even
for $s=t$ and, as a consequence, for any $s\in \mathbb{R}_{0}^{+}$ and $t\in %
\left[ s,\infty \right) $, 
\begin{equation}
\partial _{s}\{\mathfrak{V}_{t,s}\mathfrak{D}_{s}\mathfrak{V}_{t,s}^{\ast
}\}=\partial _{s}\{\mathfrak{V}_{t,s}\}\mathfrak{D}_{s}\mathfrak{V}%
_{t,s}^{\ast }+\mathfrak{V}_{t,s}\partial _{s}\{\mathfrak{D}_{s}\}\mathfrak{V%
}_{t,s}^{\ast }+\mathfrak{V}_{t,s}\mathfrak{D}_{s}\partial _{s}\{\mathfrak{V}%
_{t,s}^{\ast }\}=0\ ,  \label{infer+1}
\end{equation}%
which in turn implies that 
\begin{equation}
\mathfrak{D}_{t}=\mathfrak{V}_{t,s}\mathfrak{D}_{s}\mathfrak{V}_{t,s}^{\ast
}\ .  \label{casbounded}
\end{equation}%
If $\Upsilon _{0}\notin \mathcal{B}(\mathfrak{h})$\ then we can reproduce
this proof, provided one ensures the conservation of the domain $\mathcal{D}%
\left( \Upsilon _{0}\right) $ by $\mathfrak{V}_{t,s}^{\ast }$ while
verifying the Leibniz rule used in (\ref{infer+1}) on the domain $\mathcal{D}%
\left( \Upsilon _{0}\right) $. The fact that 
\begin{equation*}
\mathfrak{V}_{t,s}^{\ast }\mathcal{D}\left( \Upsilon _{0}\right) \subseteq 
\mathcal{D}\left( \Upsilon _{0}\right)
\end{equation*}%
is checked for $t>s$ by writing $\mathfrak{V}_{t,s}^{\ast }$ as a Dyson
series or with an integral equation. (Recall for instance that $D_{t}%
\mathfrak{h}\subseteq \mathcal{D}(\Upsilon _{0})$ for any $t>0$.) See, e.g., 
\cite[Equations (V.78)--(V.82)]{bach-bru-memo}. The justification of the
Leibniz rule used in (\ref{infer+1}) is much more involved, but it is done
exactly like in \cite[Equations (V.82)--(V.94)]{bach-bru-memo} and we thus
omit the details. Doing so, we get 
\begin{equation*}
\mathfrak{D}_{t}=\mathfrak{V}_{t,s}\mathfrak{D}_{s}\mathfrak{V}_{t,s}^{\ast }
\end{equation*}%
for any $s\in \mathbb{R}^{+}$ and $t\in \left[ s,\infty \right) $. In a
similar way, one computes from Lemma \ref{lemma constante of motion copy(5)}
combined with (\ref{sdsdsd3}) and (\ref{sdsdsd4}) that 
\begin{equation*}
\partial _{s}\left\{ \mathfrak{U}_{t,s}\mathfrak{B}_{s}\mathfrak{U}%
_{t,s}^{\ast }-4\int_{s}^{t}\mathfrak{U}_{t,\tau }D_{\tau }D_{\tau }^{\ast }%
\mathfrak{U}_{t,\tau }^{\ast }\mathrm{d}\tau \right\} =0\ ,\qquad t\in 
\mathbb{R}^{+},\ s\in \left( 0,t\right) \ .
\end{equation*}%
Note that the proof of this equality uses $\mathfrak{U}_{t,s}^{\ast }%
\mathfrak{h}\subseteq \mathcal{D}\left( \Upsilon _{0}\right) $ for $t>s>0$
and estimates showing the Leibniz rule, similar to (\ref{Leibniz rule}).
Then, it follows that, for any $t\in \mathbb{R}^{+}$ and $s,\delta \in
\left( 0,t\right) $,%
\begin{equation}
\mathfrak{U}_{t,t-\delta }\mathfrak{B}_{t-\delta }\mathfrak{U}_{t,t-\delta
}^{\ast }-4\int_{t-\delta }^{t}\mathfrak{U}_{t,\tau }D_{\tau }D_{\tau
}^{\ast }\mathfrak{U}_{t,\tau }^{\ast }\mathrm{d}\tau =\mathfrak{U}_{t,s}%
\mathfrak{B}_{s}\mathfrak{U}_{t,s}^{\ast }-4\int_{s}^{t}\mathfrak{U}_{t,\tau
}D_{\tau }D_{\tau }^{\ast }\mathfrak{U}_{t,\tau }^{\ast }\mathrm{d}\tau \ .
\label{bnm}
\end{equation}%
Since $(\mathfrak{B}_{t})_{t>0}\in C\left( \mathbb{R}^{+};\mathcal{B}(%
\mathfrak{h})\right) $ (Lemma \ref{lemma constante of motion copy(5)}) while 
$(\mathfrak{U}_{t,s})_{t\geq s}\subseteq \mathcal{B}(\mathfrak{h})$ is
strongly continuous in the parameter $s$, one uses straightforward estimates
to get 
\begin{equation}
\mathfrak{B}_{t}\varphi =\lim_{\delta \rightarrow 0^{+}}\left( \mathfrak{U}%
_{t,t-\delta }\mathfrak{B}_{t-\delta }\mathfrak{U}_{t,t-\delta }^{\ast
}-4\int_{t-\delta }^{t}\mathfrak{U}_{t,\tau }D_{\tau }D_{\tau }^{\ast }%
\mathfrak{U}_{t,\tau }^{\ast }\mathrm{d}\tau \right)  \label{bnm2}
\end{equation}%
for any $t\in \mathbb{R}^{+}$, which, combined with (\ref{bnm}), implies the
last assertion of the lemma.
\end{proof}

\begin{remark}[Bounded case]
\label{remark bounded2}\mbox{}\newline
For bounded operators $\Upsilon_{0}\in\mathcal{B}\left(\mathfrak{h}\right)$,
Proposition \ref{lemma constante of motion copy(1)} can be extended to
include the initial time $s=0$.
\end{remark}

Proposition \ref{lemma constante of motion copy(1)} directly leads to
various ellipticity properties of the operator flow. We start with the most
fundamental one. Recall that $\sigma (X)\subseteq \mathbb{C}$ denotes the
spectrum of an operator $X$.

\begin{theorem}[Elliptic operator flow -- I]
\label{theorem sympa0}\mbox{}\newline
Take a self-adjoint operator $\Upsilon _{0}=\Upsilon _{0}^{\ast }$ and $\mu
\in \mathbb{R}$ such that $\Upsilon _{0}\geq -\left( \mu -\varepsilon
\right) \mathbf{1}$ for some $\varepsilon \in \mathbb{R}^{+}$. Let $%
D_{0}=\pm D_{0}^{\top }\in \mathcal{B}\left( \mathfrak{h}\right) $ ($%
D_{0}\neq 0$). Then, for all $t\in \mathbb{R}_{0}^{+}$, $\mathfrak{D}%
_{t}\geq -2\mu \mathbf{1}$ and the real-valued function $\zeta $
well-defined on $\mathbb{R}_{0}^{+}$ by 
\begin{equation}
\zeta \left( t\right) \doteq \sup \left\{ \vartheta \in \mathbb{R}:\mathfrak{%
D}_{t}\geq \vartheta \mathbf{1}\right\} =\inf \sigma \left( \mathfrak{D}%
_{t}\right) \ ,\qquad t\in \mathbb{R}_{0}^{+}\ ,  \label{map}
\end{equation}%
is  upper  semicontinuous at $t=0$ and continuous on $\mathbb{R}%
^{+} $ as well as decreasing with $\zeta \left( \mathbb{R}_{0}^{+}\right)
\subseteq \lbrack 0,\zeta \left( 0\right) ]$ whenever $\zeta \left( 0\right)
\in \mathbb{R}_{0}^{+}$, while it is continuously increasing with $\zeta
\left( \mathbb{R}_{0}^{+}\right) \subseteq \lbrack \zeta \left( 0\right) ,0)$
whenever $\zeta \left( 0\right) \in \mathbb{R}^{-}$.
\end{theorem}

\begin{proof}
The proof is done in several steps: \medskip

\noindent \underline{Step 1:} $(\mathfrak{V}_{t,s})_{t,s\in \mathbb{R}%
_{0}^{+}}$ is a usual Dyson series obtained from the family $\mathfrak{B}$
of bounded operators. Observe in particular from (\ref{sdsdsd3}) and (\ref%
{sdsdsd4}) that 
\begin{equation*}
\partial _{x}\left\{ \mathfrak{V}_{t,x}\mathfrak{V}_{t,x}^{\ast }\right\} =16%
\mathfrak{V}_{t,x}\mathfrak{B}_{x}\mathfrak{V}_{t,x}^{\ast }\geq 0\ ,\qquad
t\in \mathbb{R}^{+},\ x\in \left[ 0,t\right] \ .
\end{equation*}%
In particular, for any $s\in \mathbb{R}_{0}^{+}$ and $t\in \left[ s,\infty
\right) $, 
\begin{equation}
\mathfrak{V}_{t,s}\mathfrak{V}_{t,s}^{\ast }\leq \mathbf{1}\ .  \label{sssss}
\end{equation}%
Note also that the (unique) solution to the non-autonomous evolution
equations 
\begin{equation}
\partial _{t}\mathfrak{V}_{s,t}=8\mathfrak{V}_{s,t}\mathfrak{B}_{t}\qquad 
\text{and}\qquad \partial _{s}\mathfrak{V}_{s,t}=-8\mathfrak{B}_{s}\mathfrak{%
V}_{s,t}  \label{DFDFDDF}
\end{equation}%
with $\mathfrak{V}_{s,s}\doteq \mathbf{1}$ for any $s\in \mathbb{R}_{0}^{+}$
and $t\in \left[ s,\infty \right) $ satisfies $\mathfrak{V}_{s,t}\mathfrak{V}%
_{t,s}=\mathbf{1}$. \medskip

\noindent \underline{Step 2:} Remark from (\ref{def D frac}) and $\Upsilon
_{0}^{\top }\geq -\mu \mathbf{1}$ that 
\begin{equation*}
\mathfrak{D}_{t}\geq -2\mu \mathbf{1}\ ,\qquad t\in \mathbb{R}_{0}^{+}\ .
\end{equation*}%
In particular, the set $\left\{ \vartheta \in \mathbb{R}:\mathfrak{D}%
_{t}\geq \vartheta \mathbf{1}\right\} $ is non-empty for any $t\in \mathbb{R}%
_{0}^{+}$. Furthermore,  for some eigenvector $\varphi _{e}\in 
\mathfrak{\mathcal{D}}(\Upsilon _{0})$ of $\Upsilon _{0}$ , observe that 
\begin{equation*}
\left\langle \varphi _{e},\mathfrak{D}_{t}\varphi _{e}\right\rangle _{%
\mathfrak{h}}=\left\langle \varphi _{e},\mathfrak{D}_{0}\varphi
_{e}\right\rangle _{\mathfrak{h}}+\left\langle \varphi _{e},\left( \Delta
_{t}+4\mathfrak{B}_{t}-4\mathfrak{B}_{0}\right) \varphi _{e}\right\rangle _{%
\mathfrak{h}}<+\infty
\end{equation*}%
for any $t\in \mathbb{R}_{0}^{+}$, because $\Delta _{t},\mathfrak{B}_{t}\in 
\mathcal{B}\left( \mathfrak{h}\right) $.  Consequently, the supremum of $%
\left\{ \vartheta \in \mathbb{R}:\mathfrak{D}_{t}\geq \vartheta \mathbf{1}%
\right\} $ is well defined for every $t\in \mathbb{R}_{0}^{+}$, and so is
the function $\zeta $ defined by (\ref{map}). In fact, by the spectral
theorem and \cite[Proposition 7.244]{Bru-Pedra-livre} applied to the bounded
operator $\mathfrak{P}_{t}\doteq \mathfrak{D}_{t}\mathbf{1}\left[ \mathfrak{D%
}_{t}<C\right] $ with $C>\inf \sigma (\mathfrak{D}_{t})$, one has that 
\begin{equation*}
\zeta \left( t\right) =\inf \sigma \left( \mathfrak{D}_{t}\right) =\inf
\sigma \left( \mathfrak{P}_{t}\right) =\inf_{\varphi \in \mathfrak{h}%
:\left\Vert \varphi \right\Vert _{\mathfrak{h}}=1}\left\langle \varphi ,%
\mathfrak{P}_{t}\varphi \right\rangle _{\mathfrak{h}}=\inf_{\varphi \in 
\mathfrak{h}:\left\Vert \varphi \right\Vert _{\mathfrak{h}}=1}\left\langle
\varphi ,\mathfrak{D}_{t}\varphi \right\rangle _{\mathfrak{h}},\qquad t\in 
\mathbb{R}_{0}^{+}\ ,
\end{equation*}%
since $\mathfrak{D}_{t}=\mathfrak{D}_{t}^{\ast }$. Additionally, by using
Theorem \ref{section extension gap} and Lemma \ref{lemma constante of motion
copy(5)}, this function is continuous on $\mathbb{R}^{+}$ and upper
semicontinuous at $t=0$, because the infimum of continuous functions is
always upper semicontinuous.\medskip

\noindent \underline{Step 3:} By Proposition \ref{lemma constante of motion
copy(1)} combined with Equations (\ref{def D frac}) and (\ref{sssss}), the
operator inequality $\mathfrak{D}_{0}\geq -\vartheta \mathbf{1}$ with $%
\vartheta \in \mathbb{R}_{0}^{+}$ directly yields 
\begin{equation}
\mathfrak{D}_{t}=\mathfrak{V}_{t,s}\mathfrak{D}_{s}\mathfrak{V}_{t,s}^{\ast
}\geq -\vartheta \mathbf{1}+4\mathfrak{V}_{t,s}\left( \mathfrak{B}_{s}-%
\mathfrak{B}_{0}\right) \mathfrak{V}_{t,s}^{\ast }+\mathfrak{V}_{t,s}\Delta
_{s}\mathfrak{V}_{t,s}^{\ast }  \label{new differential equation 3bisbis}
\end{equation}%
for any $s\in \mathbb{R}^{+}$ and $t\in \left[ s,\infty \right) $. We
proceed by taking the limit $s\rightarrow 0^{+}$ in (\ref{new differential
equation 3bisbis}). By using Theorem \ref{section extension gap}, Lemma \ref%
{lemma constante of motion copy(5)}, the fact that $\left( \mathfrak{V}%
_{t,s}^{\ast }\right) _{s\in \left[ 0,t\right] }\subseteq \mathcal{B}\left( 
\mathfrak{h}\right) $ (which can be uniformly bounded for $s\in \left[ 0,t%
\right] $) and the strong continuity (in $s$) of the operator family $\left( 
\mathfrak{V}_{t,s}^{\ast }\right) _{s\in \left[ 0,t\right] }\subseteq 
\mathcal{B}\left( \mathfrak{h}\right) $ at fixed $t\in \mathbb{R}^{+}$, we
infer from (\ref{new differential equation 3bisbis}) when $s\rightarrow
0^{+} $ that $\mathfrak{D}_{t}\geq -\vartheta \mathbf{1}$ for all $t\in 
\mathbb{R}_{0}^{+}$ as soon as $\mathfrak{D}_{0}\geq -\vartheta \mathbf{1}$
with $\vartheta \in \mathbb{R}_{0}^{+}$. \medskip

\noindent \underline{Step 4:} If $\mathfrak{D}_{t}\geq \vartheta \mathbf{1}$
with $\vartheta \in \mathbb{R}_{0}^{+}$ for some $t,\vartheta \in \mathbb{R}%
_{0}^{+}$ then we again use Proposition \ref{lemma constante of motion
copy(1)} to show that, for all $s\in (0,t]$, 
\begin{equation}
\mathfrak{D}_{s}=\mathfrak{V}_{s,t}\mathfrak{D}_{t}\mathfrak{V}_{s,t}^{\ast
}\geq \vartheta \mathfrak{V}_{s,t}\mathfrak{V}_{s,t}^{\ast }\ .  \label{fg2}
\end{equation}%
Since we have meanwhile that 
\begin{equation*}
\partial _{x}\left\{ \mathfrak{V}_{s,x}\mathfrak{V}_{s,x}^{\ast }\right\} =16%
\mathfrak{V}_{s,x}\mathfrak{B}_{x}\mathfrak{V}_{s,x}^{\ast }\geq 0\ ,\qquad
s\in \mathbb{R}_{0}^{+},\ x\in \lbrack s,\infty )\ ,
\end{equation*}%
we obtain that $\mathfrak{V}_{s,x}\mathfrak{V}_{s,x}^{\ast }\geq \mathbf{1}$
for any $x\in \lbrack s,\infty )$ and we can deduce from (\ref{fg2}) that $%
\mathfrak{D}_{t}\geq \vartheta \mathbf{1}$ for some $t,\vartheta \in \mathbb{%
R}_{0}^{+}$ implies that 
\begin{equation}
\mathfrak{D}_{x}\doteq \Upsilon _{0}+\Delta _{x}-\mu \mathbf{1}+4\mathfrak{B}%
_{x}\geq \vartheta \mathbf{1}  \label{limit x}
\end{equation}%
for all $x\in (0,t]$. See also (\ref{def D frac}). By using Lemma \ref{lemma
constante of motion copy(5)} and Theorem \ref{section extension gap}, we can
perform the limit $x\rightarrow 0$ in (\ref{limit x}) to obtain $\mathfrak{D}%
_{x}\geq \vartheta \mathbf{1}$ for all $x\in \lbrack 0,t]$ whenever $%
\mathfrak{D}_{t}\geq \vartheta \mathbf{1}$ for some $t,\vartheta \in \mathbb{%
R}_{0}^{+}$. \medskip

\noindent \underline{Step 5:} We deduce from Steps 2 and 4 that the
real-valued function $\zeta $ defined by (\ref{map}) is  upper 
semicontinuous at $t=0$ and continuous on $\mathbb{R}^{+}$ as well as
decreasing whenever $\zeta \left( 0\right) \in \mathbb{R}_{0}^{+}$. By Step
3, note that $\zeta \left( \mathbb{R}_{0}^{+}\right) \subseteq \lbrack
0,\zeta \left( 0\right) ]$ whenever $\zeta \left( 0\right) \in \mathbb{R}%
_{0}^{+}$, since in this case $\mathfrak{D}_{t}\geq 0$ for all $t\in \mathbb{%
R}_{0}^{+}$. Meanwhile, again by Steps 2 and 3, if $\zeta \left( 0\right)
\in \mathbb{R}^{-}$ then the real-valued function $\zeta $ is now
continuously increasing. Note in this last case that $\zeta \left( \mathbb{R}%
_{0}^{+}\right) \subseteq \lbrack \zeta \left( 0\right) ,0)\subseteq \mathbb{%
R}^{-}$ because the operator inequality $\mathfrak{D}_{t}\geq 0$ for some
time $t\in \mathbb{R}_{0}^{+}$ implies $\mathfrak{D}_{s}\geq 0$ for all
previous times $s\in \lbrack 0,t]$, see Step 4. This concludes the proof of\
the theorem.
\end{proof}

\begin{remark}[Bounded case]
\label{remark bounded2 copy(1)}\mbox{}\newline
For bounded operators $\Upsilon _{0}\in \mathcal{B}\left( \mathfrak{h}%
\right) $, the real-valued function $\zeta $ is always continuous on $%
\mathbb{R}_{0}^{+}$, because in this case $\mathfrak{D}\in C(\mathbb{R}%
_{0}^{+};\mathcal{B}(\mathfrak{h}))$, as explained in Remark \ref{remark
bounded1}.  We can in particular include the case $t=0$ at the end of
Step 2 in the proof of Theorem \ref{theorem sympa0}.  
\end{remark}

\begin{theorem}[Elliptic operator flow -- II]
\label{lemma constante of motion copy(3)}\mbox{}\newline
Take a self-adjoint operator $\Upsilon _{0}=\Upsilon _{0}^{\ast }$ and $\mu
\in \mathbb{R}$ such that $\Upsilon _{0}\geq -\left( \mu -\varepsilon
\right) \mathbf{1}$ for some $\varepsilon \in \mathbb{R}^{+}$. Let $%
D_{0}=\pm D_{0}^{\top }\in \mathcal{B}\left( \mathfrak{h}\right) $ ($%
D_{0}\neq 0$).

\begin{enumerate}
\item[\emph{(i)}] If $\Upsilon _{0}\in \mathcal{B}\left( \mathfrak{h}\right) 
$ then the mapping $t\mapsto \Vert \mathfrak{D}_{t}\Vert _{\mathrm{op}}$
from $\mathbb{R}_{0}^{+}$ to itself is decreasing.

\item[\emph{(ii)}] If $\mathfrak{D}_{0}\geq 0$ then the mapping $t\mapsto
\Vert \mathfrak{B}_{t}\Vert _{\mathrm{op}}$ from $\mathbb{R}_{0}^{+}$ to
itself is decreasing.
\end{enumerate}
\end{theorem}

\begin{proof}
We infer from Proposition \ref{lemma constante of motion copy(1)} in the
bounded case, i.e., for $\Upsilon _{0}\in \mathcal{B}\left( \mathfrak{h}%
\right) $, that, for any $s\in \mathbb{R}_{0}^{+}$ and $t\in \left[ s,\infty
\right) $, 
\begin{equation}
\mathfrak{D}_{t}=\mathfrak{V}_{t,s}\mathfrak{D}_{s}\mathfrak{V}_{t,s}^{\ast
}\leq \left\Vert \mathfrak{D}_{s}\right\Vert _{\mathrm{op}}\mathfrak{V}_{t,s}%
\mathfrak{V}_{t,s}^{\ast }<\infty \ .  \label{sdsdsdsd}
\end{equation}%
Note that we include here the case $s=0$, thanks to (\ref{casbounded}). See
also Remarks \ref{remark bounded1} and \ref{remark bounded2}. Using (\ref%
{sdsdsdsd}) together with (\ref{sssss}) we thus deduce that, for any $s\in 
\mathbb{R}_{0}^{+}$ and $t\in \left[ s,\infty \right) $, 
\begin{equation*}
\mathfrak{D}_{t}=\mathfrak{V}_{t,s}\mathfrak{D}_{s}\mathfrak{V}_{t,s}^{\ast
}\leq \left\Vert \mathfrak{D}_{s}\right\Vert _{\mathrm{op}}\mathfrak{V}_{t,s}%
\mathfrak{V}_{t,s}^{\ast }\leq \left\Vert \mathfrak{D}_{s}\right\Vert _{%
\mathrm{op}}\mathbf{1}<\infty \ ,
\end{equation*}%
which leads to Assertion (i). To prove the second assertion, we use again
Proposition \ref{lemma constante of motion copy(1)} to obtain the operator
inequality  
\begin{equation}
\mathfrak{B}_{t}\leq \mathfrak{U}_{t,s}\mathfrak{B}_{s}\mathfrak{U}%
_{t,s}^{\ast }\leq \mathfrak{U}_{t,s}\mathfrak{U}_{t,s}^{\ast }\left\Vert 
\mathfrak{B}_{s}\right\Vert _{\mathrm{op}}  \label{eq1}
\end{equation}%
 for any $s\in \mathbb{R}^{+}$ and $t\in \left[ s,\infty \right) $. By
doing the limit $s\rightarrow 0^{+}$ in this last inequality, note also that 
\begin{equation}
\mathfrak{B}_{t}\leq \mathfrak{U}_{t,0}\mathfrak{B}_{0}\mathfrak{U}%
_{t,0}^{\ast }\ ,  \label{eq1bis}
\end{equation}%
thanks to Lemma \ref{lemma constante of motion copy(5)} and the strong
continuity (in $s$) of the operator families $\left( \mathfrak{U}%
_{t,s}\right) _{s\in \left[ 0,t\right] }\subseteq \mathcal{B}\left( 
\mathfrak{h}\right) $ and $\left( \mathfrak{V}_{t,s}^{\ast }\right) _{s\in %
\left[ 0,t\right] }\subseteq \mathcal{B}\left( \mathfrak{h}\right) $ at
fixed $t\in \mathbb{R}^{+}$. In addition, (\ref{sdsdsd3}) and (\ref{sdsdsd4}%
) imply that 
\begin{equation}
\partial _{x}\left\{ \mathfrak{U}_{t,x}\mathrm{e}^{4\vartheta \left(
t-x\right) }\mathfrak{U}_{t,x}^{\ast }\right\} =4\mathfrak{U}_{t,x}\mathrm{e}%
^{4\vartheta \left( t-x\right) }\left( \mathfrak{D}_{x}-\vartheta \mathbf{1}%
\right) \mathfrak{U}_{t,x}^{\ast }  \label{eq2}
\end{equation}%
for any $\vartheta \in \mathbb{R}$, $t\in \mathbb{R}^{+}$ and$\ x\in \left[
0,t\right] $. If $\mathfrak{D}_{0}\geq 0$ then $\mathfrak{D}_{t}\geq 0$ for
all $t\in \mathbb{R}_{0}^{+}$, by Theorem \ref{theorem sympa0}. Therefore,
if $\mathfrak{D}_{0}\geq 0$ then  we can integrate (\ref{eq2}) for $%
\vartheta =0$, implying that 
\begin{equation*}
\mathfrak{U}_{t,s}\mathfrak{U}_{t,s}^{\ast }=\mathbf{1}-4\int_{s}^{t}%
\mathfrak{U}_{t,x}\mathfrak{D}_{x}\mathfrak{U}_{t,x}^{\ast }\mathrm{d}x\leq 
\mathbf{1}
\end{equation*}%
for any $s\in \mathbb{R}_{0}^{+}$ and $t\in \left[ s,\infty \right) $, and
combine this inequality with (\ref{eq1})  in order to get 
\begin{equation}
\left\Vert \mathfrak{B}_{t}\right\Vert _{\mathrm{op}}\leq \left\Vert 
\mathfrak{B}_{s}\right\Vert _{\mathrm{op}}<\infty \ ,\qquad s\in \mathbb{R}%
_{0}^{+},\ t\in \left[ s,\infty \right) \ .  \label{fg1}
\end{equation}
\end{proof}

In the bounded case, Theorem \ref{lemma constante of motion copy(3)} (i)
directly yields the existence of a finite constant $\mathrm{C}<\infty $ such
that 
\begin{equation*}
\Upsilon _{t}-\mu \mathbf{1+}D_{t}\left( \Upsilon _{t}^{\top }+\mu \mathbf{1}%
\right) ^{-1}D_{t}^{\ast }\leq \mathrm{C}\mathbf{1\ ,\qquad }t\in \mathbb{R}%
_{0}^{+}\ .
\end{equation*}%
In other words, the flow always stays inside the operator version of an
(unconventional) ellipse. Compare with Equation (\ref{eliptic condition})
for the one-dimensional situation. Additionally, Theorem \ref{lemma
constante of motion copy(3)} has a direct consequence in terms of the
convergence of the operator flow:

\begin{corollary}[Elliptic operator flow -- bounded case]
\label{lemma constante of motion copy(4)}\mbox{}\newline
Assume $D_{0}=\pm D_{0}^{\top }\in \mathcal{B}\left( \mathfrak{h}\right) $ ($%
D_{0}\neq 0$), $\Upsilon _{0}=\Upsilon _{0}^{\ast }\in \mathcal{B}\left( 
\mathfrak{h}\right) $ and $\mu \in \mathbb{R}$ such that $\Upsilon _{0}\geq
-\left( \mu -\varepsilon \right) \mathbf{1}$ for some $\varepsilon \in 
\mathbb{R}^{+}$. Then, for any $s\in \mathbb{R}_{0}^{+}$ and $t\in \left[
s,\infty \right) $, 
\begin{equation*}
16\int_{s}^{t}D_{\tau }D_{\tau }^{\ast }\mathrm{d}\tau +4\mathfrak{B}%
_{t}\leq \left( \left\Vert \mathfrak{D}_{s}\right\Vert _{\mathrm{op}}+2\mu
\right) \mathbf{1}\ .
\end{equation*}
\end{corollary}

\begin{proof}
 By (\ref{sdssds}) and (\ref{def D frac}), for any $s\in 
\mathbb{R}_{0}^{+}$ and $t\in \left[ s,\infty \right) $,%
\begin{equation*}
16\int_{s}^{t}D_{\tau }D_{\tau }^{\ast }\mathrm{d}\tau +4\mathfrak{B}%
_{t}=\Upsilon _{t}-\mu \mathbf{1}+4\mathfrak{B}_{t}-\Upsilon _{s}+\mu 
\mathbf{1}=\mathfrak{D}_{t}-\Upsilon _{s}+\mu \mathbf{1}\ .
\end{equation*}%
Since $\Upsilon _{s}\geq -\mu \mathbf{1}$, for any $s\in \mathbb{R}_{0}^{+}$
and $t\in \left[ s,\infty \right) $, 
\begin{equation*}
\mathfrak{D}_{t}-\Upsilon _{s}+\mu \mathbf{1}\leq \left( \left\Vert 
\mathfrak{D}_{t}\right\Vert _{\mathrm{op}}+2\mu \right) \mathbf{1}\ .
\end{equation*}%
Using these observations together with Theorem \ref{lemma constante of
motion copy(3)} (i), we then deduce the corollary. 
\end{proof}

Note that Corollary \ref{lemma constante of motion copy(4)} is also an
ellipticity property by taking now the variable $\Delta _{t}$ instead of $%
\Upsilon _{t}\doteq \Upsilon _{0}+\Delta _{t}$, since 
\begin{equation*}
16\int_{0}^{t}D_{\tau }D_{\tau }^{\ast }\mathrm{d}\tau +4\mathfrak{B}%
_{t}=\Delta _{t}+4D_{t}\left( \Delta _{t}^{\top }+\Upsilon _{0}^{\top }+\mu 
\mathbf{1}\right) ^{-1}D_{t}^{\ast }\ .
\end{equation*}

In Theorem \ref{lemma constante of motion copy(3)} (ii), we use the
assumption 
\begin{equation}
\mathfrak{D}_{0}\doteq \Upsilon _{0}-\mu \mathbf{1}+4\mathfrak{B}_{0}\geq 0
\label{condition expo decay}
\end{equation}%
with $\Upsilon _{0}=\Upsilon _{0}^{\ast }\geq -\left( \mu -\varepsilon
\right) \mathbf{1}$ for some $\mu \in \mathbb{R}$ and $\varepsilon \in 
\mathbb{R}^{+}$, which yields $\mathfrak{D}_{t}\geq 0$ for all $t\in \mathbb{%
R}_{0}^{+}$.  Equation (\ref{condition expo decay}) is another
ellipticity property. Take for instance  the simpler case for which%
\begin{equation}
K_{0}=\Upsilon _{0}D_{0}-D_{0}\Upsilon _{0}^{\top }=0\ ,  \label{commutfrist}
\end{equation}%
with $D_{0}=\pm D_{0}^{\top }\in \mathcal{B}\left( \mathfrak{h}\right) $ and 
$\Upsilon _{0}=\Upsilon _{0}^{\ast }\geq -\left( \mu -\varepsilon \right) 
\mathbf{1}$ for some $\mu \in \mathbb{R}$ and $\varepsilon \in \mathbb{R}%
^{+} $. This equality propagates for all times $t\in \mathbb{R}_{0}^{+}$
(see Lemma \ref{lemma constante of motion copy(2)}) and provides an explicit
operatorial relation given by the constant of motion (\ref%
{constmotionexplicit}), thanks to Theorem \ref{theorem constante of motion}
(ii) and Remark \ref{theorem constante of motion copy(1)}. If $K_{0}=0$ then
we can compute from Lemma \ref{lemma constante of motion copy(2)} that, for
any $t\in \mathbb{R}_{0}^{+}$,%
\begin{equation*}
\left( \Upsilon _{t}+\mu \right) ^{\frac{1}{2}}\mathfrak{D}_{t}\left(
\Upsilon _{t}+\mu \right) ^{\frac{1}{2}}=\Upsilon _{t}^{2}-\mu ^{2}\mathbf{1}%
+4D_{t}D_{t}^{\ast }\ ,
\end{equation*}%
where it exists. Using now Theorem \ref{theorem constante of motion} (ii)
and Remark \ref{theorem constante of motion copy(1)}, we deduce that 
\begin{equation}
\mathfrak{D}_{t}=\left( \Upsilon _{t}+\mu \right) ^{-\frac{1}{2}}\left(
\Upsilon _{0}^{2}-\mu ^{2}\mathbf{1}+4D_{0}D_{0}^{\ast }\right) \left(
\Upsilon _{t}+\mu \right) ^{-\frac{1}{2}}  \label{commutlast}
\end{equation}%
for any $t\in \mathbb{R}_{0}^{+}$. Therefore, to ensure the positivity of $%
\mathfrak{D}_{t}$ for all times $t\in \mathbb{R}_{0}^{+}$ when $K_{0}=0$, it
suffices that 
\begin{equation}
\Upsilon _{0}^{2}+4D_{0}D_{0}^{\ast }\geq \mu ^{2}\mathbf{1}\ .
\label{commutlast2}
\end{equation}%
This simplified example highlights the ellipticity properties of the flow
and, in particular, Condition (\ref{condition expo decay}), by recognizing
that $\Upsilon _{0}^{2}+4D_{0}D_{0}^{\ast }=\mu ^{2}\mathbf{1}$ is a
conventional ellipse for $\mu \neq 0$. Note in this special case that (\ref%
{commutlast2}) is nothing else  than  Berezin's assumption given in 
\cite[Theorem 8.2]{Berezin} (see also \cite[Theorem 3]{Kato-Mugibayashi})
with $\alpha =\mu ^{2}$, $\Upsilon _{0}=\Upsilon _{0}^{\top }$ and $%
[D_{0},\Upsilon _{0}]=0$. In fact, the assumptions of Berezin's theorem,
which gives a diagonalization of fermionic quadratic Hamiltonians (Section %
\ref{Diagonalization of Quadratic}), directly imply that $K_{0}\in \mathcal{L%
}^{2}(\mathfrak{h})$ and $\Upsilon _{0}^{2}+4D_{0}D_{0}^{\ast }\pm
2K_{0}>\alpha \mathbf{1}$.

The ellipticity condition given by (\ref{condition expo decay}) does not
require bounded initial data $\Upsilon _{0}$. By contrast, Theorem \ref%
{lemma constante of motion copy(3)} (i) and Corollary \ref{lemma constante
of motion copy(4)} are only interesting results on \emph{bounded} operators $%
\Upsilon _{0}$. This is a strong limitation as the unboundedness of $%
\Upsilon _{0}$ should not imply any issue on the convergence of the flow, $%
\Upsilon _{0}$ being bounded from below.

In fact, the ellipticity property given by Corollary \ref{lemma constante of
motion copy(4)} can be adapted by allowing unbounded operators $\Upsilon
_{0} $. This is done at the cost of additional assumptions. First, we need
the initial ellipticity property (\ref{condition expo decay}). Secondly, we
require the initial data $D_{0}$ to be Hilbert-Schmidt in order to use
properties of the trace. Having such hypotheses at our disposal,  we
show the following assertion: 

\begin{theorem}[Elliptic operator flow -- Hilbert-Schmidt case]
\label{Theorem sympa}\mbox{}\newline
Take a self-adjoint operator $\Upsilon _{0}=\Upsilon _{0}^{\ast }$ and $\mu
\in \mathbb{R}$ such that $\Upsilon _{0}\geq -\left( \mu -\varepsilon
\right) \mathbf{1}$ for some $\varepsilon \in \mathbb{R}^{+}$. Let $%
D_{0}=\pm D_{0}^{\top }\in \mathcal{L}^{2}(\mathfrak{h})$ ($D_{0}\neq 0$)
and assume that $\mathfrak{D}_{0}\geq 0$.

\begin{enumerate}
\item[\emph{(i)}] The mapping $t\mapsto \Vert \mathfrak{B}_{t}\Vert _{1}$
from $\mathbb{R}_{0}^{+}$ to itself is decreasing.

\item[\emph{(ii)}] For any $s\in \mathbb{R}_{0}^{+}$ and $t\in \left[
s,\infty \right) $, 
\begin{equation*}
16\int_{s}^{t}\left\Vert D_{\tau }\right\Vert _{2}^{2}\mathrm{d}\tau +4\Vert 
\mathfrak{B}_{t}\Vert _{1}\leq 4\Vert \mathfrak{B}_{s}\Vert _{1}\leq 4\Vert 
\mathfrak{B}_{0}\Vert _{1}<\infty \ .
\end{equation*}
\end{enumerate}
\end{theorem}

\begin{proof}
Assume all conditions of the theorem. The proof is done in several steps:
\medskip

\noindent \underline{Step 1:} First, note from (\ref{petit inequalitybisbis2}%
) and (\ref{def D frac2}) that, for any $t\in \mathbb{R}_{0}^{+}$, 
\begin{equation}
\Vert \mathfrak{B}_{t}\Vert _{1}=\mathrm{tr}\left( \mathfrak{B}_{t}\right)
\leq \varepsilon ^{-1}\mathrm{tr}\left( D_{t}D_{t}^{\ast }\right) \doteq
\varepsilon ^{-1}\left\Vert D_{t}\right\Vert _{2}^{2}\leq \varepsilon ^{-1}%
\mathrm{e}^{8\mu t}\left\Vert D_{0}\right\Vert _{2}^{2}<\infty \ ,
\label{dfdfdfdfd}
\end{equation}%
keeping in mind that $\mathfrak{B}_{t}\geq 0$. Now, if $\mathfrak{D}_{0}\geq
0$ then $\mathfrak{D}_{t}\geq 0$ for all $t\in \mathbb{R}_{0}^{+}$, thanks
to Theorem \ref{theorem sympa0}. Therefore, Equations (\ref{sdsdsd3sdsdsd3})
and (\ref{sdsdsd3sdsdsd3+1}) yield 
\begin{equation*}
\partial _{x}\left\{ \mathfrak{U}_{x,s}^{\ast }\mathfrak{U}_{x,s}\right\} =-4%
\mathfrak{U}_{x,s}^{\ast }\mathfrak{D}_{x}\mathfrak{U}_{x,s}\leq 0
\end{equation*}%
for any $s\in \mathbb{R}_{0}^{+}$ and $x\in \left( s,\infty \right) $. It
follows that 
\begin{equation*}
\mathfrak{U}_{t,s}^{\ast }\mathfrak{U}_{t,s}\leq \mathfrak{U}_{x,s}^{\ast }%
\mathfrak{U}_{x,s}
\end{equation*}%
for any $s\in \mathbb{R}_{0}^{+}$ and $x\in \left( s,t\right) $. Using the
above operator inequality in the limit $x\rightarrow s$, the cyclicity of
the trace and the positivity of operators $\mathfrak{B}_{t}$, we infer from (%
\ref{eq1}) that%
\begin{equation}
\mathrm{tr}\left( \mathfrak{B}_{t}\right) \leq \mathrm{tr}\left( \mathfrak{U}%
_{t,s}\mathfrak{B}_{s}\mathfrak{U}_{t,s}^{\ast }\right) =\mathrm{tr}\left( 
\mathfrak{B}_{s}^{1/2}\mathfrak{U}_{t,s}^{\ast }\mathfrak{U}_{t,s}\mathfrak{B%
}_{s}^{1/2}\right) \leq \mathrm{tr}\left( \mathfrak{B}_{s}\right)
\label{dfdfddf}
\end{equation}%
for any $s\in \mathbb{R}^{+}$ and $t\in \left[ s,\infty \right) $. In
particular, the use of $\mathfrak{D}_{0}\geq 0$ strongly improves (\ref%
{dfdfdfdfd}) because the above inequality obviously yields 
\begin{equation*}
\sup_{t\in \mathbb{R}^{+}}\Vert \mathfrak{B}_{t}\Vert _{1}\leq \varepsilon
^{-1}\left\Vert D_{0}\right\Vert _{2}^{2}<\infty \ .
\end{equation*}%
Assertion (i) is also a direct consequence of (\ref{dfdfddf}). \medskip

\noindent \underline{Step 2:} Note from the Cauchy-Schwarz inequality that 
\begin{equation*}
\left\Vert D_{t}D_{t}^{\ast }-D_{s}D_{s}^{\ast }\right\Vert _{1}\doteq 
\mathrm{tr}\left( \left\vert D_{t}D_{t}^{\ast }-D_{s}D_{s}^{\ast
}\right\vert \right) \leq \left\Vert D_{t}-D_{s}\right\Vert _{2}\left(
\left\Vert D_{t}\right\Vert _{2}+\left\Vert D_{s}\right\Vert _{2}\right)
\end{equation*}%
for any $s,t\in \mathbb{R}_{0}^{+}$. By\ Lemma \ref{Lemma HS1}, the operator
family $(D_{t}D_{t}^{\ast })_{t\in \mathbb{R}^{+}}$ is thus continuous in $%
\mathcal{L}^{1}(\mathfrak{h})$. Additionally, similar to Lemma \ref{Lemma
HS1}, the operator families $(\mathfrak{D}_{t}\mathfrak{B}_{t})_{t\in 
\mathbb{R}^{+}}$ and $(\mathfrak{B}_{t}\mathfrak{D}_{t})_{t\in \mathbb{R}%
^{+}}$ are continuous in $\mathcal{L}^{1}(\mathfrak{h})$. To see this,
remark from (\ref{def D frac})--(\ref{def D frac2}) that, for any $t\in 
\mathbb{R}_{0}^{+}$, 
\begin{multline*}
\mathfrak{D}_{t}\mathfrak{B}_{t}=\Upsilon _{t}D_{t}\left( \Upsilon
_{t}^{\top }+\mu \mathbf{1}\right) ^{-1}D_{t}^{\ast }-\mu D_{t}\left(
\Upsilon _{t}^{\top }+\mu \mathbf{1}\right) ^{-1}D_{t}^{\ast } \\
+4D_{t}\left( \Upsilon _{t}^{\top }+\mu \mathbf{1}\right) ^{-1}D_{t}^{\ast
}D_{t}\left( \Upsilon _{t}^{\top }+\mu \mathbf{1}\right) ^{-1}D_{t}^{\ast }\
.
\end{multline*}%
By (\ref{D_explicit_solutionbis}) and (\ref{hhhhh}) together with $D_{0}\in 
\mathcal{L}^{2}(\mathfrak{h})$, one gets that $\mathfrak{D}_{t}\mathfrak{B}%
_{t}\in \mathcal{L}^{1}(\mathfrak{h})$ for any $t\in \mathbb{R}^{+}$.
Additionally, by combining (\ref{hhhhh}) and (\ref{derivative2}) with $D\in
C(\mathbb{R}_{0}^{+};\mathcal{L}^{2}(\mathfrak{h}))$ (Lemma \ref{Lemma HS1})
and Lebesgue's dominated convergence theorem one checks that $(\mathfrak{D}%
_{t}\mathfrak{B}_{t})_{t\in \mathbb{R}^{+}}\in C(\mathbb{R}^{+};\mathcal{L}%
^{1}(\mathfrak{h}))$. The same holds true for the operator family $(%
\mathfrak{B}_{t}\mathfrak{D}_{t})_{t\in \mathbb{R}^{+}}\subseteq \mathcal{L}%
^{1}(\mathfrak{h})$.\medskip

\noindent \underline{Step 3:} By using Lemma \ref{lemma constante of motion
copy(5)}, one verifies that, for any $t\in \mathbb{R}^{+}$\ and $\delta \in 
\mathbb{R}^{+}$, 
\begin{multline*}
\left\Vert \delta ^{-1}\left\{ \mathfrak{B}_{t+\delta }-\mathfrak{B}%
_{t}\right\} +2\mathfrak{D}_{t}\mathfrak{B}_{t}+2\mathfrak{B}_{t}\mathfrak{D}%
_{t}+4D_{t}D_{t}^{\ast }\right\Vert _{1} \\
\leq 2\delta ^{-1}\int_{t}^{t+\delta }\left\{ \left\Vert \mathfrak{D}_{t}%
\mathfrak{B}_{t}-\mathfrak{D}_{\tau }\mathfrak{B}_{\tau }\right\Vert
_{1}+\left\Vert \mathfrak{B}_{t}\mathfrak{D}_{t}-\mathfrak{B}_{\tau }%
\mathfrak{D}_{\tau }\right\Vert _{1}\right\} \ \mathrm{d}\tau \\
+4\delta ^{-1}\int_{t}^{t+\delta }\left\Vert D_{t}D_{t}^{\ast }-D_{\tau
}D_{\tau }^{\ast }\right\Vert _{1}\ \mathrm{d}\tau \ .
\end{multline*}%
Mutadis mutandis for sufficiently small $\delta <0$ such that $t+\delta >0 $%
. Therefore, using Step 2 and the limit $\delta \rightarrow 0$, we conclude
that the first derivative of Lemma \ref{lemma constante of motion copy(5)}
holds true in the trace norm topology, which allows us to use Lebesgue's
dominated convergence theorem and thus arrive at the derivative 
\begin{equation*}
\partial _{t}\Vert \mathfrak{B}_{t}\Vert _{1}=-2\mathrm{tr}\left( \mathfrak{D%
}_{t}\mathfrak{B}_{t}+\mathfrak{B}_{t}\mathfrak{D}_{t}\right) -4\mathrm{tr}%
\left( D_{t}D_{t}^{\ast }\right) \in \mathbb{R}
\end{equation*}%
for all $t\in \mathbb{R}^{+}$. \medskip

\noindent \underline{Step 4:} Using this last derivative, we now easily
compute that 
\begin{equation*}
\partial _{t}\left( 4\int_{s}^{t}\Vert D_{\tau }\Vert _{2}^{2}\mathrm{d}\tau
+\Vert \mathfrak{B}_{t}\Vert _{1}\right) =-2\mathrm{tr}\left( \mathfrak{D}%
_{t}\mathfrak{B}_{t}+\mathfrak{B}_{t}\mathfrak{D}_{t}\right) \ .
\end{equation*}%
Through the cyclicity of the trace \cite[Lemma 100]{bach-bru}, we observe
that, for all $t\in \mathbb{R}^{+}$, 
\begin{equation*}
\partial _{t}\left( 4\int_{s}^{t}\left\Vert D_{\tau }\right\Vert _{2}^{2}%
\mathrm{d}\tau +\Vert \mathfrak{B}_{t}\Vert _{1}\right) =-2\mathrm{tr}\left( 
\mathfrak{B}_{t}^{1/2}\mathfrak{D}_{t}\mathfrak{B}_{t}^{1/2}\right) \leq 0\ ,
\end{equation*}%
thanks again to Theorem \ref{theorem sympa0}. Assertion (ii) then follows
for $s\in \mathbb{R}^{+}$, which can be extended by continuity to $s\in 
\mathbb{R}_{0}^{+}$, see Theorem \ref{Corollary existence}.
\end{proof}

\subsection{Asymptotics of the Elliptic Flow\label{Section tech6}}

Before analyzing concrete asymptotic results, we start by deriving basic
properties on the limit operators, provided it exists at least in the weak
operator topology. To present this preliminary result, we define the real
number $\zeta _{\infty }$ as the limit%
\begin{equation}
\zeta _{\infty }\doteq \lim_{t\rightarrow \infty }\zeta \left( t\right) \in 
\left[ -\left\vert \zeta \left( 0\right) \right\vert ,\left\vert \zeta
\left( 0\right) \right\vert \right] \subseteq \left[ -2\mu ,\infty \right)
\label{seeddd}
\end{equation}%
of the continuous, bounded and monotone function $\zeta :\mathbb{R}%
^{+}\rightarrow \mathbb{R}$ of Theorem \ref{theorem sympa0}. Clearly, this
limit is well-defined because of Theorem \ref{theorem sympa0} and the
monotone convergence theorem. Note that $\zeta \left( t\right) \geq -2\mu \ $%
\ for all $t\in \mathbb{R}_{0}^{+}$, thanks again to Theorem \ref{theorem
sympa0}. As above, $\Delta \equiv (\Delta _{t})_{t\geq 0}$ and $D\equiv
(D_{t})_{t\geq 0}$ are always given by Theorem \ref{section extension gap}
in all this section. Recall finally that the evolution family $%
(W_{t,s})_{t\geq s\geq 0}$ is defined by (\ref{definition of W}).

\begin{lemma}[Asymptotics properties of the flow]
\label{lemma asymptotics1}\mbox{}\newline
Take a self-adjoint operator $\Upsilon _{0}=\Upsilon _{0}^{\ast }$ and $\mu
\in \mathbb{R}$ such that $\Upsilon _{0}\geq -\left( \mu -\varepsilon
\right) \mathbf{1}$ for some $\varepsilon \in \mathbb{R}^{+}$. Let $%
D_{0}=\pm D_{0}^{\top }\in \mathcal{B}\left( \mathfrak{h}\right) $ ($%
D_{0}\neq 0$). If $t\mapsto \Vert D_{t}\Vert _{\mathrm{op}}$ is bounded on $%
\mathbb{R}_{0}^{+}$ and $\Delta \equiv (\Delta _{t})_{t\geq 0}$ converges in
the weak operator topology to some bounded operator $\Delta _{\infty }\in 
\mathcal{B}\left( \mathfrak{h}\right) $, as $t\rightarrow \infty $, then the
following assertions hold true:

\begin{enumerate}
\item[\emph{(i)}] $D\equiv (D_{t})_{t\geq 0}$ converges in the strong
operator topology to zero, along a subsequence.

\item[\emph{(ii)}] $\Delta _{\infty }$ satisfies the operator inequality 
\begin{equation*}
\Upsilon _{\infty }\doteq \Upsilon _{0}+\Delta _{\infty }\geq \left( \mu
+\zeta _{\infty }\right) \mathbf{1}\ .
\end{equation*}

\item[\emph{(iii)}] If additionally 
\begin{equation}
\liminf_{t\rightarrow 0}\Vert \mathfrak{B}_{t}\Vert _{\mathrm{op}}=0
\label{fghfghfghgf0}
\end{equation}%
then, for any $\delta \in \mathbb{R}^{+}$, there is $s\in \mathbb{R}_{0}^{+}$
such that%
\begin{equation*}
\left\Vert W_{t,s}\right\Vert _{\mathrm{op}}=\left\Vert W_{t,s}^{\top
}\right\Vert _{\mathrm{op}}\leq \mathrm{e}^{-2\left( \alpha -\delta \right)
\left( t-s\right) }\ ,\qquad t\in \lbrack s,\infty )\ ,
\end{equation*}%
with $\alpha \doteq \zeta _{\infty }+\mu $.
\end{enumerate}
\end{lemma}

\begin{proof}
Note that the assumptions of the lemma include all the conditions of Theorem %
\ref{section extension gap}. If we additionally assume from now that $\Delta
\equiv (\Delta _{t})_{t\geq 0}$ converges in the weak operator topology to $%
\Delta _{\infty }\in \mathcal{B}\left( \mathfrak{h}\right) $, then we extend
the definition (\ref{def D frac}) of $(\mathfrak{D}_{t})_{t\geq 0}$ to $%
t=\infty $ via 
\begin{equation}
\mathfrak{D}_{\infty }\doteq \Upsilon _{\infty }-\mu \mathbf{1}\doteq
\Upsilon _{0}+\Delta _{\infty }-\mu \mathbf{1}\ ,  \label{ertertrt}
\end{equation}%
defined on the same domain $\mathcal{D}(\Upsilon _{0})$. In this case we
also observe from the flow (\ref{flow equation-quadratic deltabis}) and the
weak convergence of $\Delta \equiv (\Delta _{t})_{t\geq 0}$ to some bounded
operator $\Delta _{\infty }\in \mathcal{B}\left( \mathfrak{h}\right) $ that
the mapping $t\mapsto \left\langle \varphi ,D_{t}D_{t}^{\ast }\varphi
\right\rangle _{\mathfrak{h}}$ defined on $\mathbb{R}_{0}^{+}$ is continuous
and integrable and, for any $\varphi \in \mathfrak{h}$, 
\begin{equation*}
\liminf_{t\rightarrow 0}\Vert D_{t}\varphi \Vert _{\mathfrak{h}}=0\ ,
\end{equation*}%
otherwise the positive quantity 
\begin{equation}
\left\langle \varphi ,\Delta _{t}\varphi \right\rangle _{\mathfrak{h}%
}=16\int_{0}^{t}\left\langle \varphi ,D_{\tau }D_{\tau }^{\ast }\varphi
\right\rangle _{\mathfrak{h}}\mathrm{d}\tau  \label{positive quantity 0}
\end{equation}%
would diverge to $\infty $ as $t\rightarrow \infty $. Above, one uses
Fubini's theorem to put the scalar product in the integrand as well as the
flow (\ref{flow equation-quadratic deltabis}). In particular, for any $%
\varphi \in \mathfrak{h}$, there is a subsequence $\left( t_{n}\right)
_{n\in \mathbb{N}}$\ such that%
\begin{equation*}
\lim_{n\rightarrow \infty }\Vert D_{t_{n}}\varphi \Vert _{\mathfrak{h}}=0\ .
\end{equation*}%
If $\mathfrak{h}$ is separable then we can take an orthonormal basis $%
\left\{ \psi _{k}\right\} _{k=1}^{\infty }\subseteq \mathfrak{h}$\ and by a
standard argument with a so-called diagonal subsequence, we can choose the
sequence $(t_{n})_{n\in \mathbb{N}}$ such that, for any $k\in \mathbb{N}$, 
\begin{equation}
\lim_{n\rightarrow \infty }\Vert D_{t_{n}}\psi _{k}\Vert _{\mathfrak{h}}=0\ .
\label{sdsdsdsfdgdf}
\end{equation}%
We define the linear span 
\begin{equation*}
\mathcal{D}\doteq \mathrm{span}\left\{ \psi _{k}\right\} _{k=1}^{\infty
}\subseteq \mathfrak{h}
\end{equation*}%
of the family $\left\{ \psi _{k}\right\} _{k=1}^{\infty }$. It is a dense
set of $\mathfrak{h}$. By assumption, we have 
\begin{equation*}
\sup_{n\in \mathbb{N}}\Vert D_{t_{n}}\varphi \Vert _{\mathfrak{h}}<\infty \
,\qquad \varphi \in \mathfrak{h}\ ,
\end{equation*}%
which, combined with (\ref{sdsdsdsfdgdf}) and the density of $\mathcal{D}$,
in turn implies that%
\begin{equation}
\lim_{n\rightarrow \infty }\Vert D_{t_{n}}\varphi \Vert _{\mathfrak{h}}=0\
,\qquad \varphi \in \mathfrak{h}\ .  \label{sdasdasdsad}
\end{equation}%
Additionally, since 
\begin{equation}
\Upsilon _{t}\geq \Upsilon _{0}\geq -\left( \mu -\varepsilon \right) \mathbf{%
1}\text{ },\qquad t\in \mathbb{R}_{0}^{+}\ ,  \label{tttrrrr}
\end{equation}%
we deduce from (\ref{def D frac})--(\ref{def D frac2}) and (\ref{ertertrt})
that, for any $\varphi \in \mathfrak{h}$ and $n\in \mathbb{N}$, 
\begin{equation}
\left\vert \left\langle \varphi ,\left( \mathfrak{D}_{\infty }-\mathfrak{D}%
_{t_{n}}\right) \varphi \right\rangle _{\mathfrak{h}}\right\vert \leq
\left\langle \varphi ,\left( \Delta _{\infty }-\Delta _{t_{n}}\right)
\varphi \right\rangle _{\mathfrak{h}}+4\varepsilon ^{-1}\left\langle \varphi
,D_{t_{n}}D_{t_{n}}^{\ast }\varphi \right\rangle _{\mathfrak{h}}\ .
\label{trytry}
\end{equation}%
Since the sequence $(t_{n})_{n\geq 0}$ satisfies (\ref{sdsdsdsfdgdf}), for
any fixed $\varphi \in \mathfrak{h}$ the last upper bound vanishes as ${n}%
\rightarrow \infty $, by weak and strong convergence of $\Delta $ and $D$
(along the subsequence $(t_{n})_{n\geq 0}$, see (\ref{sdasdasdsad})) towards 
$\Delta _{\infty }$ and zero, respectively. Together with Theorem \ref%
{theorem sympa0}, Assertion (ii) then follows.

Finally, we observe from (\ref{def D frac}) and (\ref{map}) that%
\begin{equation*}
\Upsilon _{t}=\mathfrak{D}_{t}+\mu \mathbf{1}-4\mathfrak{B}_{t}\geq \left(
\zeta \left( t\right) +\mu -4\Vert \mathfrak{B}_{t}\Vert _{\mathrm{op}%
}\right) \mathbf{1}\ .
\end{equation*}%
In particular, if $\alpha \doteq \zeta _{\infty }+\mu $ and (\ref%
{fghfghfghgf0}) holds true then, for any $\delta \in \mathbb{R}^{+}$, there
is $s\in \mathbb{R}_{0}^{+}$ such that $\Upsilon _{s}=\Upsilon _{s}^{\ast
}\geq \left( \alpha -\delta \right) \mathbf{1}$, which, combined with (\ref%
{petit inequalitybisbis}) for $\mu =\delta -\alpha $ and $T_{\max }=\infty $
(Theorem \ref{section extension gap}), yields Assertion (iii).
\end{proof}

Lemma \ref{lemma asymptotics1} is based on the preliminary knowledge that
the flow converges in the weak sense. Such a property can be delicate to
analyze, as it is demonstrated in the hyperbolic case \cite{bach-bru-memo}.
This is particularly true here when the positivity of $\Upsilon _{0}$ does
not hold. The key arguments to understand the infinite-time limit of the
flow result from its elliptic nature proven in Section \ref{Section tech5}.
For instance, Corollary \ref{lemma constante of motion copy(4)} already
gives an asymptotics on the operator-valued function $D$ in the bounded
case. When $\mathfrak{D}_{0}\geq 0$ and $D_{0}\in \mathcal{L}^{2}(\mathfrak{h%
})$, this can even be strongly strengthened. In fact, one has the following
integrability of $D$:

\begin{corollary}[Square-integrability of the operator-valued function $D$]
\label{Corollaire sympa}\mbox{}\newline
Take a self-adjoint operator $\Upsilon _{0}=\Upsilon _{0}^{\ast }$ and $\mu
\in \mathbb{R}$ such that $\Upsilon _{0}\geq -\left( \mu -\varepsilon
\right) \mathbf{1}$ for some $\varepsilon \in \mathbb{R}^{+}$. Let $%
D_{0}=\pm D_{0}^{\top }\in \mathcal{B}\left( \mathfrak{h}\right) $ ($%
D_{0}\neq 0$).

\begin{enumerate}
\item[\emph{(i)}] If $\Upsilon _{0}\in \mathcal{B}\left( \mathfrak{h}\right) 
$ then, for any $\varphi \in \mathfrak{h}$,%
\begin{equation*}
\int_{0}^{\infty }\left\langle \varphi ,D_{\tau }D_{\tau }^{\ast }\varphi
\right\rangle _{\mathfrak{h}}\mathrm{d}\tau \leq 4^{-2}\left( \left\Vert
\Upsilon _{0}\right\Vert _{\mathrm{op}}+4\left\Vert \mathfrak{B}%
_{0}\right\Vert _{\mathrm{op}}+\left\vert \mu \right\vert +2\mu \right)
<\infty \ .
\end{equation*}

\item[\emph{(ii)}] If $\mathfrak{D}_{0}\geq 0$ and $D_{0}\in \mathcal{L}^{2}(%
\mathfrak{h})$ then%
\begin{equation*}
\int_{0}^{\infty }\left\Vert D_{\tau }\right\Vert _{2}^{2}\mathrm{d}\tau
\leq 4^{-1}\Vert \mathfrak{B}_{0}\Vert _{1}<\infty \ .
\end{equation*}
\end{enumerate}
\end{corollary}

\begin{proof}
Assertion (i) is a consequence of Corollary \ref{lemma constante of motion
copy(4)}, by observing from (\ref{def D frac}) and (\ref{upper bound toto0})
that 
\begin{equation*}
\left\Vert \mathfrak{D}_{0}\right\Vert _{\mathrm{op}}\leq \left\Vert
\Upsilon _{0}\right\Vert _{\mathrm{op}}+4\left\Vert \mathfrak{B}%
_{0}\right\Vert _{\mathrm{op}}+\left\vert \mu \right\vert <\infty \ ,
\end{equation*}%
provided of course that $\Upsilon _{0}\in \mathcal{B}\left( \mathfrak{h}%
\right) $. Note that 
\begin{equation}
\int_{0}^{\infty }\left\langle \varphi ,D_{\tau }D_{\tau }^{\ast }\varphi
\right\rangle _{\mathfrak{h}}\mathrm{d}\tau \doteq \lim_{t\rightarrow \infty
}\int_{0}^{t}\left\langle \varphi ,D_{\tau }D_{\tau }^{\ast }\varphi
\right\rangle _{\mathfrak{h}}\mathrm{d}\tau =\sup_{t\in \mathbb{R}%
_{0}^{+}}\int_{0}^{t}\left\langle \varphi ,D_{\tau }D_{\tau }^{\ast }\varphi
\right\rangle _{\mathfrak{h}}\mathrm{d}\tau <\infty \ ,\qquad \varphi \in 
\mathfrak{h}\ ,  \label{sdsdsds}
\end{equation}%
thanks to the monotone convergence theorem. Assertion (ii) results from
Theorem \ref{Theorem sympa} (ii). Again, by the monotone convergence
theorem, observe that 
\begin{equation*}
\int_{0}^{\infty }\left\Vert D_{\tau }\right\Vert _{2}^{2}\mathrm{d}\tau
\doteq \lim_{t\rightarrow \infty }\int_{0}^{t}\left\Vert D_{\tau
}\right\Vert _{2}^{2}\mathrm{d}\tau =\sup_{t\in \mathbb{R}%
_{0}^{+}}\int_{0}^{t}\left\Vert D_{\tau }\right\Vert _{2}^{2}\mathrm{d}\tau
<\infty \ .
\end{equation*}
\end{proof}

Corollary \ref{Corollaire sympa} implies the existence of a positive limit
operator 
\begin{equation}
\Upsilon _{\infty }\doteq \Upsilon _{0}+\Delta _{\infty }\doteq \Upsilon
_{0}+16\int_{0}^{+\infty }D_{\tau }D_{\tau }^{\ast }\mathrm{d}\tau \ ,
\label{limit operator}
\end{equation}%
defined on the domain $\mathcal{D}(\Upsilon _{0})$ and using additionally
Lemma \ref{lemma asymptotics1} we finally arrive at the following results,
presented in three different propositions:

\begin{proposition}[Limit operators -- Bounded case]
\label{limit prop1}\mbox{}\newline
Take a self-adjoint operator $\Upsilon _{0}=\Upsilon _{0}^{\ast }$ and $\mu
\in \mathbb{R}$ such that $\Upsilon _{0}\geq -\left( \mu -\varepsilon
\right) \mathbf{1}$ for some $\varepsilon \in \mathbb{R}^{+}$. Let $%
D_{0}=\pm D_{0}^{\top }\in \mathcal{B}\left( \mathfrak{h}\right) $ ($%
D_{0}\neq 0$). If $\Upsilon _{0}\in \mathcal{B}\left( \mathfrak{h}\right) $
then, as $t\rightarrow \infty $, $D$ and $\Delta $ converge respectively in
the strong operator topology to zero and in the weak operator topology to an
operator $\Delta _{\infty }\in \mathcal{B}\left( \mathfrak{h}\right) $ such
that 
\begin{equation*}
\Upsilon _{\infty }\doteq \Upsilon _{0}+\Delta _{\infty }\geq \left( \mu
+\zeta _{\infty }\right) \mathbf{1}\geq -\mu \mathbf{1}\ ,
\end{equation*}%
the real number $\zeta _{\infty }$ being defined by (\ref{seeddd}).
\end{proposition}

\begin{proof}
From (a version of) the polarization identity for (the sesquilinear form) $%
\left\langle \cdot ,\Delta _{t}(\cdot )\right\rangle $, for all $\varphi
,\psi \in \mathfrak{h}$ and $t\in \mathbb{R}_{0}^{+}$, 
\begin{equation*}
\left\langle \varphi ,\Delta _{t}\psi \right\rangle =\frac{1}{4}%
\sum\limits_{n=1}^{4}(-i)^{n}\left\langle \varphi +i^{n}\psi ,\Delta
_{t}\left( \varphi +i^{n}\psi \right) \right\rangle _{\mathfrak{h}}\text{ }.
\end{equation*}%
Therefore, it suffices to combine Corollary \ref{Corollaire sympa} (i) and 
\cite[Theorem VI.1]{ReedSimon} with the last equality to get the weak
convergence of the operator-valued function $\Delta \subseteq \mathcal{B}%
\left( \mathfrak{h}\right) $ to some bounded operator $\Delta _{\infty }\in 
\mathcal{B}\left( \mathfrak{h}\right) $. Moreover, by the Cauchy-Schwarz
inequality, for any $\varphi \in \mathfrak{h}$ and $t\in \mathbb{R}_{0}^{+}$%
, 
\begin{equation*}
\left\vert \partial _{t}\left\langle \varphi ,D_{t}D_{t}^{\ast }\varphi
\right\rangle _{\mathfrak{h}}\right\vert \leq 8\left( \left\Vert \Delta
_{\infty }\right\Vert _{\mathrm{op}}+\left\Vert \Upsilon _{0}\right\Vert _{%
\mathrm{op}}\right) \left\Vert D_{t}\right\Vert _{\mathrm{op}}^{2}\ .
\end{equation*}%
Since one has from Corollary \ref{lemma constante of motion copy(4)} that%
\begin{equation*}
\left\Vert D_{t}\right\Vert _{\mathrm{op}}^{2}=\left\Vert D_{t}D_{t}^{\ast
}\right\Vert _{\mathrm{op}}\leq \left( \left\Vert \Delta _{\infty
}\right\Vert _{\mathrm{op}}+\left\Vert \Upsilon _{0}\right\Vert _{\mathrm{op}%
}+\left\vert \mu \right\vert \right) \sup_{t\in \mathbb{R}%
_{0}^{+}}\left\Vert \mathfrak{B}_{t}\right\Vert _{\mathrm{op}}<\infty \ ,
\end{equation*}%
we conclude that the mapping $t\mapsto \left\langle \varphi
,D_{t}D_{t}^{\ast }\varphi \right\rangle _{\mathfrak{h}}$ defined on $%
\mathbb{R}_{0}^{+}$ is uniformly continuous and integrable, thanks to (\ref%
{flow equation-quadratic deltabis}) and the weak convergence of $\Delta
\equiv (\Delta _{t})_{t\geq 0}$ to some bounded operator $\Delta _{\infty
}\in \mathcal{B}\left( \mathfrak{h}\right) $. As a consequence, for any $%
\varphi \in \mathfrak{h}$, 
\begin{equation*}
\lim_{t\rightarrow 0}\Vert D_{t}\varphi \Vert _{\mathfrak{h}}=0\ ,
\end{equation*}%
otherwise the positive quantity (\ref{positive quantity 0}) would diverge to 
$\infty $ as $t\rightarrow \infty $. The proposition then follows, thanks to
Lemma \ref{lemma asymptotics1} (ii).
\end{proof}

\begin{proposition}[Limit operators -- Positive case with gap]
\label{limit prop2}\mbox{}\newline
Take a self-adjoint operator $\Upsilon _{0}=\Upsilon _{0}^{\ast }$ and $\mu
\in \mathbb{R}^{-}$ such that $\Upsilon _{0}\geq -\left( \mu -\varepsilon
\right) \mathbf{1}$ for some $\varepsilon \in \mathbb{R}^{+}$. Let $%
D_{0}=\pm D_{0}^{\top }\in \mathcal{B}\left( \mathfrak{h}\right) $ ($%
D_{0}\neq 0$). Then, as $t\rightarrow \infty $, $D$ and $\Delta $
exponentially converge in the norm topology respectively to zero and an
operator $\Delta _{\infty }\in \mathcal{B}\left( \mathfrak{h}\right) $ such
that 
\begin{equation}
\Upsilon _{\infty }\doteq \Upsilon _{0}+\Delta _{\infty }\geq \left\vert \mu
\right\vert \mathbf{1}\ .  \label{semibounded20}
\end{equation}
\end{proposition}

\begin{proof}
The proof of this proposition, which refers to the case of positive,
possibly unbounded, operators $\Upsilon _{0}$ with spectral gap, is
straightforward: By (\ref{sdsdsdsdsdsdsdsdsdsdsd}) for $T_{\max }=\infty $
(Theorem \ref{section extension gap}), $D$ exponentially converges in the
norm topology to zero. Using the triangle inequality and (\ref{flow
equation-quadratic deltabis}), one directly obtains that 
\begin{equation}
\left\Vert \Delta _{t}-\Delta _{s}\right\Vert _{\mathrm{op}}\leq
16\int_{s\wedge t}^{s\vee t}\left\Vert D_{\tau }\right\Vert _{\mathrm{op}%
}^{2}\mathrm{d}\tau \ ,\qquad s,t\in \mathbb{R}_{0}^{+}\ ,  \label{ssss}
\end{equation}%
where $s\wedge t$\ and $s\vee t$ stand respectively for the minimum and the
maximum of real numbers $s,t$. Therefore, being a Cauchy sequence in the
Banach space $\mathcal{B}\left( \mathfrak{h}\right) $, $\Delta $ converges
in the norm topology to some operator $\Delta _{\infty }\in \mathcal{B}(%
\mathfrak{h})$. This convergence is exponentially fast, thanks again to (\ref%
{sdsdsdsdsdsdsdsdsdsdsd}) and (\ref{ssss}) for $t=T_{\max }=\infty $. Note
that $\Upsilon _{\infty }\geq \left\vert \mu \right\vert \mathbf{1}$ results
from Lemma \ref{lemma asymptotics1} (ii) , since 
\begin{equation*}
\zeta _{\infty }\doteq \lim_{t\rightarrow \infty }\zeta \left( t\right) \geq
-2\mu \ .
\end{equation*}%
(Thanks to the convergence of $D$ in the norm topology, $t\mapsto \Vert
D_{t}\Vert _{\mathrm{op}}$ is of course bounded on $\mathbb{R}_{0}^{+}$.)
\end{proof}

\begin{proposition}[Limit operators -- Non-positive case]
\label{limit prop3}\mbox{}\newline
Take a self-adjoint operator $\Upsilon _{0}=\Upsilon _{0}^{\ast }$ and $\mu
\in \mathbb{R}\backslash \{0\}$ such that $\Upsilon _{0}\geq -\left( \mu
-\varepsilon \right) \mathbf{1}$ for some $\varepsilon \in \mathbb{R}^{+}$.
Let $D_{0}=\pm D_{0}^{\top }\in \mathcal{L}^{2}(\mathfrak{h})$ ($D_{0}\neq 0$%
) and assume that $\mathfrak{D}_{0}\geq 0$. Then, as $t\rightarrow \infty $, 
$D$ exponentially converges in the Hilbert-Schmidt topology (i.e., in $%
\mathcal{L}^{2}(\mathfrak{h})$) to zero, while $\Delta $ exponentially
converges in the trace norm topology\footnote{%
Note that the trace norm topology is always stronger than the
Hilbert-Schmidt topology. See, e.g., \cite[Theorem VI.22 (d)]{ReedSimon}.}
(i.e., in $\mathcal{L}^{1}(\mathfrak{h})$) to an operator $\Delta _{\infty
}\in \mathcal{L}^{1}(\mathfrak{h})$ satisfying 
\begin{equation*}
\Upsilon _{\infty }\doteq \Upsilon _{0}+\Delta _{\infty }\geq \left\vert \mu
\right\vert \mathbf{1}\ .
\end{equation*}
\end{proposition}

\begin{proof}
Under the assumptions of the proposition, $\Delta $ and $D$ can be given by
Theorem \ref{Corollary existence}. Then, Corollary \ref{Corollaire sympa}
(ii) implies the convergence of $D$ to zero in the Hilbert-Schmidt topology,
at least along a subsequence, i.e., 
\begin{equation}
\liminf_{t\rightarrow 0}\Vert D_{t}\Vert _{2}=0\ .  \label{pppp}
\end{equation}%
Since Fubini's theorem and the flow (\ref{flow equation-quadratic deltabis})
yield 
\begin{equation}
\left\Vert \Delta _{t}-\Delta _{s}\right\Vert _{1}=16\int_{s\wedge t}^{s\vee
t}\left\Vert D_{\tau }\right\Vert _{2}^{2}\mathrm{d}\tau \ ,\qquad s,t\in 
\mathbb{R}_{0}^{+}\ ,  \label{ssdssdsdssds}
\end{equation}%
(where $s\wedge t$\ and $s\vee t$ stand respectively for the minimum and the
maximum of real numbers $s,t$), we also deduce from Corollary \ref%
{Corollaire sympa} (ii) that $\Delta $ converges in the trace norm topology
to $\Delta _{\infty }\in \mathcal{L}^{1}(\mathfrak{h})$. By (\ref{upper
bound toto0}) and (\ref{pppp}) note in this case that 
\begin{equation*}
\liminf_{t\rightarrow 0}\left\Vert \mathfrak{B}_{t}\right\Vert _{\mathrm{op}%
}=0\ .
\end{equation*}%
(In fact one can even use the trace norm). So, by Lemma \ref{lemma
asymptotics1} (iii), for any $\delta \in \mathbb{R}^{+}$, there is $s\in 
\mathbb{R}_{0}^{+}$ such that 
\begin{equation}
\left\Vert W_{t,s}\right\Vert _{\mathrm{op}}=\left\Vert W_{t,s}^{\top
}\right\Vert _{\mathrm{op}}\leq \mathrm{e}^{-2\left( \alpha -\delta \right)
\left( t-s\right) }\ ,\qquad t\in \lbrack s,\infty )\ ,  \label{fdgdfg}
\end{equation}%
with $\alpha \doteq \zeta _{\infty }+\mu $. If $\mathfrak{D}_{0}\geq 0$ then
Theorem \ref{theorem sympa0} leads to%
\begin{equation*}
\zeta _{\infty }\doteq \lim_{t\rightarrow \infty }\zeta \left( t\right) \geq
\sup \left\{ 0,-2\mu \right\} \ ,
\end{equation*}%
which in turn yields the inequality $\alpha \geq \left\vert \mu \right\vert
>0$. In particular, $\Upsilon _{\infty }\geq \left\vert \mu \right\vert 
\mathbf{1}$, see Lemma \ref{lemma asymptotics1} (ii) observing that (\ref%
{fdgdfg}) ensures the uniform boundedness of $\Vert D_{t}\Vert _{\mathrm{op}%
} $ on $\mathbb{R}_{0}^{+}$. Moreover, the combination of (\ref%
{D_explicit_solutionbis}) and (\ref{fdgdfg}) tells us that, for any $\delta
\in (0,\left\vert \mu \right\vert )$, there is $s\in \mathbb{R}_{0}^{+}$
such that 
\begin{equation}
\left\Vert D_{t}\right\Vert _{2}\leq \mathrm{e}^{-4\left( \left\vert \mu
\right\vert -\delta \right) \left( t-s\right) }\ ,\qquad t\in \lbrack
s,\infty )\ .  \label{dddddddddddd}
\end{equation}%
In other words, $D$ exponentially converges in the Hilbert-Schmidt topology
to zero and, as a consequence, $\Delta $ exponentially converges in the
trace norm topology to some trace-class operator $\Delta _{\infty }$, thanks
to (\ref{ssdssdsdssds}).
\end{proof}

 Under the conditions of Proposition \ref{limit prop3}, the limit
operator is always positive with a spectral gap if $\mu \neq 0$.  This
can be used to obtain the limit operator $\Upsilon _{\infty }$\ in the
special case where $\Upsilon _{0}D_{0}=D_{0}\Upsilon _{0}^{\top }$, see
Theorem \ref{theorem constante of motion} (ii) and Remark \ref{theorem
constante of motion copy(1)}.

\begin{lemma}[Explicit form of the limit operator -- Commutative case]
\label{lemma square root}\mbox{}\newline
Take a self-adjoint operator $\Upsilon _{0}=\Upsilon _{0}^{\ast }$ and $\mu
\in \mathbb{R}\backslash \{0\}$ such that $\Upsilon _{0}\geq -\left( \mu
-\varepsilon \right) \mathbf{1}$ for some $\varepsilon \in \mathbb{R}^{+}$.
Let $D_{0}=\pm D_{0}^{\top }\in \mathcal{B}\left( \mathfrak{h}\right) $ ($%
D_{0}\neq 0$).  Assume $\Upsilon _{0}D_{0}=D_{0}\Upsilon _{0}^{\top }$
and that  $D$ and $\Delta $ converge in the norm topology respectively
to zero and an operator $\Delta _{\infty }\in \mathcal{B}\left( \mathfrak{h}%
\right) $, as $t\rightarrow \infty $.  Then, $\Upsilon _{\infty
}\doteq \Upsilon _{0}+\Delta _{\infty }$ satisfies on the domain $\mathcal{D}%
(\Upsilon _{0}^{2})$ the equality  
\begin{equation*}
\Upsilon _{\infty }^{2}=\Upsilon _{0}^{2}+4D_{0}D_{0}^{\ast }\ .
\end{equation*}%
Additionally, if $\Upsilon _{\infty }\geq 0$ then 
\begin{equation*}
\Upsilon _{\infty }=\sqrt{\Upsilon _{0}^{2}+4D_{0}D_{0}^{\ast }}\ .
\end{equation*}
\end{lemma}

\begin{proof}
 If $\Upsilon _{0}D_{0}=D_{0}\Upsilon _{0}^{\top }$ then we
deduce from Theorem \ref{theorem constante of motion} (ii) that, for any $%
t\in \mathbb{R}_{0}^{+}$, 
\begin{equation}
\Upsilon _{t}D_{t}=D_{t}\Upsilon _{t}^{\top }\quad \text{and}\quad \left(
\Upsilon _{t}^{2}+4D_{t}D_{t}^{\ast }-\Upsilon _{s}^{2}-4D_{s}D_{s}^{\ast
}\right) \left( \Upsilon _{0}^{\top }+\mu \right) ^{-1}=0\ ,
\label{integrability norm0}
\end{equation}%
where $\Upsilon _{t}\doteq \Upsilon _{0}+\Delta _{t}$. It follows that, for
any $t\in \mathbb{R}_{0}^{+}$,%
\begin{equation*}
\Upsilon _{\infty }\Upsilon _{t}-\left( \Upsilon _{0}^{2}+4D_{0}D_{0}^{\ast
}\right) =\left( \Delta _{\infty }-\Delta _{t}\right) \Upsilon
_{t}-4D_{t}D_{t}^{\ast }
\end{equation*}%
on the domain $\mathcal{D}(\Upsilon _{0}^{2})$.  If $D$ and $\Delta $
converge in the norm topology respectively to zero and an operator $\Delta
_{\infty }\in \mathcal{B}\left( \mathfrak{h}\right) $ then, for any $\varphi
\in \mathcal{D}(\Upsilon _{0}^{2})$, 
\begin{equation}
\lim_{t\rightarrow \infty }\left\Vert \left( \Upsilon _{\infty }\Upsilon
_{t}-\left( \Upsilon _{0}^{2}+4D_{0}D_{0}^{\ast }\right) \right) \varphi
\right\Vert =0\ .  \label{sdsdsdsds}
\end{equation}%
Meanwhile, since $\Upsilon _{0}$ is a self-adjoint operator that is bounded
from below, $\mathcal{D}(\Upsilon _{0}^{2})\subseteq \mathcal{D}(\Upsilon
_{0})$ and for any $\varphi \in \mathcal{D}(\Upsilon _{0}^{2})$, $\Upsilon
_{t}\varphi $ also converges to $\Upsilon _{\infty }\varphi $, as $%
t\rightarrow \infty $, since $\Delta $ converges by assumption in the norm
topology to $\Delta _{\infty }\in \mathcal{B}\left( \mathfrak{h}\right) $.
Combining this last limit with (\ref{sdsdsdsds}) we conclude that 
\begin{equation*}
\Upsilon _{\infty }^{2}=\Upsilon _{0}^{2}+4D_{0}D_{0}^{\ast }\ ,
\end{equation*}%
on the domain $\mathcal{D}(\Upsilon _{0}^{2})$, because $\Upsilon _{\infty
}^{2}$ is a closed operator (being self-adjoint). Using now the functional
calculus, if $\Upsilon _{\infty }\geq 0$ then 
\begin{equation}
\Upsilon _{\infty }\doteq \Upsilon _{0}+\Delta _{\infty }=\sqrt{\Upsilon
_{0}^{2}+4D_{0}D_{0}^{\ast }}  \label{sdddfdf}
\end{equation}%
at least on the domain $\mathcal{D}(\Upsilon _{0}^{2})\subseteq \mathcal{D}%
(\Upsilon _{0})$, which is a core for the self-adjoint operator $\Upsilon
_{\infty }$, thanks to the assumption $\Delta _{\infty }\in \mathcal{B}%
\left( \mathfrak{h}\right) $ as well as \cite[Chapter II, Proposition 1.8]%
{EngelNagel} and \cite[Theorem VIII.7]{ReedSimon}. Observe that $\Upsilon
_{0}D_{0}=D_{0}\Upsilon _{0}^{\top }$ yields $[\Upsilon
_{0}^{2},D_{0}D_{0}^{\ast }]=0$, which implies by \cite[Proposition 5.27]%
{Konrad} that their so-called bounded transforms $\Upsilon _{0}^{2}\left( 
\mathbf{1}+\Upsilon _{0}^{4}\right) ^{-1/2}$ and $D_{0}D_{0}^{\ast }\left( 
\mathbf{1}+(D_{0}D_{0}^{\ast })^{2}\right) ^{-1/2}$ commute. By using the
general functional calculus for commuting self-adjoint operators, which
results in this case to \cite[Theorem 5.23]{Konrad}, we can use the
real-valued function $\left( x,y\right) \mapsto \sqrt{x+y}$ on $\mathbb{R}%
^{2}$ to define the operator $\sqrt{\Upsilon _{0}^{2}+4D_{0}D_{0}^{\ast }}$.
In addition, one verifies from \cite[Proposition 4.17 (ii)]{Konrad} that it
is self-adjoint and from \cite[Theorem 4.13 (iii)]{Konrad} that $\mathcal{D}%
(\Upsilon _{0}^{2})$ is a core of this operator. Consequently, (\ref{sdddfdf}%
) is an operator equality.
\end{proof}

Assumptions (\ref{integrability norm0}) are verified when $\Upsilon
_{0}D_{0}=D_{0}\Upsilon _{0}^{\top }$, see Theorem \ref{theorem constante of
motion} (ii) and Remark \ref{theorem constante of motion copy(1)}. The norm
convergence of the operator-valued functions $D$ and $\Delta $ is satisfied
under Conditions of Propositions \ref{limit prop2} and \ref{limit prop3}. In
this case we even prove the positivity of the limit operator $\Upsilon
_{\infty }$.

\subsection{Illustration of the Flow on Scalar Fields\label{Flows on Scalar
Fields}}

Given $\alpha ,\beta \in \mathbb{R}$, we introduce the two functions $f:%
\mathbb{R}_{0}^{+}\rightarrow \mathbb{R}$ and $g:\mathbb{R}%
_{0}^{+}\rightarrow i\mathbb{R}$ satisfying the system of differential
relations 
\begin{equation}
\left\{ 
\begin{array}{l}
\partial _{t}f\left( t\right) =16\left\vert g\left( t\right) \right\vert ^{2}
\\ 
\partial _{t}g\left( t\right) =4\left( \alpha -f\left( t\right) \right)
g\left( t\right)%
\end{array}%
\right.  \label{eq:flownum}
\end{equation}%
for any $t\in \mathbb{R}_{0}^{+}$, with initial conditions $f(0)=0$ and $%
g(0)=i\beta $. Existence and uniqueness of the solution is standard. For $%
\beta \neq 0$, it suffices here to invoke Theorem \ref{section extension gap}
with $\mathfrak{h}=\mathbb{C}$, $\Upsilon _{0}=-\alpha $ and $D_{0}=i\beta $%
. It is straightforward to check that 
\begin{equation*}
g\left( t\right) =i\beta \exp \left( 4\alpha t-4\int_{0}^{t}f\left( \tau
\right) \mathrm{d}\tau \right) \ ,\qquad t\in \mathbb{R}_{0}^{+}\ .
\end{equation*}%
See for instance (\ref{D_explicit_solutionbis}). Additionally observe that 
\begin{equation*}
\partial _{t}\left( \left( f\left( t\right) -\alpha \right) ^{2}+4\left\vert
g\left( t\right) \right\vert ^{2}\right) =0\ ,
\end{equation*}%
which leads to the (scalar) constant of motion: 
\begin{equation}
\left( f\left( t\right) -\alpha \right) ^{2}+4\left\vert g\left( t\right)
\right\vert ^{2}=\alpha ^{2}+4\beta ^{2}\ ,\qquad t\in \mathbb{R}_{0}^{+}\ .
\label{constant motion scalar}
\end{equation}%
Compare with Theorem \ref{theorem constante of motion}. It means that, for
all times $t\in \mathbb{R}_{0}^{+}$, $x=f\left( t\right) -\alpha $ and $%
y=\left\vert g\left( t\right) \right\vert $ always belong to the ellipse 
\begin{equation}
x^{2}+4y^{2}=\alpha ^{2}+4\beta ^{2}\ .  \label{constant motion scalar2}
\end{equation}%
In fact, assume without loss of generality that $\beta >0$. Then, one can
compute that the solution to (\ref{eq:flownum}) with initial conditions $%
f(0)=0$ and $g(0)=i\beta $ is explicitly given for any time $t\in \mathbb{R}%
_{0}^{+}$ by%
\begin{eqnarray*}
f\left( t\right) &=&\alpha +c\frac{\left( c+\alpha \right) \left( 1-\mathrm{e%
}^{-8ct}\right) -2\alpha }{\left( c+\alpha \right) \left( 1+\mathrm{e}%
^{-8ct}\right) -2\alpha }\ , \\
g\left( t\right) &=&\frac{ic}{2}\sqrt{1-\left( \frac{\left( c+\alpha \right)
\left( 1-\mathrm{e}^{-8ct}\right) -2\alpha }{\left( c+\alpha \right) \left(
1+\mathrm{e}^{-8ct}\right) -2\alpha }\right) ^{2}}\ ,
\end{eqnarray*}%
where $c\doteq \sqrt{\alpha ^{2}+4\beta ^{2}}$.

\begin{figure}[tbp]
\centering
\par
\begin{tikzpicture}
\begin{axis}[
    axis lines = left,
    xlabel = \(x\),
    ylabel = {\(f(x)\)},
]
\addplot [
    domain=-0:2.5,
    samples=70,
    color=red,
]
 {0.99+1*(1.99*(1-exp(-8*x))-2*0.99)/(1.99*(1+exp(-8*x))-2*0.99};
\addlegendentry{\(f(x)\)}
\addplot [
    domain=.:2.5,
    samples=63,
    color=blue,
    ]
   {0.5*sqrt(1-((1.99*(1-exp(-8*1*\x))-2*0.99)/(1.99*(1+exp(-8*1*\x))-2*0.99))*((1.99*(1-exp(-8*1*\x))-2*0.99)/(1.99*(1+exp(-8*1*\x))-2*0.99)))};
\addlegendentry{\(-ig(x)\)}

\end{axis}
\end{tikzpicture}
\caption{Time evolution for $f$ and $g$, with parameters $\protect\alpha%
=0.99 $ and $\protect\beta = 0.07$.}
\label{fig}
\end{figure}
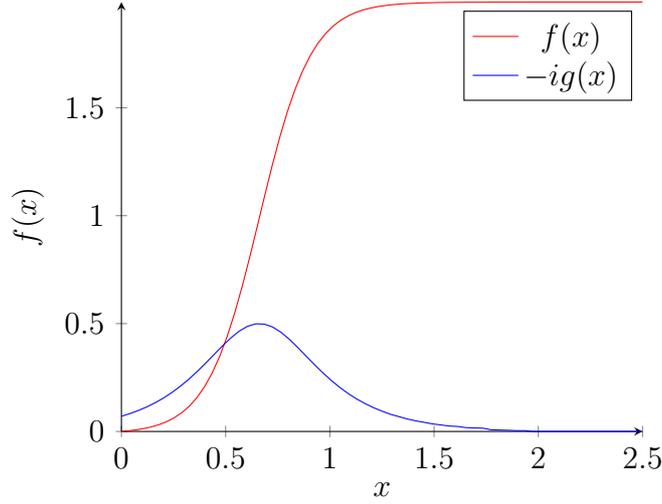

To illustrate the elliptic flows on scalar fields, one can take a look at
Figure \ref{fig}, where the parameters have been chosen to exhibit an
interesting behavior, as compared to the hyperbolic case studied in \cite%
{bach-bru-memo}: the function $g$ is not monotonous, but it still decays
exponentially to zero for sufficiently large times.

\bigskip

\noindent \textit{Acknowledgments:}This work is supported by the Basque
Government through the grant IT1615-22 and the BERC 2022-2025 program, by
the grant PID2020-112948GB-I00 funded by MCIN/AEI/10.13039/501100011033 and
by \textquotedblleft ERDF A way of making Europe\textquotedblright . We
thank Walter de Siquiera Pedra for valuable comments. We are also very
grateful to the two reviewers for their constructive suggestions and
interest in improving the paper. For instance, the generalization of the
result for Hilbert-Schmidt operators to (noncommutative) $L^{p}$--spaces
done in Theorem \ref{Corollary existence copy(1)} is a consequence of one
reviewer's observations.

\end{document}